\pdfoutput=1
\documentclass[11pt,a4paper]{article}

\usepackage{jheppub}
\usepackage{graphicx}
\usepackage{amsmath}
\usepackage{xspace}
\usepackage{xcolor}
\usepackage{float}
\usepackage{comment}
\usepackage{subcaption}
\usepackage{textcomp}
\usepackage{placeins}
\usepackage{booktabs,latexsym,amssymb,amsmath,bm,epsfig,psfrag,float,slashed}
\usepackage{color}
\usepackage[normalem]{ulem}

\newcommand{\eq}[1]{eq.~\eqref{eq:#1}}
\newcommand{\eqs}[2]{eqs.~\eqref{eq:#1} and \eqref{eq:#2}}
\renewcommand{\sec}[1]{sec.~\ref{sec:#1}}

\newcommand{\subsec}[1]{sec.~\ref{subsec:#1}}
\newcommand{\fig}[1]{Fig.~\ref{fig:#1}}
\newcommand{\figs}[2]{Figs.~\ref{fig:#1} and \ref{fig:#2}}

\newcommand{\ord}[1]{\mathcal{O}(#1)}

\newcommand{\eg}{\textit{e.g.}\ }
\newcommand{\ie}{\textit{i.e.}\ }
\newcommand{\df}{\mathrm{d}}

\newcommand{\as}{\alpha_{\rm s}}

\newcommand{\Tau}{\mathcal{T}}

\newcommand{\GeV}{\,\mathrm{GeV}}

\newcommand{\nn}{\nonumber}

\newcommand{\cP}{\mathcal{P}}

\newcommand{\cut}{\mathrm{cut}}

\newcommand{\FO}{\mathrm{FO}}

\newcommand{\NLO}{\mathrm{NLO}}

\newcommand{\nons}{\mathrm{nons}}

\newcommand{\NS}{\mathrm{NS}}

\newcommand{\one}{{(1)}}

\newcommand{\zj}{\mbox{$0$-jettiness}\xspace}

\newcommand{\nj}{\mbox{$N$-jettiness}\xspace}

\newcommand{\dsigMC}{\df\sigma^\textsc{mc}}

\newcommand{\obs}{X}

\newcommand{\geneva}{\textsc{Geneva}\xspace}
\newcommand{\mcatnlo}{\textsc{MC@NLO}\xspace}
\newcommand{\powheg}{\textsc{Powheg}\xspace}

\newcommand{\pythiaEight}{\textsc{Pythia8}\xspace}
\newcommand{\herwig}{\textsc{Herwig}\xspace}

\newcommand{\rivet}{\textsc{Rivet}\xspace}

\newcommand{\openloops}{\textsc{OpenLoops}\xspace}

\renewcommand{\matrix}{\textsc{Matrix}\xspace}
\newcommand{\diphox}{\texttt{DIPHOX}\xspace}
\newcommand{\form}{\texttt{FORM}\xspace}
\newcommand{\twogammamc}{\texttt{2gammaMC}\xspace}
\newcommand{\mcfm}{\texttt{MCFM}\xspace}
\newcommand{\cutemcfm}{\texttt{CuTe-MCFM}\xspace}
\newcommand{\twogammannlo}{\texttt{2$\gamma$NNLO}\xspace}
\newcommand{\resbos}{\texttt{RESBOS}\xspace}
\newcommand{\twogammares}{\texttt{2$\gamma$Res}\xspace}
\newcommand{\resolve}{\texttt{reSolve}\xspace}
\newcommand{\sherpa}{\textsc{Sherpa}\xspace}

\def\pt{\ensuremath{p_\mathrm{T}}\xspace}

\def\thetaPSiso{\ensuremath{\Theta^{\mathrm{PS}}}\xspace}

\def\thetaProj{\ensuremath{\Theta^{\mathrm{proj}}\xspace}}

\def\thetaBarProj{\ensuremath{\overline{\Theta}^{\mathrm{proj}}\xspace}}

\renewcommand{\thesection}{\arabic{section}}

\allowdisplaybreaks[2]

\definecolor{caribbeangreen}{rgb}{0.0, 0.8, 0.6}

\newcommand{\rescaleoneplot}{0.49\columnwidth}
\newcommand{\rescaletwoplots}{0.49\textwidth}
\newcommand{\hspacebetweentwoplots}{\hspace*{.02\textwidth}}
\newcommand{\vspacebetweentwoplots}{2ex}
\newcommand{\rescalethreeplots}{0.32\textwidth}
\newcommand{\hspacebetweenthreeplots}{\hspace*{.02\textwidth}}
\newcommand{\vspacebetweenthreeplots}{2ex}
\newcommand{\spaceabovefigurecaption}{\vspace*{-2ex}}
\newcommand{\spacebelowfigurecaption}{\vspace*{0ex}}
\makeatletter
\g@addto@macro\bfseries{\boldmath}
\makeatother

\begin{document}


\title{Precise predictions for photon pair production matched to parton showers in GENEVA}

\author[a]{Simone Alioli,}

\author[a]{Alessandro Broggio,}

\author[a]{Alessandro Gavardi,}

\author[a]{Stefan Kallweit,}

\author[a]{\quad\quad Matthew A.~Lim,}

\author[a]{Riccardo Nagar,}

\author[a]{Davide Napoletano}

\author[b]{and Luca Rottoli}

\affiliation[a]{Universit\`{a} degli Studi di Milano-Bicocca \& INFN
  Sezione di Milano-Bicocca, Piazza della Scienza 3, Milano 20126,
  Italy\vspace{0.5ex}}

\affiliation[b]{Physik Institut, Universit\"at Z\"urich,
  Winterthurerstrasse 190, 8057 Z\"urich, Switzerland}

\emailAdd{simone.alioli@unimib.it, alessandro.broggio@unimib.it,
  a.gavardi@campus.unimib.it, stefan.kallweit@unimib.it,
  matthew.lim@unimib.it, riccardo.nagar@unimib.it,
  davide.napoletano@unimib.it, luca.rottoli@physik.uzh.ch}

\preprint{ZU-TH 37/20}

\date{\today}

\abstract{ We present a new calculation for the production of isolated
  photon pairs at the LHC with NNLL$^\prime_{\Tau_0}$+NNLO
  accuracy. This is the first implementation within the \geneva Monte
  Carlo framework of a process with a nontrivial Born-level definition
  which suffers from QED singularities. Throughout the computation we
  use a smooth-cone isolation algorithm to remove such
  divergences. The higher-order resummation of the \zj resolution
  variable $\Tau_0$ is based on a factorisation formula derived within
  Soft-Collinear Effective Theory which predicts all of the singular,
  virtual and real NNLO corrections.  Starting from this precise
  parton-level prediction and by employing the \geneva method, we
  provide fully showered and hadronised events using \pythiaEight,
  while retaining the NNLO QCD accuracy for observables which are
  inclusive over the additional radiation. We compare our final
  predictions to LHC data at 7 TeV and find good agreement.  }

\maketitle
\flushbottom

\section{Introduction}
\label{sec:intro}

The production of two isolated photons is one of the most interesting
processes to study at the Large Hadron Collider (LHC), both to further
test the Standard Model (SM) and to search for new exotic signatures,
\eg by looking for heavy resonances in the diphoton invariant mass
spectrum~\cite{TheATLAScollaboration:2015mdt,CMS:2015dxe}.  Precise
experimental measurements are available due to the clean final state
and the relatively high rate of production for this process.  Efforts
in the study of the diphoton final state were especially boosted by
the discovery of a Higgs boson decaying into two photons at the
LHC~\cite{Aad:2012tfa,Chatrchyan:2012ufa}. For all these reasons,
recent experimental analyses of diphoton production have been carried
out by the ATLAS experiment both at 7
TeV~\cite{Aad:2011mh,Aad:2012tba}, 8 TeV~\cite{Aaboud:2017vol} and 13
TeV~\cite{ATLAS:2020qqv}, and by CMS at 7
TeV~\cite{Chatrchyan:2011qt,Chatrchyan:2014fsa}.  Given the relevance
of this process for the LHC experimental analyses, precise
calculations including higher-order QCD corrections are required from
the theoretical side.

In this paper we study the production of ``prompt" photon pairs which
are directly produced in the hard scattering interaction. A second
production mechanism involves the radiation of photons in a jet during
the fragmentation process. This second mechanism can be suppressed by
isolating the photons from the final-state hadrons with either the
fixed-cone or the smooth-cone (\textit{a.k.a.}\ Frixione) isolation
procedure~\cite{Frixione:1998jh}.\footnote{A summary of the different
  isolation procedures can be found in \eg Ref.~\cite{Chen:2019zmr}.}
The fixed-cone isolation is widely used in the experimental analyses
due to the simplicity of its implementation. It cannot, however,
completely eliminate the fragmentation contribution without spoiling
infrared (IR) safety. The smooth-cone isolation can achieve this in an
IR-safe manner, thus simplifying the computation of the radiative
corrections.

Next-to-leading-order (NLO) QCD corrections for the process
$pp/p\bar{p}\to \gamma \gamma +X$ were first computed and implemented
in the fully differential Monte Carlo program
\diphox~\cite{Binoth:1999qq} which included the calculation of both
the ``direct" and the fragmentation contributions to the cross
section.  The leading-order~(LO) calculation~\cite{Dicus:1987fk} of
the gluon channel $gg\to \gamma \gamma$ contribution, which is
formally a next-to-next-to-leading-order (NNLO) effect, was also
implemented in the \diphox code.  The inclusion of this channel has a
relatively large impact on the cross section, due to the enhanced
gluon parton distributions at the LHC.  The NLO corrections to the
gluon channel (which amount to N$^{3}$LO effects relative to the LO
process $q \bar{q} \to \gamma \gamma $) were implemented in the
parton-level Monte Carlo programs \twogammamc~\cite{Bern:2002jx} and
\mcfm~\cite{Campbell:2011bn}.  Recently, mass effects due to the
top-quark loops were studied in the gluon fusion channel at NLO
accuracy in Refs.~\cite{Maltoni:2018zvp,Chen:2019fla}.

In the context of smooth-cone isolation, where fragmentation functions
are not required, the first NNLO calculation for diphoton production
was carried out at the fully differential level using the $q_T$
subtraction method of Ref.~\cite{Catani:2007vq}. It has been
implemented in the numerical codes \twogammannlo~\cite{Catani:2011qz}
and \matrix~\cite{Grazzini:2017mhc}. An independent calculation at
NNLO accuracy for diphoton production was also performed in
Ref.~\cite{Campbell:2016yrh} and implemented in the \mcfm
program~\cite{Boughezal:2016wmq}.  This calculation uses the \nj
subtraction method~\cite{Boughezal:2015dva,Gaunt:2015pea} derived
within Soft-Collinear Effective Theory (SCET).  Phenomenological
studies of photon isolation at NNLO accuracy were carried out in
Ref.~\cite{Catani:2018krb} by using the aforementioned programs
\twogammannlo and \matrix.  More recently, an independent NNLO
analysis using the antenna subtraction
method~\cite{GehrmannDeRidder:2005cm,Daleo:2006xa,Currie:2013vh} was
presented in Ref.~\cite{Gehrmann:2020oec}, highlighting the importance
of the choice of the isolation criterion and of the scales in
estimating the theoretical uncertainties.  Electroweak (EW)
corrections for the diphoton production process at the LHC were
computed in Refs.~\cite{Bierweiler:2013dja,Chiesa:2017gqx}. Their
effect for inclusive observables is at the level of a few percent, and
increases as one imposes higher transverse-momentum cuts on the
photons.

Resummed calculations in the region of small transverse momentum of
the photon pair have been carried out at next-to-next-to-leading
logarithmic (NNLL) accuracy and are available in the programs
\resbos~\cite{Balazs:2006cc,Nadolsky:2007ba,Balazs:2007hr},
\twogammares~\cite{Cieri:2015rqa} and
\resolve~\cite{Coradeschi:2017zzw}. Very recently, the resummation was
pushed to N$^3$LL accuracy in \cutemcfm~\cite{Becher:2020ugp},
extending the matching to NNLO accuracy.
The process is also available at this accuracy in the public
interface \textsc{MATRIX+RadISH}~\cite{Kallweit:2020gva}.

Shower Monte Carlo (SMC) simulations for this process at NLO accuracy
matched to the \sherpa~\cite{Gleisberg:2008ta} parton shower program
were presented in Ref.~\cite{Hoeche:2009xc}.  A similar accuracy
obtained via matching to the \herwig~\cite{Corcella:2000bw} parton
shower using the \powheg approach~\cite{Nason:2004rx,Frixione:2007vw}
was presented in Ref.~\cite{DErrico:2011cgc}.

In this paper we implement a new fully differential NNLO event
generator for the diphoton production process $pp\to \gamma \gamma
+X$. We improve on the existing fixed-order~(FO) calculations by
including the NNLL$^\prime_{\Tau_0}$ resummation of the \zj resolution
variable $ \Tau_0$ (beam thrust) within the \geneva Monte Carlo
framework~\cite{Alioli:2012fc,Alioli:2013hqa,Alioli:2015toa,Alioli:2016wqt}.
The resummation of the \zj variable is carried out by using a
factorisation formula~\cite{Stewart:2009yx,Stewart:2010pd} valid at
small $\Tau_0$ derived in SCET. This is the first time that such a
resummation has been presented in the literature for this process. We
employ the smooth-cone isolation algorithm to define IR-finite events,
ensuring that the application of this cut is compatible with the
$\Tau_0$ resummation.  In addition, we provide fully showered and
hadronised events by matching partonic events to the
\pythiaEight~\cite{Sjostrand:2014zea} parton shower while retaining
the FO accuracy for inclusive observables. Diphoton production extends
the list of NNLO+PS processes, such as
Drell--Yan~\cite{Alioli:2015toa} and associated Higgs
production~\cite{Alioli:2019qzz} and decay~\cite{Alioli:2020fzf},
which were previously implemented in the \geneva Monte Carlo program.
Alternative approaches to reach NNLO+PS accuracy are actively being
developed
\cite{Hamilton:2012rf,Hamilton:2013fea,Monni:2019whf,Monni:2020nks,Hoeche:2014aia,Hoche:2014dla}.

This paper is organised as follows. We discuss the process definition
and the photon isolation criteria in \sec{PhotonIso}. In
\sec{TheoreticalFramework} we recap the theoretical background for the
\zj resummation in SCET, while in \sec{Genevapart} we provide the
details of the implementation in the \geneva framework.  We present
our results in \sec{results}, including the comparison to 7 TeV LHC
data.  Finally we give our conclusions and outlook in \sec{conc}.

\section{Process definition and photon isolation}
\label{sec:PhotonIso}

In this paper we consider the process
\begin{align}
\label{eq:processdef}
p \, p \to \gamma \, \gamma+X \nonumber \, ,
\end{align}
where the photon pair is produced through the hard scattering
interaction (\ie direct photon production).  In order to avoid QED
singularities when a photon is collinear to an initial-state parton,
one needs to impose kinematic cuts on the transverse momenta of the
photons.  We prefer to use asymmetric cuts on the harder ($\gamma_h$)
and the softer ($\gamma_s$) photon, \ie $p^{\gamma_h}_{T} \geq
p^{\gamma_h}_{T,\textrm{cut}}$ and $p^{\gamma_s}_{T} \geq
p^{\gamma_s}_{T,\textrm{cut}}$ (where $p^{\gamma_h}_{T} \geq
p^{\gamma_s}_{T}$) which avoids issues in the fixed-order predictions
with symmetric cuts~\cite{Frixione:1997ks,Banfi:2003jj,Alioli:2012tp}.
These cuts are also employed in experimental analyses to exclude
events in which the photons are emitted in the uninstrumented region
parallel to the beam line.

As mentioned in the introduction, final-state photons can also be
produced via the fragmentation of a quark or a gluon into a
photon. Usually, photons produced in a fragmentation process are
distinguishable from the direct photons since they lie inside hadronic
jets. Direct photons can therefore be separated from the rest of the
hadrons in the event through an isolation procedure.  A possible
isolation mechanism is to construct a cone with fixed radius
$R_{\textrm{iso}}$ around the direction of the photon candidate. One
then restricts the amount of transverse hadronic (partonic) energy
$E^{\textrm{had}}_T(R_{\textrm{iso}})$ inside the cone.  In
particular, a photon is considered to be isolated if
$E^{\textrm{had}}_T(R_{\textrm{iso}})$ is smaller than a certain value
$E^{\textrm{max}}_T$, usually parameterised as a (linear) function of
the transverse energy $E^\gamma_T$ of the photon and a fixed numerical
value $E^{\textrm{thres}}_T$~\cite{Catani:2002ny}:
\begin{align}
E^{\textrm{had}}_T(R_{\textrm{iso}}) < E^{\textrm{max}}_T \equiv \varepsilon\, E^\gamma_T + E^{\textrm{thres}}_T \, .
\end{align}
Due to its simplicity, the fixed-cone isolation procedure is currently
used in all experimental measurements of processes involving
photons. This method has the theoretical drawback of being sensitive
to the fragmentation contributions since configurations with a photon
collinear to a final-state parton are still allowed.  Indeed, to
completely remove the collinear QED divergences, one should require
that absolutely no hadronic energy is allowed within the
cone. Unfortunately, this condition is not IR safe since it forbids
emissions of soft partons inside the cone, therefore spoiling the
cancellation of QCD divergences.

The smooth-cone isolation procedure instead overcomes this problem by
considering a continuous series of smaller sub-cones with radius
$r \leq R_{\textrm{iso}}$, in addition to the outer cone of radius
$R_{\textrm{iso}}$.  The isolation condition then requires that
\begin{align} E^{\textrm{had}}_T(r) \leq E^{\textrm{max}}_T \,
\chi(r;R_{\textrm{iso}})\, ,\quad \textrm{for all sub-cones with} \,\,
r \leq R_{\textrm{iso}}\, ,
\end{align} where the isolation function $\chi(r;R_{\textrm{iso}})$
must be a smooth and monotonic function which  vanishes when $r\to 0$.
This requirement implies that the hadronic
activity is reduced in a smooth way when approaching the photon
direction, and  the parton radiation collinear
to the photon is completely absent at  $r= 0$. Hence, the fragmentation
component is eliminated while  soft radiation is still permitted in
a finite region away from the collinear limit, making the cross section
IR safe.  The standard choice for the $\chi$-function is
\begin{align}
\chi(r;R_{\text{iso}}) = \left(\frac{1-\cos r}{1-\cos{R_{\text{iso}}}}\right)^{n}\!,
\end{align}
where the exponent parameter $n$ is usually set to $n=1$.  Other
isolation functions satisfying these criteria are possible and have been
employed in the literature~\cite{Catani:2018krb}.

Though the use of a smooth-cone isolation procedure has the positive
effect of simplifying theoretical calculations, its implementation in
the experimental analyses is complicated by the granular nature of the
detector. The direct comparison of the theoretical predictions with
data is therefore challenging.  One possible solution is to adjust the
free parameters of the smooth-cone isolation algorithm to reproduce
the effects of the fixed-cone procedure, making a comparison at least
feasible. A potentially better approach, which has been recently
investigated in Refs.~\cite{Siegert:2016bre,Chen:2019zmr}, is the
introduction of a hybrid-cone isolation procedure. In this case the
theoretical calculation is initially carried out using the smooth-cone
isolation with a very small radius parameter $R_{\text{iso}}$, such
that only a tiny slice of phase space around the photon direction is
removed. In a second step, the fixed-cone isolation procedure with a
larger radius $R \gg R_{\text{iso}}$ is applied to the events which
passed the smooth-cone criterion. In other words, one initially
applies very loose smooth-cone isolation cuts which are then tightened
by the fixed-cone procedure. This makes the resulting events directly
suitable for experimental analyses.\footnote{In this case, however,
  one must ensure that the predictions are not too strongly dependent
  on the detailed tuning of the inner smooth-cone parameters.} In this
paper we investigate both the smooth-cone and the hybrid-cone
isolation procedures. We use the former to compare to results obtained
with the \matrix code~\cite{Grazzini:2017mhc} in
\subsec{genevavalidation}, while the latter is used for a direct
comparison to LHC data in \sec{results}.

\section{Resummation in Soft-Collinear Effective Theory}
\label{sec:TheoreticalFramework}

In this work, we  use the \nj~\cite{Stewart:2010tn} resolution
variable to discriminate between resolved emissions with different jet
multiplicities. Given an $M$-particle phase space point $\Phi_M$ with
$M \ge N$, the \nj is defined as
\begin{align}\label{eq:TauN} \Tau_N(\Phi_M) = \sum_k \textrm{min}
\big\{\hat{q}_a\cdot p_k, \hat{q}_b\cdot p_k, \hat{q}_1\cdot
p_k,\ldots, \hat{q}_N\cdot p_k \big\} \, ,
\end{align} where the index $k$ runs over all
strongly-interacting final-state particles and where
$\hat{q}_i=n_i=(1,\vec{n}_i)$ are light-like reference vectors
parallel to the beam and jet directions, defined in the rest frame of
the $\gamma \gamma$ colour-singlet state. The limit $\Tau_N\to 0$
describes an event with $N$ pencil-like hard jets, where the
unresolved emissions can either be soft or collinear to the
final-state jets or to the beams.  In the case of colour-singlet
production (\eg Drell--Yan, $VH$, $\gamma\gamma$, \ldots) the relevant
resolution variable is the \zj (beam thrust). Starting from the
general definition in \eq{TauN}, the expression for the \zj can be
considerably simplified to
\begin{align} \Tau_0 = \sum_k |\vec{p}_{kT} | \, e^{-|\eta_k - Y|} \,
,
\end{align} where $|\vec{p}_{kT} |$ and $\eta_k$ are the transverse
momentum and the rapidity of the emission $p_k$ and $Y$ is the
rapidity of the colour-singlet state.

The introduction of a jet resolution variable sets a new dynamical
energy scale in the problem, which can in principle differ from the
other physical scales of the process. As such, large logarithms of the
ratios of these scales may appear in the cross section which spoil the
convergence of the perturbative expansion and must therefore be
resummed. For \zj this happens in the small $\Tau_0/Q$ region, with
$Q$ being a typical hard scale of the process, \eg the invariant mass
of the colour-singlet final state.  In this region, the cross section
differential in the Born phase space $\Phi_0$ and $\Tau_0$ obeys a
factorisation formula which was originally derived in
Refs.~\cite{Stewart:2009yx,Stewart:2010pd} for the Drell--Yan process,
but which can be extended for diphoton production as
\begin{align}\label{eq:factorization}
\frac{\textrm{d} \sigma^{\text{SCET}}}{\textrm{d} \Phi_0 \textrm{d}
  \Tau_0} = \thetaPSiso (\Phi_0) \sum_{ij} H^{\gamma
  \gamma}_{ij}(Q^2,t,\mu)\int\! \df r_a\, \df r_b \,& B_i(r_a,x_a,\mu)
B_j(r_b,x_b,\mu) \, S(\Tau_0-\tfrac{r_a+r_b}{Q},\mu)\, .
\end{align}
The sum in the equation above runs over all possible $q\bar{q}$ pairs
$ij=\{ u\bar{u}, \bar{u} u, d \bar{d}, \bar{d} d,\ldots\}$.  For this
process we need to impose process-defining phase space restrictions
$\thetaPSiso (\Phi_0)$ in order to have a finite cross section. In
particular, we apply $p_T$ cuts on each of the photons as well as
isolation cuts to eliminate final-state collinear QED
singularities.\footnote{Notice that any isolation procedure acting on
  final-state particles has no effect on partonic $\Phi_0$ events.}
Other cuts, such as on the photons' rapidities or on the invariant
mass of the pair, can be imposed at the analysis level but are not
needed to define IR-finite cross sections.

The factorisation formula depends on the hard $H^{\gamma \gamma}_{ij}$,
soft $S$ and beam $B_{i,j}$ functions which describe the square of the
hard interaction Wilson coefficients, the soft real emissions between
external partons and the hard emissions collinear to the beams
respectively.

The hard functions $H^{\gamma \gamma}_{ij}(Q^2,t,\mu)$
are process-dependent objects and encode information about the Born
and virtual squared matrix elements. In order to achieve NNLL$^\prime$
accuracy, they are needed up to two-loop order. They are regular
functions of the Mandelstam invariants $Q^2=s$ and $t$ and can be
extracted from the two-loop squared amplitude
expressions~\cite{Anastasiou:2002zn}, after subtracting the IR poles
(as explained in detail in appendix \ref{app:hardfunction}).

The $B_i(r,x,\mu)$ are the inclusive (anti-)quark beam
functions~\cite{Stewart:2009yx}. They depend on the virtualities $r_{a,b}$ of
the initial-state partons $i$ and $j$ annihilated in the hard
interaction and on the momentum fractions $x_{a,b}$. These
can be written in terms of the diphoton rapidity $Y_{\gamma
\gamma}$ and  invariant mass $M_{\gamma \gamma}$ as
\begin{align}
x_a = \frac{M_{\gamma \gamma}}{E_{\textrm{cm}}} e^{Y_{\gamma \gamma}}
, \quad x_b = \frac{M_{\gamma \gamma}}{E_{\textrm{cm}}} e^{-Y_{\gamma
    \gamma}} \, ,
\end{align}
where $E_{\textrm{cm}}$ is the hadronic centre-of-mass energy.
The beam functions are calculated via an operator product expansion
and are defined as
\begin{align}
B_i(r_a,x_a,\mu) = \sum_k \int_{x_a}^{1} \frac{\df \xi_a}{\xi_a}\,
\mathcal{I}_{ik}\bigg(r_a,\frac{x_a}{\xi_a},\mu\bigg) \,
f_k(\xi_a,\mu)\, .
\end{align}
The perturbatively computable parts of the above equation are the
matching coefficients $\mathcal{I}_{ik}(t_a,z_a,\mu)$ which describe
the collinear virtual and real initial-state radiation (ISR)
emissions. The function $ f_k(\xi_a,\mu)$ represents the usual parton
distribution function (PDF) for parton $k$ with momentum fraction
$\xi_a$. The matching coefficients $\mathcal{I}_{ik}(t_a,z_a,\mu)$
were computed to NNLO accuracy in Ref.~\cite{Gaunt:2014xga}.
$S(k,\mu)$ is the quark hemisphere soft function for beam thrust and
has been computed to the required NNLO accuracy including the scale
independent terms in Refs.~\cite{Kelley:2011ng,Monni:2011gb}.

The hard, beam and soft functions which appear in \eq{factorization}
are single-scale objects.  This means that when they are evaluated at
their own characteristic scales
\begin{align}
\label{eq:canscaling}
\mu_H = Q,\quad \mu_B=\sqrt{Q \Tau_0},\quad \mu_S = \Tau_0\, ,
\end{align}
no large logarithmic corrections are present in their fixed-order
perturbative expansions.  However, \eq{factorization} must be
evaluated at a single common scale $\mu$. In order to do this, the
separate components must be evolved via renormalisation group (RG)
equations from their own characteristic scales to the common scale
$\mu$, which results in the large logarithms being resummed.  This
proceeds via convolutions of the single scale factors with the
evolution functions $U_i(\mu_i,\mu)$.  The resummed formula for the
$\Tau_0$ spectrum is then given by
\begin{align}
\label{eq:standardresum}
\frac{\textrm{d} \sigma^{\text{NNLL}^\prime}}{\textrm{d} \Phi_0 \textrm{d} \Tau_0} =\;& \thetaPSiso (\Phi_0) \sum_{ij} H^{\gamma \gamma}_{ij}(Q^2,t,\mu_H) \, U_H(\mu_H,\mu)\,\big\{\big[ B_i(t_a,x_a,\mu_B)\otimes U_B(\mu_B,\mu)\big]\, \nonumber \\
& \times \big[B_j(t_b,x_b,\mu_B)\otimes U_B(\mu_B,\mu)\big] \big\}\, \otimes \big[ S(\mu_s)\otimes U_S(\mu_S,\mu)\big]\, ,
\end{align}
where the convolution between the different functions is written in a
schematic form. The scale setting procedure will be explained in the
next section, where we will introduce the profile functions which are
employed to switch off resummation outside its kinematical range of
validity. In order to reach NNLL$^\prime$ accuracy, we need to know
the boundary conditions of the evolution, namely the hard, beam and
soft functions up to NNLO accuracy, and the cusp (non-cusp) anomalous
dimensions up to three-(two-)loop order. The expressions for the
anomalous dimensions up to the required order can be found in
Refs.~\cite{Idilbi:2006dg,Becher:2006mr,Hornig:2011iu,Kang:2015moa,Gaunt:2015pea}.

\section[Implementation within the G{\scriptsize ENEVA} framework]{Implementation within the G{\footnotesize ENEVA} framework}
\label{sec:Genevapart}

In this section we review the \geneva framework and present the
implementation of the diphoton production process, highlighting the
main differences compared to the processes that have previously been
implemented. More details on the general features of the \geneva
method are given in
Refs.~\cite{Alioli:2013hqa,Alioli:2015toa,Alioli:2019qzz}.

For all numerical results in this paper we evaluate all the matrix
elements up to the one-loop level with the amplitudes provided by the
\openloops~2 package~\cite{Cascioli:2011va,Buccioni:2017yxi,Buccioni:2019sur}.
We include the top-quark mass dependence in the NLO calculation of the diphoton plus jet production process and set $m_t=173.2$~GeV.
In the present implementation, however, we do not include the
diagrams with a closed top-quark loop in the hard function. These
effects begin to contribute at
$\mathcal{O}(\alpha_s^2)$  and  are currently unknown
in the $q \bar{q}$ channel.\footnote{Partial numerical results for the two-loop master
integrals were computed in Ref.~\cite{Mandal:2018cdj}.
We leave their implementation and the study of their phenomenological effect, \eg  at large $M_{\gamma \gamma}$, to future work.}

If not stated otherwise, in this section we use the following settings:
we use the
\texttt{MMHT2014nnlo68cl} PDF set~\cite{Harland-Lang:2014zoa} from
\texttt{LHAPDF}~\cite{Buckley:2014ana} and the $\alpha(0)$ input scheme to fix EW couplings and set
$\alpha^{-1}(0)=137$.  We apply the kinematic cuts
\begin{align}
  \label{eq:ptcuts} p^{\gamma_h}_T \ge 25\,\, \mathrm{GeV},
  \quad p^{\gamma_s}_T \ge 22\,\, \mathrm{GeV}, \quad M_{\gamma \gamma}
  \ge 25\,\, \mathrm{GeV}\, ,
\end{align}
and the smooth-cone isolation algorithm with parameters
\begin{align}
  \label{eq:isocuts} E^{\mathrm{max}}_T = 4 \,\,
  \mathrm{GeV}, \quad R_{\mathrm{iso}} = 0.4 ,\quad \mathrm{and} \quad n
  = 1 \, .
\end{align}
We set $\mu_\FO=M_{\gamma\gamma}$ or $\mu_\FO=M_{\gamma\gamma}^T\equiv \sqrt{M^2_{\gamma\gamma}+p^2_{T,\gamma\gamma}}$ and indicate the respective
choice with the results.
We consider only the $q \bar{q}$ channel contribution up to and including
\subsec{genevashower} and take into account also
the loop-induced $gg$ channel contribution thereafter.

\subsection[The \geneva method]{The G{\scriptsize ENEVA} method}
\label{sec:Genevapart1}
The final aim of a fully exclusive event generator is to produce
physical events, where all of the IR divergences are cancelled on an
event-by-event basis.  Following the \geneva method, these events are
assigned a Monte Carlo (MC) cross section $\df \sigma^{\mathrm{MC}}_N$
according to the value of the \nj resolution parameter $\Tau_N$. This
encodes the probability of producing a phase space point $\Phi_N$ with
$N$ partonic jets: all contributions from unresolved emissions below a
certain resolution cutoff $\Tau_N < \Tau^{\text{cut}}_N$ are included
in $\df \sigma^{\mathrm{MC}}_N$.  For the process at hand at NNLO, the
exclusive cross sections for events with $0$ and $1$ jet, and the
$2$-jet inclusive cross section are defined by the cutoffs on the
$\Tau_0$ and $\Tau_1$ resolution variables as
\begin{align}
\label{eq:NNLOevents}
\text{$\Phi_0$ events: }
& \qquad \frac{\dsigMC_0}{\df\Phi_0}(\Tau_0^\cut)
\,,\nn\\
\text{$\Phi_1$ events: }
& \qquad
\frac{\dsigMC_{1}}{\df\Phi_{1}}(\Tau_0 > \Tau_0^\cut; \Tau_{1}^\cut)
\,,
\nn\\
\text{$\Phi_2$ events: }
& \qquad
\frac{\dsigMC_{\ge 2}}{\df\Phi_{2}}(\Tau_0 > \Tau_0^\cut, \Tau_{1} > \Tau_{1}^\cut)
\,.
\end{align}
Since one integrates over the unresolved emissions below the cutoffs,
the definitions of the partonic jets used above depend on the phase space
maps $\Phi_N(\Phi_M)$ (with $N\le M$) which project the unresolved $M$-body phase
space points onto $\Phi_N$.  Using
\eq{NNLOevents} the cross section for a generic observable $X$
is written as
\begin{align}
\sigma(\obs) &= \int\!\df\Phi_0\,
\frac{\dsigMC_0}{\df\Phi_0}(\Tau_0^\cut)\, M_\obs(\Phi_0) \nn \\ &
\quad + \int\!\df\Phi_{1}\, \frac{\dsigMC_{1}}{\df\Phi_{1}}(\Tau_0 >
\Tau_0^\cut; \Tau_{1}^\cut)\, M_\obs(\Phi_{1}) \nn\\ & \quad +
\int\!\df\Phi_{2}\, \frac{\dsigMC_{\ge 2}}{\df\Phi_{2}}(\Tau_0 >
\Tau_0^\cut, \Tau_{1} > \Tau_{1}^\cut)\, M_\obs(\Phi_{2}) \,,
\end{align}
where $M_X(\Phi_N)$ is the measurement function that computes the
observable $X$ for the \mbox{$N$-parton} final-state point
$\Phi_N$.

The cross section defined above is not equivalent to that obtained from a fixed-order calculation.
Indeed, for any unresolved emission, the observable is
computed on the projected point $\Phi_N(\Phi_M)$ rather than the exact
$\Phi_M$ point. Since the resulting difference vanishes in the limit
\mbox{$\Tau^{\textrm{cut}}_N\to 0$}, it is advisable to choose this cutoff to
be as small as possible. This, however, introduces large logarithms of
$\Tau_N$ and $\Tau^{\textrm{cut}}_N$, which need to be resummed in
order to obtain physically meaningful results.

We perform the resummation of $\Tau_0$  at  NNLL$^\prime$
accuracy, matching it to a NNLO$_0$\footnote{Here and in the following we use the subscript notation to specify the jet multiplicity of the FO final state. In this case the NNLO$_0$ acronym indicates the next-to-next-to-leading order corrections to diphoton production with no additional jets.} calculation.
This has the positive feature of correctly describing the spectrum both in the
small-$\Tau_0$ region, where the resummation dominates, and also in the
large-$\Tau_0$ region, where the correct dependence is given by the fixed-order
result.

Depending on the final-state jet multiplicity, the \geneva method
requires one to evaluate the resummed and resummed-expanded terms in
the cross sections on projected phase space points of lower
multiplicity. Even these projected configurations are required to
satisfy the cuts which avoid the QED singularities.  We use the symbol
$\thetaProj(\widetilde{\Phi}_N)$ (and
$\thetaBarProj(\widetilde{\Phi}_N)$ for its complement) to indicate
this set of phase space restrictions acting on the higher dimensional
$\Phi_{N+1}$ phase space due to the cuts on the projected
configuration $\widetilde{\Phi}_N$. In practice, this means that when
a term in the cross section, evaluated at a $\Phi_{N+1}$ phase space
point, is multiplied by $\thetaProj(\widetilde{\Phi}_N)$, the
$\Phi_{N+1}$ phase space point is projected onto a
$\widetilde{\Phi}_{N}$ point and the cuts are applied on this lower
dimensional space. If the projected configuration does not pass the
cuts, the initial $\Phi_{N+1}$ configuration is excluded.  Notice that
the separation realised by the introduction of $\thetaProj$ and
$\thetaBarProj$ is not usually required in a fixed-order calculation.
Nonetheless, we choose to perform it even when evaluating the FO
contribution and split this into two parts. Of these, the first,
singular part enters in the $0$-jet cumulant, \eq{0full}, while the
second, nonsingular part enters in the $1$-jet cross section below the
$\Tau_0^{\rm cut}$, \eq{tau0smaller}.

Since the resummation for  \zj is carried out at NNLL$^\prime$
accuracy in \geneva, meaning that it contains all of the singular
corrections in $\Tau_0$ up to $\mathcal{O}(\alpha^2_s)$, we can write
the 0- and 1-jet cross sections as
\begin{align}
\label{eq:0master}
\frac{\dsigMC_0}{\df\Phi_0}(\Tau_0^\cut)
=&\; \frac{\df\sigma^{\rm NNLL'}}{\df\Phi_0}(\Tau_0^\cut)
+ \frac{\df\sigma_0^{\rm nons}}{\df\Phi_0}(\Tau_0^\cut)
\,,\\[1ex]
\label{eq:1incmaster}
\frac{\dsigMC_{\geq 1}}{\df\Phi_{1}}(\Tau_0 > \Tau_0^\cut)
=& \left[\!\frac{\df\sigma^{\rm NNLL'}}{\df\Phi_{0}\df \Tau_0} \, \cP(\Phi_1) \, \thetaPSiso(\Phi_1) \thetaProj(\widetilde{\Phi}_0)\!+\! \frac{\df\sigma_{\ge 1}^{\rm nons}}{\df\Phi_1}(\Tau_0 > \Tau_0^\cut) \!\right]\! \theta(\Tau_0 > \Tau_0^\cut),
\end{align}
where $\df\sigma^{\rm NNLL'}/\df\Phi_{0}\df \Tau_0$ is the resummed
$\Tau_0$ spectrum and $\df\sigma^{\rm NNLL'}/\df\Phi_0(\Tau_0^\cut)$
is its integral. In the above equation we introduced a splitting
probability function $\mathcal{P}(\Phi_1)$ which satisfies the
normalisation condition
\begin{align}
\label{eq:Pnorm}
\int \! \frac{\df\Phi_1}{\df \Phi_{0} \df \Tau_0} \, \cP(\Phi_1) = 1
\end{align}
to make the $\Tau_0$ spectrum fully differential in $\Phi_1$.  The
nonsingular contributions are given by
\begin{align}
\frac{\df\sigma_0^\nons}{\df\Phi_{0}}(\Tau_0^\cut)
\label{eq:0nons}
=&\; \frac{\df\sigma_0^{{\rm NNLO_0}}}{\df\Phi_{0}}(\Tau_0^\cut)
  - \biggl[\frac{\df\sigma^{\rm NNLL'}}{\df\Phi_{0}}(\Tau_0^\cut) \biggr]_{\rm NNLO_0}
\,,\\[1ex]
\frac{\df\sigma_{\ge 1}^\nons}{\df\Phi_{1}}(\Tau_0 > \Tau_0^\cut) =&\;  \frac{\df\sigma_{\ge 1}^{{\rm NLO_1}}}{\df\Phi_{1}}(\Tau_0 > \Tau_0^\cut) -  \biggl[\frac{\df\sigma^{\rm NNLL'}}{\df\Phi_0 \df \Tau_0}\cP(\Phi_1) \biggr]_{\rm NLO_1} \!\!\!\thetaPSiso(\Phi_1)\,  \thetaProj(\widetilde{\Phi}_0)\,.
    \label{eq:1nons}
\end{align}
The terms in squared brackets are the expanded expressions to
$\mathcal{O}(\alpha^2_s)$ of the resummed cumulant and spectrum. The
NLO$_1$ term refers to the NLO corrections to the diphoton plus jet
production process. The projected $\Phi_1\to\widetilde{\Phi}_0$ point
in the above equation is obtained through a FKS projection.\footnote{To be precise we use the projective map described in section 5.1.1 of Ref.~\cite{Frixione:2007vw}.} After
explicitly writing the FO contributions to the cross sections we
obtain
\begin{align}
\label{eq:0full}
\frac{\dsigMC_0}{\df\Phi_0}(\Tau_0^\cut)
=&\; \frac{\df\sigma^{\rm NNLL'}}{\df\Phi_0}(\Tau_0^\cut)\, - \biggl[\frac{\df\sigma^{\rm NNLL'}}{\df\Phi_{0}}(\Tau_0^\cut) \biggr]_{\rm NNLO_0} \, \nn\\
&+(B_0+V_0+W_0)(\Phi_0) \, \thetaPSiso(\Phi_0) \, \nn \\
&+  \int \frac{\mathrm{d} \Phi_1}{\mathrm{d} \Phi_0} (B_1 + V_1)(\Phi_1)\, \thetaPSiso (\Phi_1)\,  \thetaProj(\widetilde{\Phi}_0)\,\theta\big( \Tau_0(\Phi_1)< \Tau_0^{\mathrm{cut}}\big)\, \nn \\
&+  \int \frac{\mathrm{d} \Phi_2}{\mathrm{d} \Phi_0} \,B_2 (\Phi_2)\, \thetaPSiso (\Phi_2) \,\theta\big( \Tau_0(\Phi_2)< \Tau_0^{\mathrm{cut}}\big)\,,\\[1ex]
\label{eq:sigma>=1}
\frac{\dsigMC_{\geq 1}}{\df\Phi_{1}}(\Tau_0 > \Tau_0^\cut)
=&\; \Bigg\{\Bigg(\frac{\df\sigma^{\rm NNLL'}}{\df\Phi_{0}\df \Tau_0} -  \biggl[\frac{\df\sigma^{\rm NNLL'}}{\df\Phi_0 \df \Tau_0} \biggr]_{\rm NLO_1}\Bigg)\, \cP(\Phi_1)\, \thetaPSiso(\Phi_1) \thetaProj(\widetilde{\Phi}_0) \, \nn \\
&+ (B_1+V_1)(\Phi_1) \, \thetaPSiso (\Phi_1) + \!\int  \frac{\df\Phi_{2}}{\df\Phi^\Tau_1} \, B_2(\Phi_2) \thetaPSiso (\Phi_2)\Bigg\}\, \theta(\Tau_0 > \Tau_0^\cut),
\end{align}
where $B_1$ and $B_2$ are the 1-parton and 2-parton tree-level
contributions respectively, $V_0$ and $V_1$ correspond instead to the
0-parton and 1-parton one-loop contributions while $W_0$ is the
two-loop contribution. Notice that the matrix elements appearing in the formula
above are implicitly assumed to be UV-renormalised and IR-subtracted such that
they are finite. We refrain from explicitly writing the subtraction terms
in order to make the notation less cumbersome.
We also introduced the notation
\begin{align}
\label{eq:dPhiRatio}
\frac{\df \Phi_{M}}{\df \Phi^{\mathcal{O}}_N} = \df \Phi_{M} \, \delta[ \Phi_N - \Phi^{\mathcal{O}}_N(\Phi_M)]\Theta^{\mathcal{O}}(\Phi_N)\,,
\end{align}
to indicate that the integration over a region of the $M$-body phase
space is done keeping the $N$-body phase space and the value of some
specific observable $\mathcal{O}$ fixed, with $N\leq M$.  The
$\Theta^\mathcal{O}(\Phi_N)$ term in the previous equation limits the
integration to the phase space points included in the singular
contribution for the given observable $\mathcal{O}$.  Since the
resummed and resummed-expanded contributions are differential in
$\Tau_0$, the phase space integration of the 2-parton contribution in
\eq{sigma>=1} should be parameterised in such a way
that it is also differential in $\Tau_0$. Indeed the projection
$\Phi_{2}\to\widetilde{\Phi}_1\equiv \Phi^{\Tau}_1$ must use a map
which preserves $\Tau_0$:
\begin{equation} \label{eq:Tau0map}
\Tau_0(\Phi_1^\Tau(\Phi_2)) = \Tau_0(\Phi_2)\,.
\end{equation}
In this way all of the terms in the inclusive 1-jet cross section
(\eq{sigma>=1}) can be evaluated at the same value of $\Tau_0$, and
the point-wise cancellation of the singular $\Tau_0$ contributions is
achieved.  The symbol $\Theta^\Tau(\Phi_2)$ defines the region of
$\Phi_2$ which can be projected onto the $\Phi_1$ phase space through
the map $\Phi_1^\Tau(\Phi_2)$. The non-projectable region of $\Phi_2$,
which corresponds to nonsingular events, will be included in the
2-parton event cross section below.  To this end we introduce the
following notation. We identify the set of cuts due to the particular
map used by $\Theta_{\mathrm{map}}$ (while its complement is defined
as $\overline{\Theta}_{\mathrm{map}}$).  Nonsingular contributions
coming from non-projectable regions (either because they fail the
isolation cuts or because they result in an invalid flavour
projection) are also present in the case of a
$\Phi_1\to\widetilde{\Phi}_0$ projection. We explicitly include them
in our formula for the cross section below the $\Tau^\cut_0$ as
\begin{align}\label{eq:tau0smaller}
\frac{\dsigMC_{ 1}}{\df\Phi_{1}}(\Tau_0 \leq \Tau_0^\cut) &= \thetaPSiso (\Phi_1)  \, (B_1+V_1)\, (\Phi_1)\,\theta(\Tau_0<\Tau^{\mathrm{cut}}_0)\,  \big[ \thetaBarProj(\widetilde{\Phi}_0)  + \overline{\Theta}^{\mathrm{FKS}}_{\mathrm{map}}(\Phi_1)  \big]  \, ,
\end{align}
where we indicate the failure to obtain a valid flavour projection
with the symbol $\overline{\Theta}^{\mathrm{FKS}}_{\mathrm{map}}$ and
the failure to satisfy the isolation cuts with
$\thetaBarProj(\widetilde{\Phi}_0)$.

Starting at $\mathcal{O}(\alpha^2_s)$ a new channel opens up and
contributes to the diphoton production process at the LHC. This is the
loop-induced $gg$ channel ``box" contribution, which is finite and
included at the order we are working, keeping the full dependence on
the top-quark mass.  In the present work we do not implement any
$\Tau_0$ resummation for this channel and only add its effects at
fixed order in the $\delta (\Tau_0)$ term.  We leave the resummation
of this channel, which formally also starts contributing at NNLL$^\prime$ accuracy, to future work.

\subsection{Profile functions}
\label{subsec:genevaprofiles}

In this section we describe the scale choices for the hard, beam and
soft functions in \eq{standardresum}. The cross section for the
diphoton production process is affected by large logarithms of
$\Tau_0/Q$ when $\Tau_0\ll Q$ and $Q=M_{\gamma \gamma}$ is the hard
scale of the process. However, for larger values of $\Tau_0$, the
logarithmic terms become numerically small, and the cross section is
dominated by the FO result in that region of phase space. Therefore,
we need to switch off the resummation before reaching this region. To
this end, it is helpful to investigate the numerical relevance of the
different terms which contribute to the cross section as a function of
$\Tau_0$.
\begin{figure}[tp]
\begin{center}
\begin{tabular}{ccc}
\includegraphics[width=\rescaletwoplots]{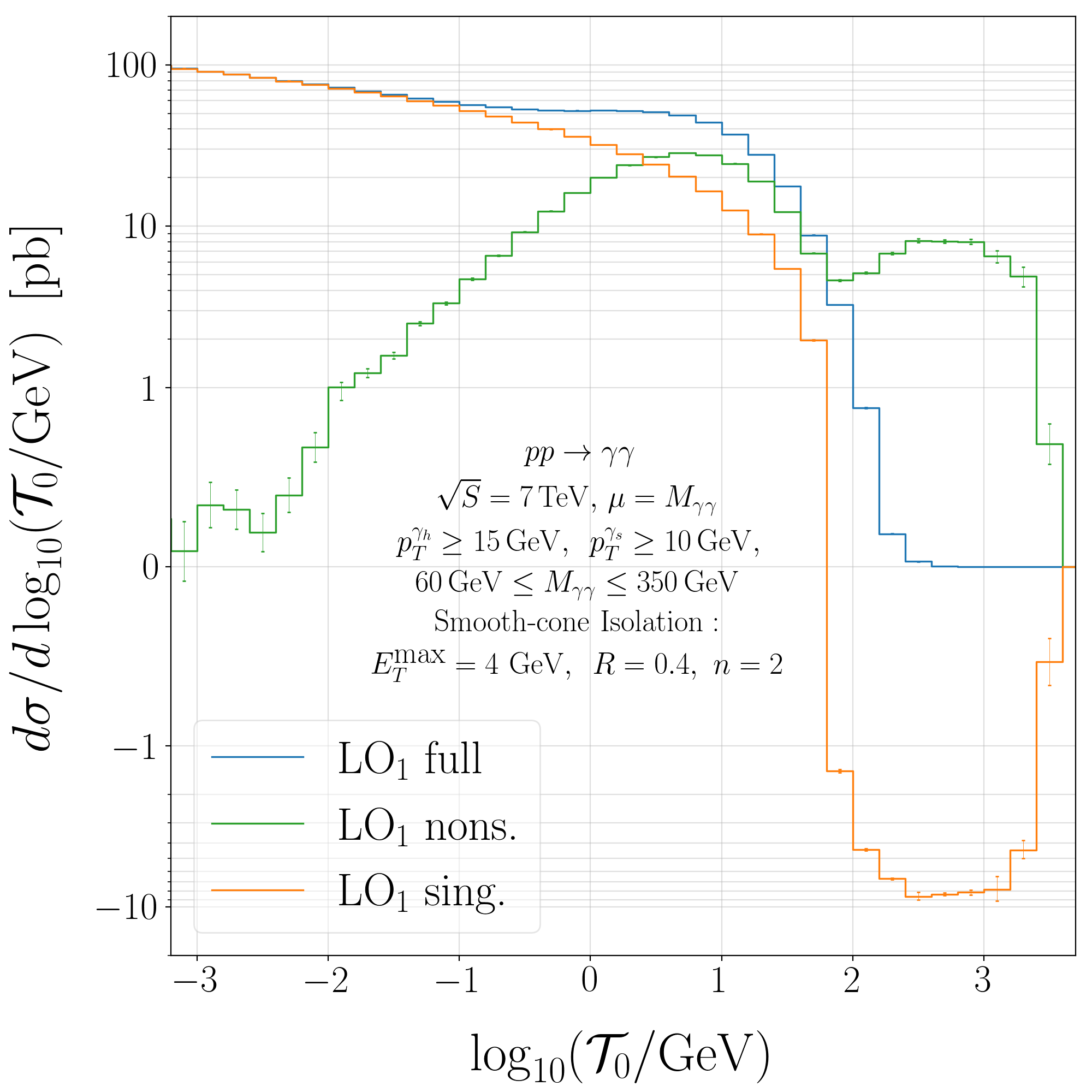} &\hspacebetweentwoplots&
\includegraphics[width=\rescaletwoplots]{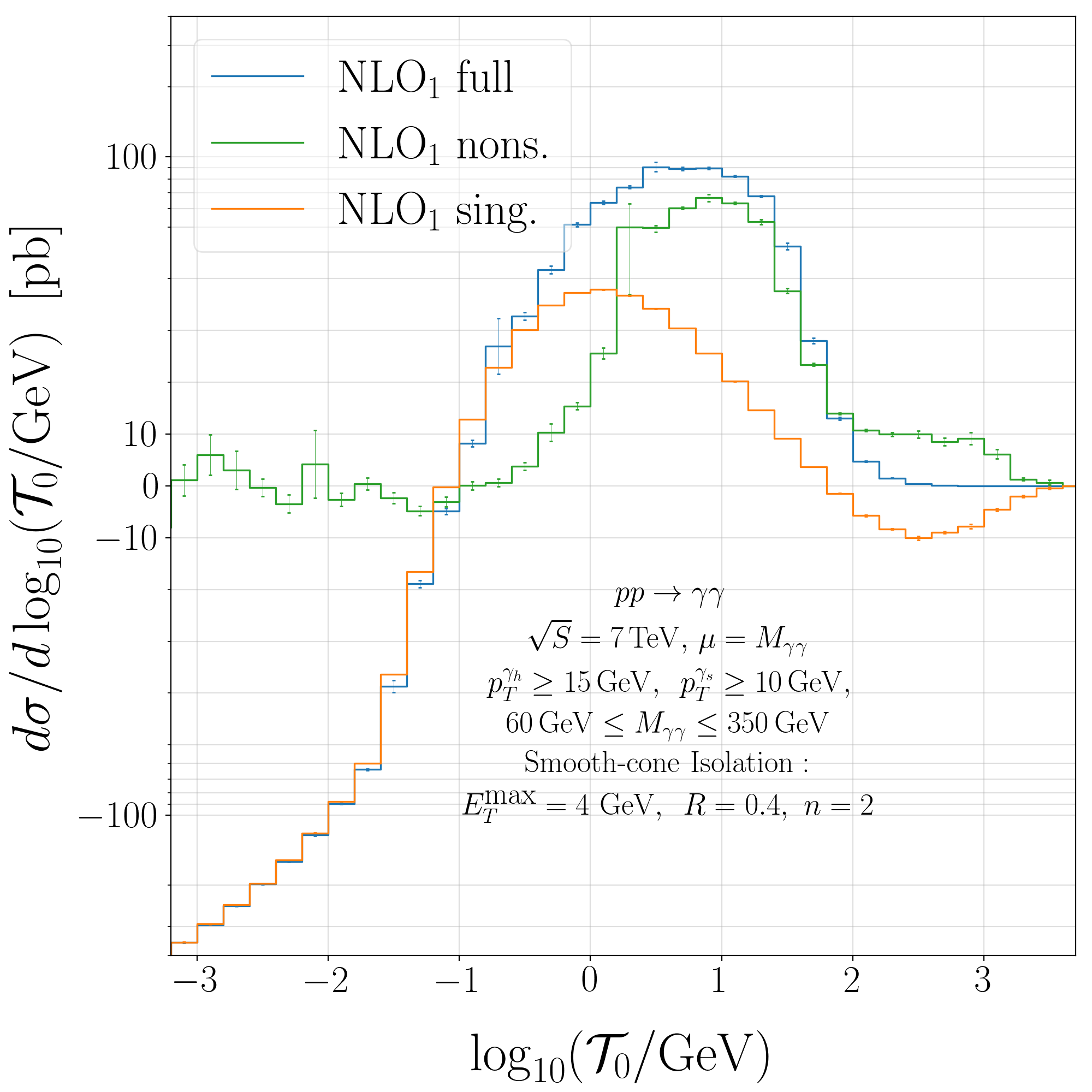}
\end{tabular}
\end{center}
\spaceabovefigurecaption
\caption{Singular and nonsingular contributions to the diphoton production cross section as a function of $\Tau_0$ at NLO (left) and NNLO (right).
\label{fig:Tau0SingvsNSing}
}
\spacebelowfigurecaption
\end{figure}
In \fig{Tau0SingvsNSing} we perform this comparison at
different fixed orders.\footnote{Here and in all the following plots the error bars represent the statistical error coming from the Monte Carlo integration.} In both panels we show the FO calculation, the
leading singular terms from the expansion of the resummed formula to
the relevant fixed order and the difference between the two. The
latter amounts to a nonsingular contribution to the cross section in
the limit $\Tau_0\to 0$. As expected from the leading-power
factorisation formula, each resummed-expanded contribution perfectly
reproduces the singular behaviour of the corresponding FO term and
acts as a subtraction to remove the IR divergences. Indeed, the
nonsingular contributions vanish on a logarithmic scale as
$\Tau_0 \to 0$.  One may also notice that the nonsingular terms become
of similar size to the singular contributions at small values of
$\Tau_0$, in the range of a few GeV. This behaviour is something of a
novelty compared to the previous Drell--Yan and $VH$ production
calculations and appears to be peculiar to the diphoton production
process. Although the nonsingular terms are power suppressed for small
$\Tau_0$ values, the photon isolation procedure might become the
primary source of power corrections, enhancing their contribution and
making them numerically
large~\cite{Ebert:2019zkb,Balsiger:2018ezi,Becher:2020ugp,Wiesemann:2020gbm}.

Resummation is carried out via RGE evolution and can be
switched off by setting all of the scales equal to a common
nonsingular scale $\mu_\NS = \mu_S = \mu_B = \mu_H$.  We do not evolve
the hard function, which is always evaluated at
$\mu_H=\mu_\NS$~\cite{Ligeti:2008ac, Abbate:2010xh}. Instead we evolve
the soft and the beam functions from their characteristic scales up to
the hard scale. This is achieved by employing profile scales
$\mu_B(\Tau_0)$ and $\mu_S(\Tau_0)$ which ensure a smooth transition
between the resummation and the FO regimes. Explicitly,
\begin{align} \label{eq:centralscale}
\mu_H &= \mu_\NS\nn\,,  \\
\mu_S(\Tau_0) & = \mu_\NS\ f_{\rm run}(\Tau_0/Q)\,, \nn \\
\mu_B(\Tau_0) &=  \mu_\NS\ \sqrt{f_{\rm run}(\Tau_0/Q)} \,,
\end{align}
where the common profile function $f_{\rm run}(x)$ is given by~\cite{Stewart:2013faa}
\begin{align}
f_{\rm run}(x) &=
\begin{cases} x_0 \bigl[1+ (x/x_0)^2/4 \bigr] & x \le 2x_0\,,
\\ x & 2x_0 \le x \le x_1\,,
\\ x + \frac{(2-x_2-x_3)(x-x_1)^2}{2(x_2-x_1)(x_3-x_1)} & x_1 \le x \le x_2\,,
\\  1 - \frac{(2-x_1-x_2)(x-x_3)^2}{2(x_3-x_1)(x_3-x_2)} & x_2 \le x \le x_3\,,
\\ 1 & x_3 \le x\,.
\end{cases}
\label{eq:frun}
\end{align}
This functional form ensures the canonical scaling behaviour as in
\eq{canscaling} for values below $x_1$ and turns off resummation above
$x_3$. After considering that the invariant mass distribution peaks in
the range $50$-$80$ GeV (depending on the specific cuts that are
applied) and that the nonsingular corrections in \fig{Tau0SingvsNSing}
become of the same size of the singular at $\Tau_0\sim 1-3$~GeV, we
choose the following parameters for the profile functions:
\begin{align} \label{eq:TauBprofile}
x_0 = 2.5\GeV/Q\,, \quad
\{x_1,x_2, x_3\} = \{0.1, 0.5, 0.8\}\,.
\end{align}
In the resummation region the nonsingular scale $\mu_\NS$ must be of
the same order as the hard scale of the process $M_{\gamma \gamma}$,
while in the FO region it can be chosen to be any fixed or dynamical
scale $\mu_\FO$. One can, for
example, set it either to $M_{\gamma \gamma}$ or to the transverse mass
of the photon pair $M^T_{\gamma\gamma}$.

We estimate the theoretical uncertainties for the FO predictions by
varying the central choice for $\mu_\NS$ up and down by a factor of
two and take for each observable the maximal
absolute deviation from the central result as the FO uncertainty. For
the resummation uncertainties, we vary the central choices for the
profile scales $\mu_B$ and $\mu_S$ independently while keeping
$\mu_H=\mu_\NS$ fixed. This gives us four independent variations. In
addition, we consider two more profile functions where we shift all
the $x_i$ transition points together by $\pm 0.05$ while keeping all
of the scales fixed at their central values. Hence we
obtain in total six profile variations. We consider the maximal
absolute deviation in the results with respect to the central
prediction as the resummation uncertainty.  The total perturbative
uncertainty is then calculated by adding the FO and the resummation
uncertainties in quadrature.
\begin{figure}[tp]
\begin{center}
\begin{tabular}{ccc}
\includegraphics[width=\rescaletwoplots]{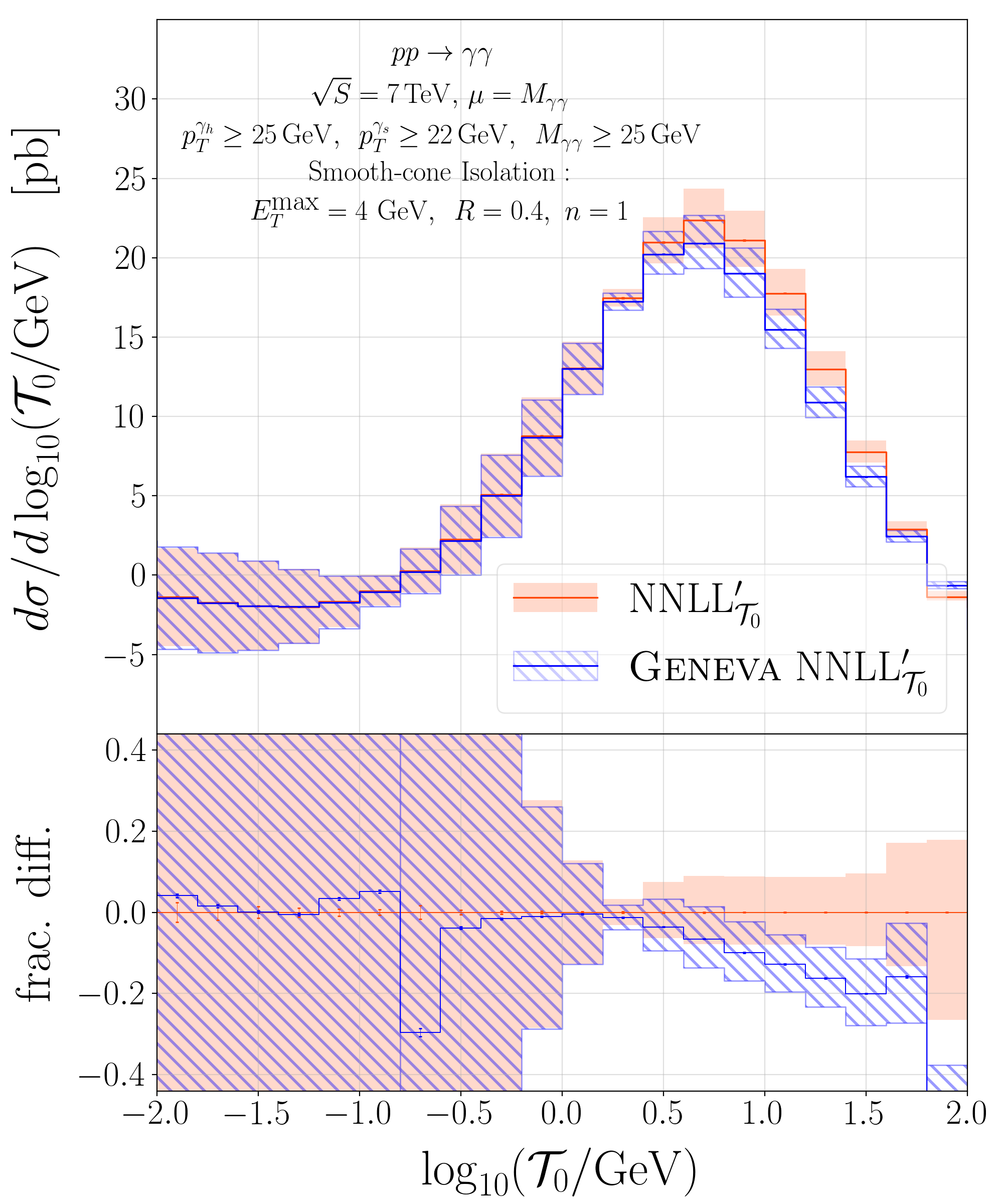} &\hspacebetweentwoplots&
\includegraphics[width=\rescaletwoplots]{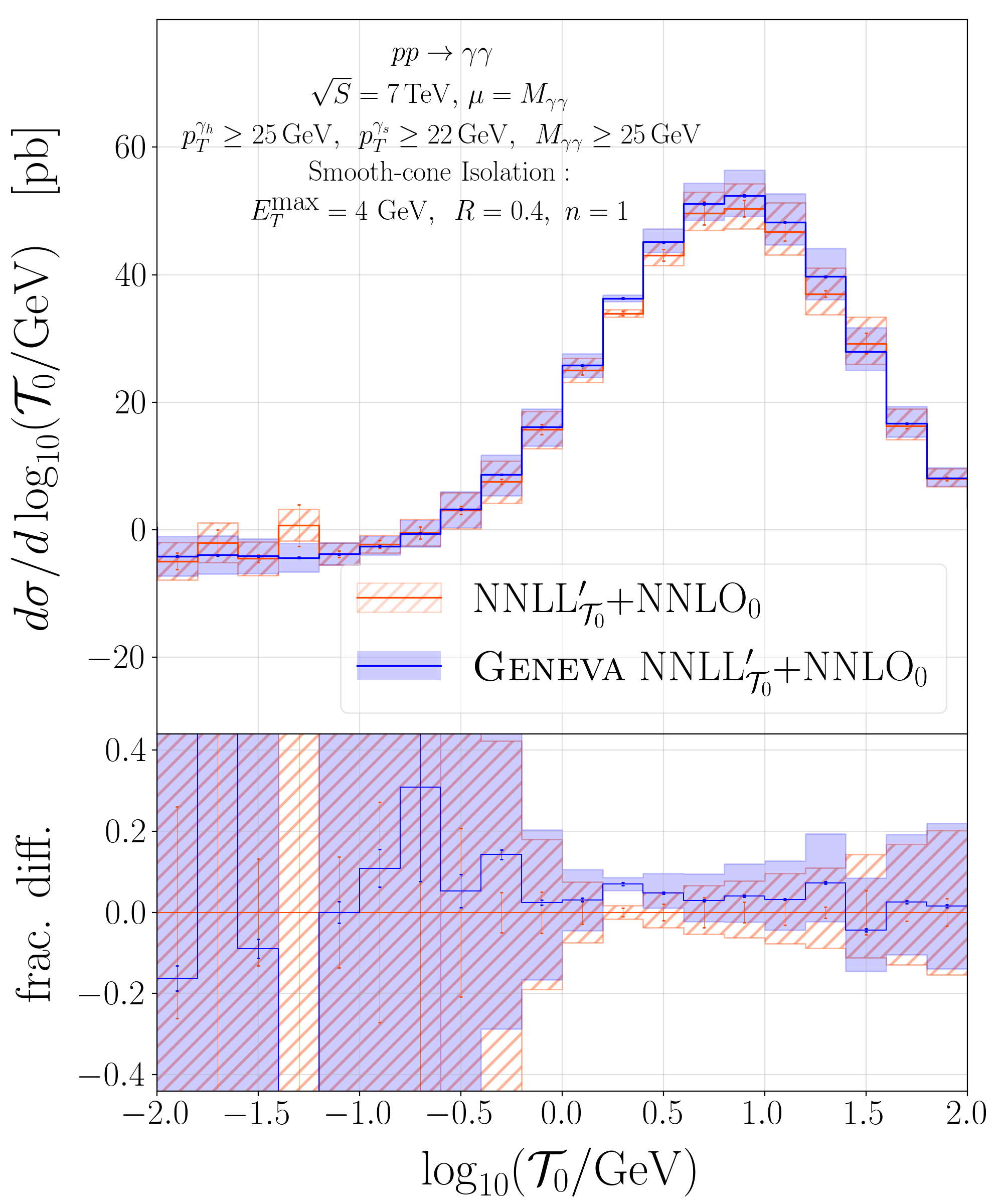}
\end{tabular}
\end{center}
\spaceabovefigurecaption
\caption{Comparison between standard resummation and event generation
  in \geneva in the presence of projection cuts. We show the resummed
  contribution alone (left) and the result matched to NNLO (right).
\label{fig:stdresumVsGVA}
}
\spacebelowfigurecaption
\end{figure}

As explained in detail in Refs.~\cite{Alioli:2015toa,Alioli:2019qzz},
the $\Tau_0$ integration of the resummation formula
(\eq{standardresum}) and the procedure of choosing the scales are
operations which do not commute with each other.  The expression for
the cumulant is not, therefore, exactly the same as the integral of
the $\Tau_0$ spectrum, since the profile scales have a functional
dependence on $\Tau_0$. To obtain an expression for the resummed
cumulant instead, one must first integrate the expression in
\eq{standardresum} for the resummed $\Tau_0$ distribution and then
choose the scales using the same profile scales but with the $\Tau_0$
replaced by $\Tau^{\mathrm{cut}}_0$. For example the canonical scales
assume the values
\begin{align}
\mu_H = Q,\quad \quad \mu_B= \sqrt{Q \Tau^{\mathrm{cut}}_0}, \quad \quad \mu_S=\Tau^{\mathrm{cut}}_0\, .
\end{align}
The difference between the two results for inclusive quantities is
formally beyond NNLL$^\prime$ accuracy~\cite{Almeida:2014uva}.
However, the numerical effect can be large
and, in order to preserve the value of the NNLO$_0$ cross section, we
follow the prescription for fully inclusive quantities discussed in
Refs.~\cite{Alioli:2015toa,Alioli:2019qzz} which amounts to adding the
contribution
\begin{align}
\label{eq:term}
\kappa(\Tau_0) \left[ \frac{\df}{\df \Tau_0} \frac{\df \sigma^{\rm
      NNLL'}}{\df \Phi_0}(\Tau_0, \mu_h(\Tau_0)) - \frac{\df
    \sigma^{\rm NNLL'}}{\df \Phi_0 \df \Tau_0} (\mu_h(\Tau_0))
  \right],
\end{align}
where $\kappa(\Tau_0)$ and $\mu_h(\Tau_0)$ are smooth functions.  The
$\mu_h(\Tau_0)$ are new profile scales which are chosen to turn off
resummation earlier than the normal profile scales in the resummed
calculation, in order to maintain the accurate
description of the tail of the $\Tau_0$ spectrum. Indeed, in the FO
region we have that $\mu_h(\Tau_0)=Q$ and the difference in the
brackets of \eq{term} is zero, since it is proportional to $\df
\mu_h/\df \Tau_0$.  The $\kappa(\Tau_0)$ function interpolates from a
constant $\mathcal{O}(1)$ value at $\Tau_0 \ll Q$ reaching zero when
$\mu_h(\Tau_0)=Q$.  The value of $\kappa(\Tau_0\to 0)$ can be tuned
such that after integration of the $\Tau_0$ spectrum together with
\eq{term},  the correct inclusive cross section is obtained. A
more detailed explanation on the precise structure of this additional
term can be found in
Refs.~\cite{Alioli:2015toa,Alioli:2019qzz}.

\subsection{Comparison with standard resummation and matching}
\label{subsec:stdresvsgeneva}

The implementation of resummation in the \geneva framework takes a
very different perspective compared to the usual resummation
approach. The latter is normally carried out with the purpose of
improving the description of a single particular observable in the
limit where it approaches very small or very large values compared to
other scales present in the process.  In practice, this involves a
scan over the $\Tau_0$ spectrum by directly evaluating the resummation
formula (\eq{standardresum}) on a $\Phi_0$ phase space point. Hence
the information about the physical event which generates a particular
value of the resummed variable is not retained in the calculation.  In
an event generator such as \geneva, however, one starts by generating
\eg $\Phi_1$ events and calculating the value of the resummed
variable, in this case $\Tau_0$, resulting from that particular event
configuration. At this point one has to perform a projection to the
lower-multiplicity phase space to evaluate the resummation
formula. This second method is more flexible because it allows one to
access the event information in a fully differential way and to
associate to each event a resummed weight, also allowing the matching
to a FO calculation.

Infrared safety ensures that even in the presence of process-defining
cuts, which are applied either to $\Phi_0$ in the standard resummation
or to $\Phi_1$ in \geneva, these two approaches are identical in the
limit $\Tau_0\to 0$.  Away from the limit, the two procedures will
give the same result only if the quantities upon which the cuts are
imposed are preserved by the $\Phi_1\to \Phi_0$ projection.  For the
Drell--Yan and $VH$ production processes previously studied in
\geneva, the only process-defining cut was applied on the invariant
mass of the colour-singlet system. Since this quantity was preserved
by every projection, the problem was avoided.  In the case of diphoton
production instead, the $p_T$ cuts on each photon are not preserved by
our mappings.\footnote{In the case of a $\Phi_2\to \Phi_1$ projection
  the complicated cuts due to the photon isolation procedure can also
  not be preserved.} Hence, these cuts applied to the projected
$\widetilde{\Phi}_0$ configurations will effectively remove some
contributions to the resummed and resummed-expanded terms of the cross
section formula in \eq{sigma>=1} which are present in the usual
resummed results (\ie without any recoil considered).

The difference between the two procedures is shown in the left plot of
\fig{stdresumVsGVA} for the resummed contribution alone, while
the right plot shows the same comparison after
matching to fixed order.
We observe good compatibility between the
two curves even at large $\Tau_0$, meaning that the difference between
the two approaches is eliminated at FO by the matching procedure.  We
also observe larger fluctuations of the standard result at small
$\Tau_0$ values, likely due to combining the histogram bins of
the matched calculation rather than combining the contributions on an
event-by-event basis as is done in \geneva.

We are now in a position to compare the effects of the $\Tau_0$
resummation matched to FO calculations by evaluating the cross section
formulae in \eq{0full} and \eq{sigma>=1} at different accuracies.
\begin{figure}[tp]
\begin{center}
\begin{tabular}{ccc}
\includegraphics[width=\rescaletwoplots]{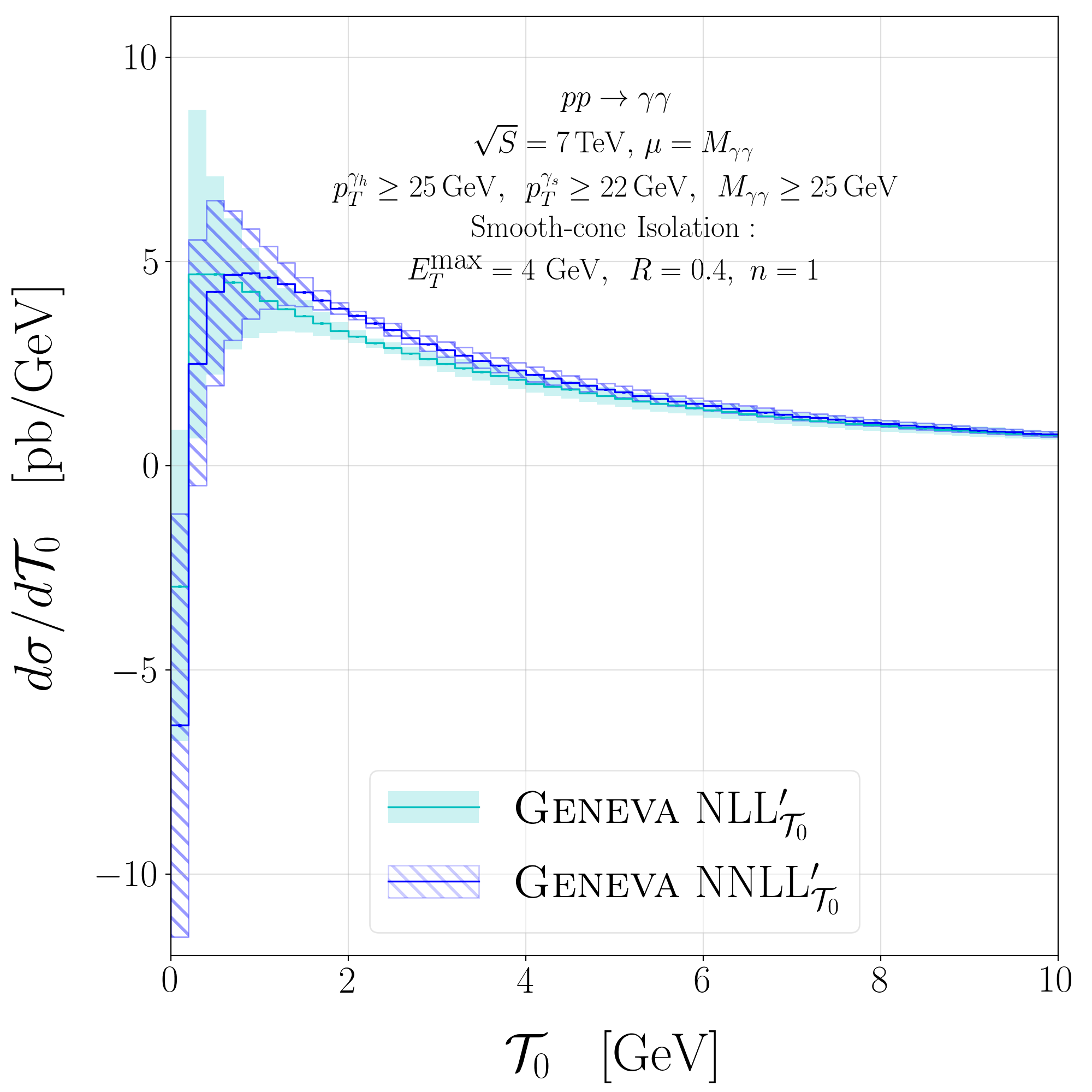} &\hspacebetweentwoplots&
\includegraphics[width=\rescaletwoplots]{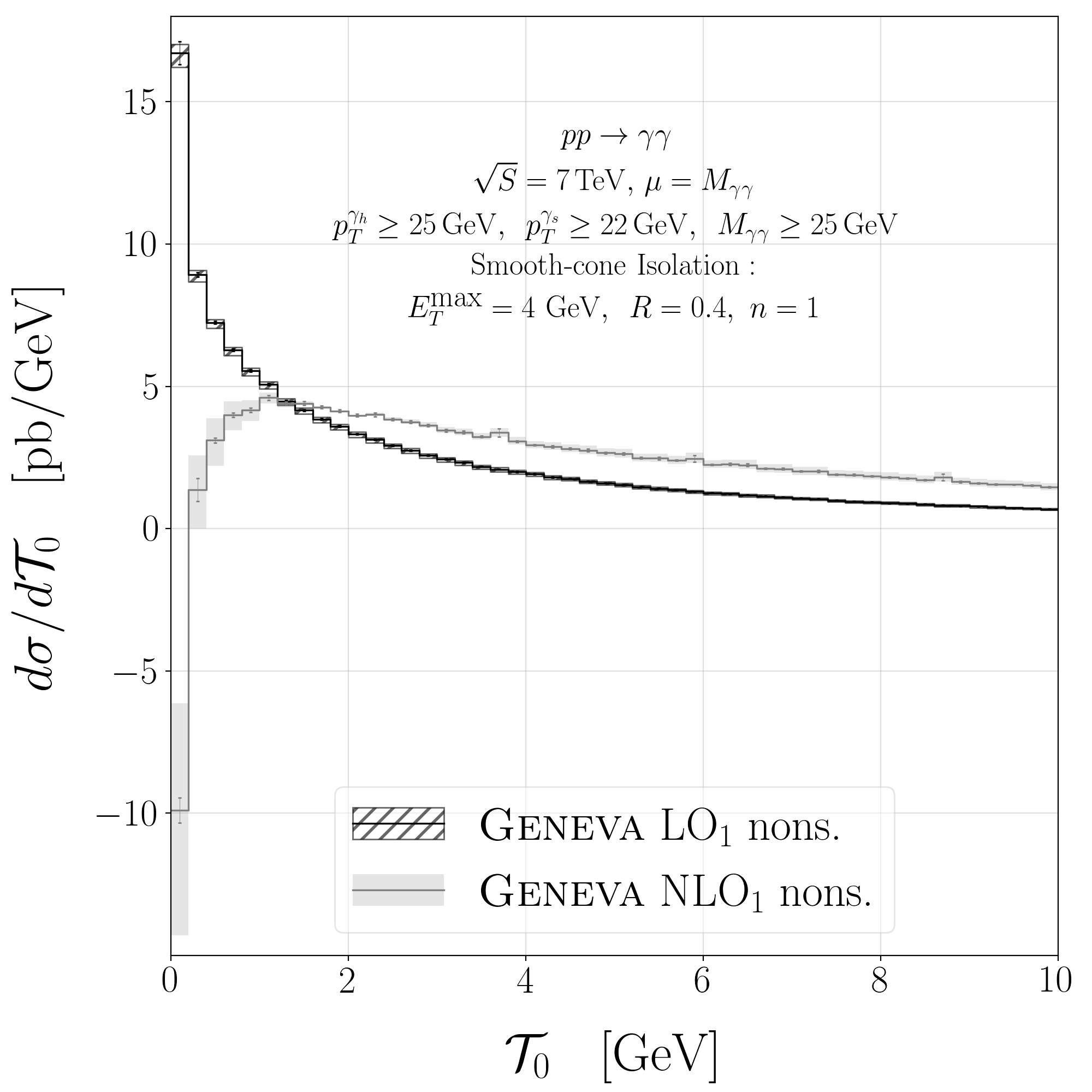}
\end{tabular}
\end{center}
\spaceabovefigurecaption
\caption{Comparison between the NLL$^\prime$ and NNLL$^\prime$ resummed $\Tau_0$ distributions in the peak region (left) and the corresponding nonsingular contribution (right).
\label{fig:resNLLvsNNLLsep}
}
\spacebelowfigurecaption
\end{figure}
In the left panel of  \fig{resNLLvsNNLLsep} we compare the NLL$^\prime$ and
NNLL$^\prime$ results for the $\Tau_0$ distribution in the peak region.
In the same figure we also show the nonsingular contribution at NLO and NNLO in the same range of $\Tau_0$ on the right.

In the peak region, the two results at different resummation
accuracies do overlap, but we do not observe a substantial reduction
of the resummation uncertainties.  We also notice that the nonsingular
contribution at very small $\Tau_0$ values takes opposite signs at the
different orders. Its size also looks particularly large when plotted
on a linear scale, as is done in this plot (\textit{c.f.}
\fig{Tau0SingvsNSing}).

\begin{figure}[tp]
\begin{center}
\begin{tabular}{ccccc}
\includegraphics[width=\rescalethreeplots]{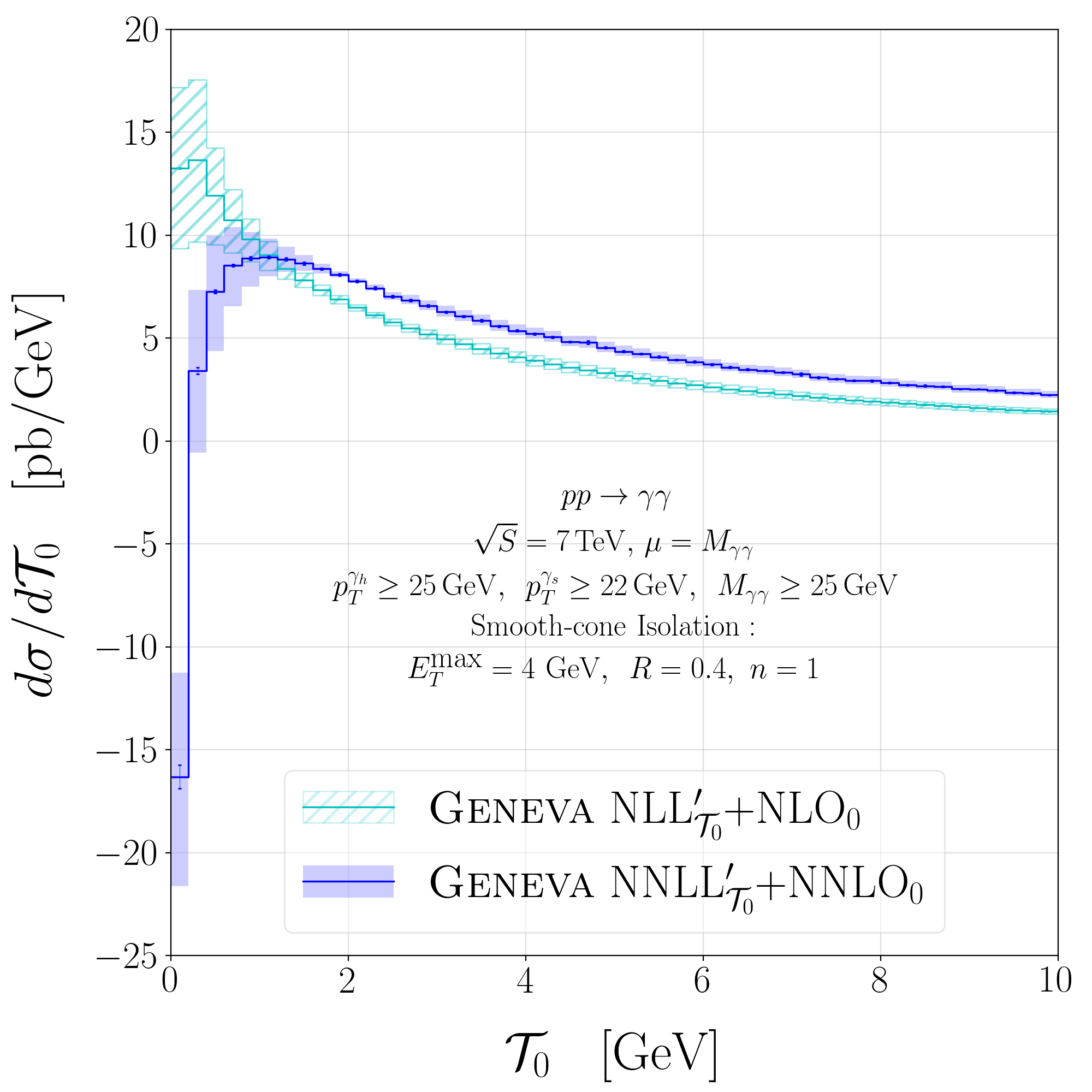} &\hspacebetweenthreeplots
\includegraphics[width=\rescalethreeplots]{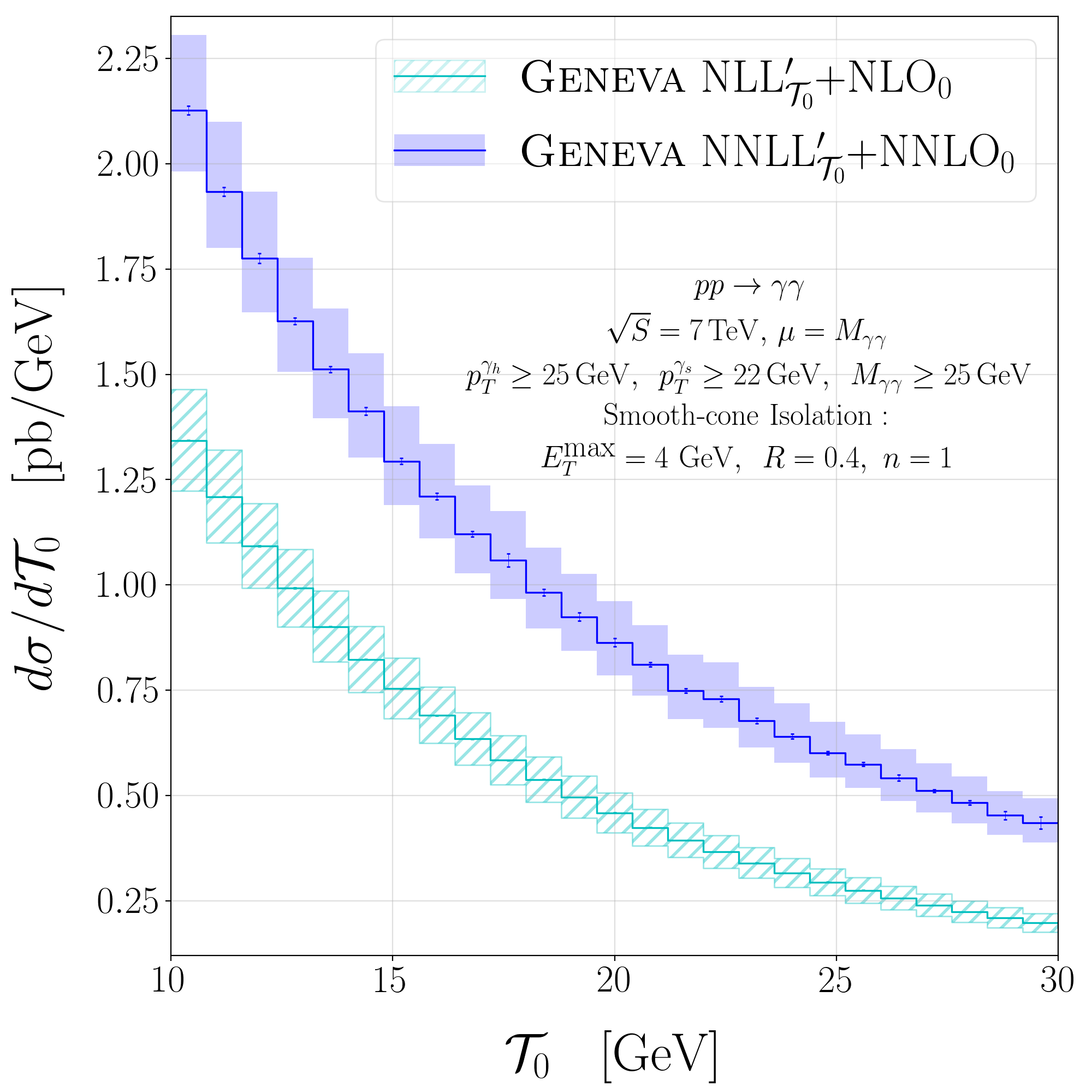} &\hspacebetweenthreeplots
\includegraphics[width=\rescalethreeplots]{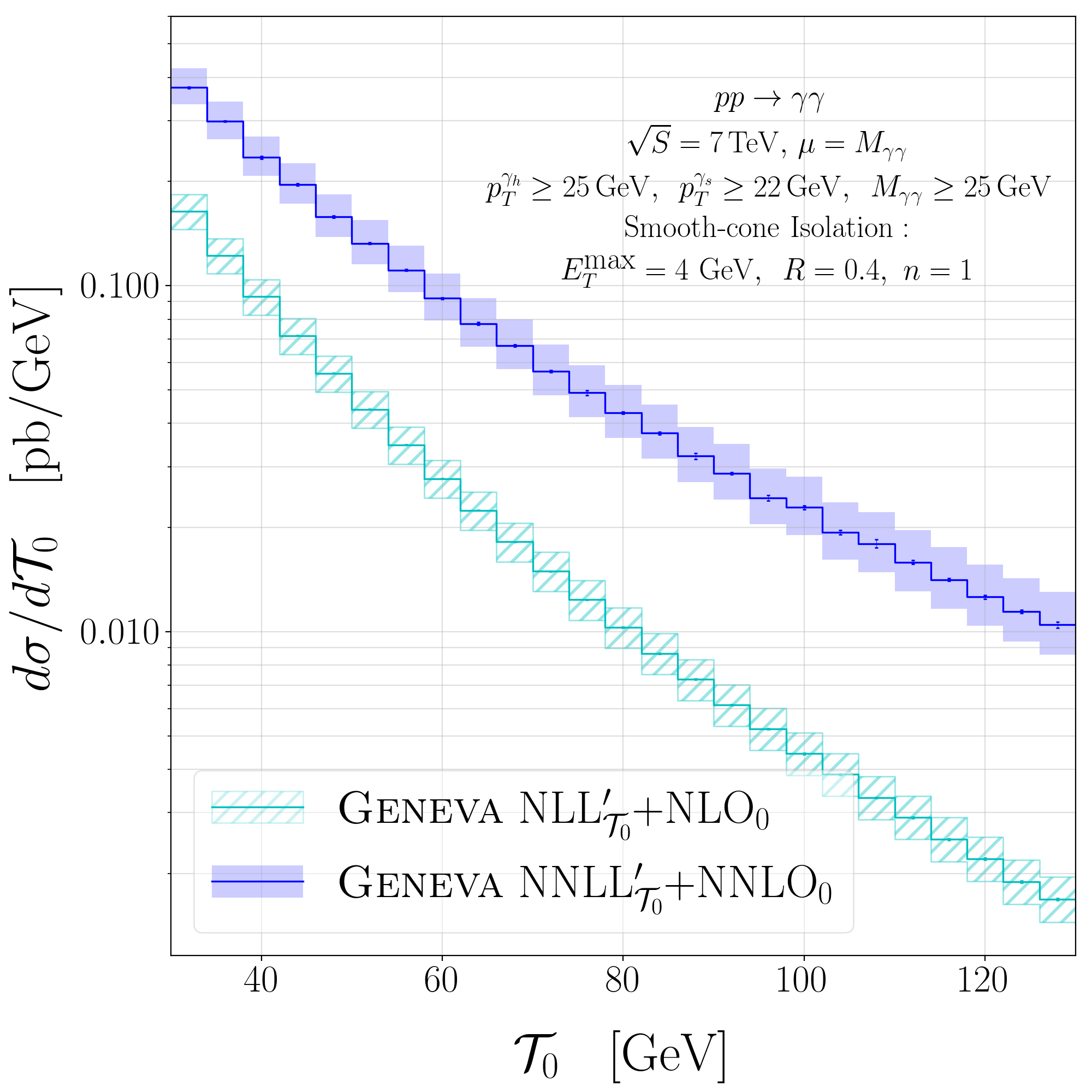}
\end{tabular}
\end{center}
\spaceabovefigurecaption
\caption{Comparison between the NLL$^\prime$+NLO and NNLL$^\prime$+NNLO matched $\Tau_0$ distributions in the peak (left), transition (centre) and tail (right) regions.
\label{fig:resNLLvsNNLL}
}
\spacebelowfigurecaption
\end{figure}
In \fig{resNLLvsNNLL} we instead plot the same resummed results
matched to the appropriate FO calculation, in the peak, transition and
tail regions.  As a consequence of the size of the nonsingular
corrections, the two curves only partially overlap for $1 <\Tau_0<
2$~GeV, close to the peak. A similarly poor convergence was also
observed for the $p^{\gamma \gamma}_T$ distribution after performing
the $q_T$ resummation (see Ref.~\cite{Cieri:2015rqa}).  Also in the
tail and transition regions, the effect of including the NNLO
corrections is large and the uncertainty bands do not overlap with
those at lower order. This was previously noticed in
Refs.~\cite{Catani:2011qz,Catani:2018krb,Campbell:2016yrh}.

\subsection{Subleading power corrections}
\label{subsec:powercorr}

In order to express the 0-jet cross section as in \eq{0full}, \ie
fully differential in the $\Phi_0$ phase space one would need to implement a local NNLO subtraction
method. However, if power corrections below the resolution cutoff are kept negligible by a careful choice of the cutoff, a local subtraction is not explicitly needed. This is the case of the \geneva approach, which is based on the \nj subtraction~\cite{Boughezal:2015dva,Gaunt:2015pea}.

Moreover, even if a local
subtraction were provided, the predictions
of an event generator would be inherently correct only for the total cross
section and for observables which are left unchanged by the
$\Phi_1\to\Phi_0$ and $\Phi_2\to\Phi_1\to \Phi_0$ projections like,
for example, the diphoton invariant mass. Hence, the
presence of power corrections in $\Tau^{\mathrm{cut}}_0$ cannot be
avoided for generic observables that depend on the $\Phi_0$
kinematics. We therefore replace the formula for
the 0-jet cross section in \eq{0full} with
\begin{align}
\frac{\widetilde{\dsigMC_0}}{\df\Phi_0}(\Tau_0^\cut) =&\;
\frac{\df\sigma^{\rm NNLL'}}{\df\Phi_0}(\Tau_0^\cut)\, -
\biggl[\frac{\df\sigma^{\rm NNLL'}}{\df\Phi_{0}}(\Tau_0^\cut)
  \biggr]_{\rm NLO_0} \nn\\ &+(B_0+V_0)(\Phi_0) \, \thetaPSiso(\Phi_0)
\, \nn \\ &+ \int \frac{\mathrm{d} \Phi_1}{\mathrm{d} \Phi_0}
\,\,B_1(\Phi_1)\, \thetaPSiso (\Phi_1)\,
\thetaProj(\widetilde{\Phi}_0)\,\theta\big( \Tau_0(\Phi_1)<
\Tau_0^{\mathrm{cut}}\big)\, ,
\label{eq:0widetilde}
\end{align}
where the local subtraction and the expansion of the resummation
formula are only needed up to $\mathcal{O}(\alpha_s)$. This formula
assumes that there is an exact cancellation between the FO and the
resummed-expanded contribution at $\mathcal{O}(\alpha^2_s)$ below the
$\Tau^\cut_0$. This holds for the singular contributions due to the
NNLL$^\prime$ accuracy of our resummation formula.  However, this
formula is only accurate at leading power in the SCET expansion
parameter and fails to capture the nonsingular contributions in
$\Tau^{\mathrm{cut}}_0$. These can be expressed as
\begin{align} \label{eq:Sigmanons}
\frac{\df\sigma_0^\nons}{\df\Phi_{0}}(\Tau_0^\cut)
&= \bigl[ \as f_1(\Tau_0^\cut, \Phi_0) + \as^2 f_2(\Tau_0^\cut, \Phi_0) \bigr]\, \Tau_0^\cut \, ,
\end{align}
while their integral over the phase space can be written
as
\begin{align}
\Sigma_{\mathrm{ns}} (\Tau_0^\cut) & = \int \df \Phi_0 \,
\frac{\df\sigma_0^\nons}{\df\Phi_{0}}(\Tau_0^\cut) \, .
\end{align}
Since the functions $f_i(\Tau_0^\cut, \Phi_0)$ contain at worst
logarithmic divergences, the nonsingular cumulant vanishes in the
limit $\Tau_0^\cut\to 0$. In our calculation we include the term
$f_1(\Tau_0^\cut, \Phi_0)$ exactly by means of the NLO$_1$ FKS local
subtraction. The $f_2(\Tau_0^\cut, \Phi_0)$ term is instead completely
neglected in \eq{Sigmanons}. This is acceptable as long as we choose
$\Tau_0^\cut$ to be very small.
\begin{figure}[tp]
\begin{center}
\includegraphics[width=\rescaleoneplot]{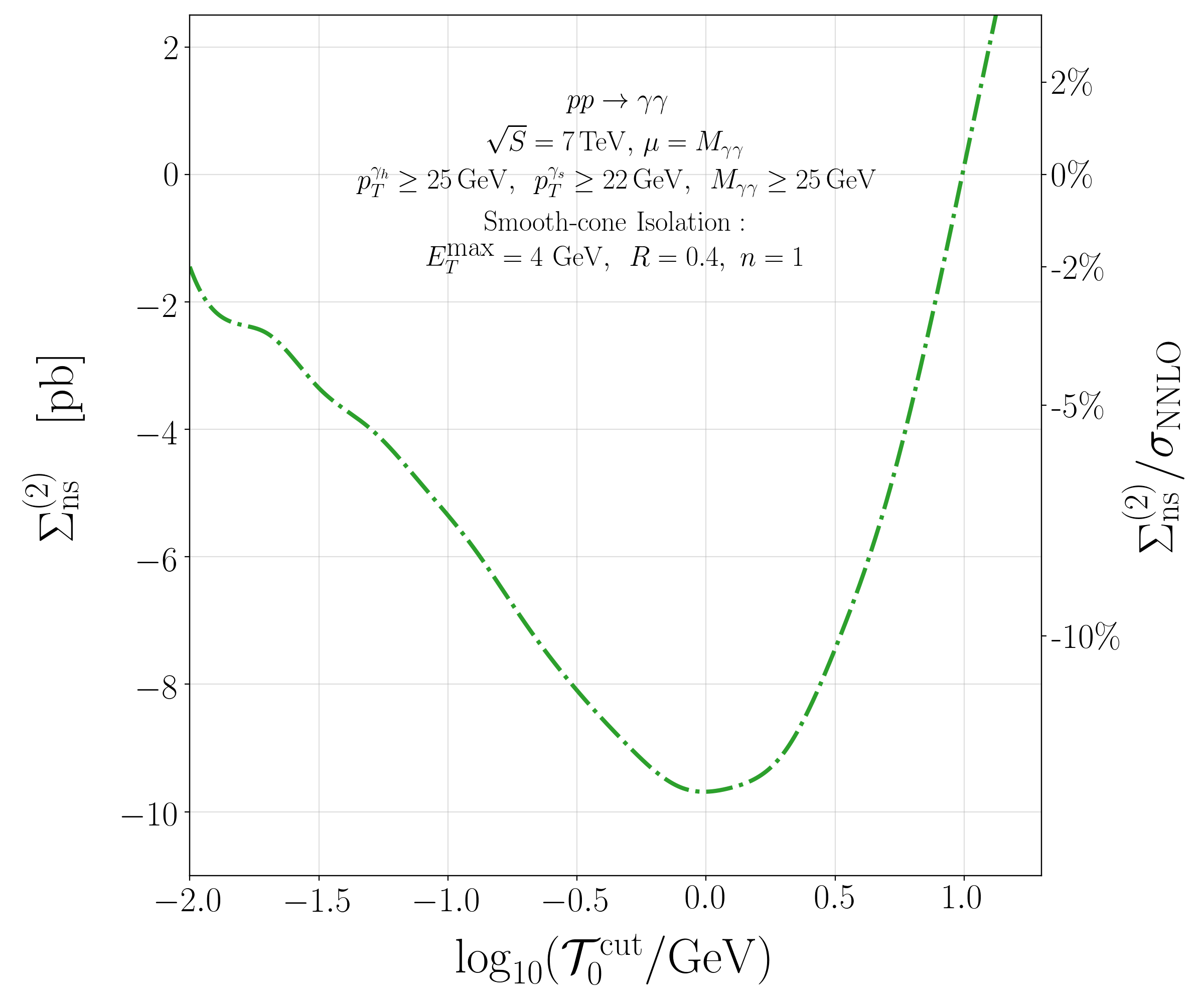}
\end{center}
\spaceabovefigurecaption
\caption{The neglected $\mathcal{O}(\alpha^2_s)$ nonsingular
  contribution to the $\Tau_0$ cumulant, $\Sigma_{\mathrm{ns}}^{(2)}$,
  as a function of $\Tau^{\mathrm{cut}}_0$.
\label{fig:Tau0NSingCum}
}
\spacebelowfigurecaption
\end{figure}
The effect of our approximation is shown in \fig{Tau0NSingCum}, where
we plot the size of the neglected pure $\ord{\as^2}$ terms in
$\Sigma_{\mathrm{ns}} (\Tau_0^\cut)$ as a function of
$\Tau^\cut_0$. The size of the missing contributions is not completely
negligible and to reduce their impact we run with a default cut value
of $\Tau^\cut_0=0.01$ GeV. The magnitude of the missing corrections
for such value of the cut is around $1.45$~pb (which corresponds to
$\sim 2\%$ of the total cross section for the particular set of cuts
chosen).  Comparing this result to the previous Drell--Yan and $VH$
calculations, we notice that in the diphoton case the relative size of
the nonsingular corrections below the cut is larger.

One could improve on this by systematically calculating the subleading
terms in the expansion parameter using a SCET formalism. Presently,
only the first terms in the expansion are known, for a limited set of
processes~\cite{Moult:2016fqy,Boughezal:2016zws,Moult:2017jsg}.

We eventually provide the missing nonsingular
$\mathcal{O}(\alpha^2_s)$ contributions from an independent NNLO
calculation obtained with \matrix~\cite{Grazzini:2017mhc}, by simply
rescaling the weights of the $\Phi_0$ events by the cross section
ratios. We remind the reader that the \matrix calculation is based on
the $q_T$ subtraction method, which is very similar in spirit to \zj
subtraction and thus in principle affected by the same issue.
However, \matrix uses an extrapolation procedure for $q^\cut_T \to 0$
which provides an estimate of the cross section and its numerical
error. Therefore, reweighting the $\Phi_0$ events using the total
cross section inputs from \matrix provides us with NNLO accuracy.

\begin{figure}[tp]
\begin{center}
\begin{tabular}{ccc}
\includegraphics[width=\rescaletwoplots]{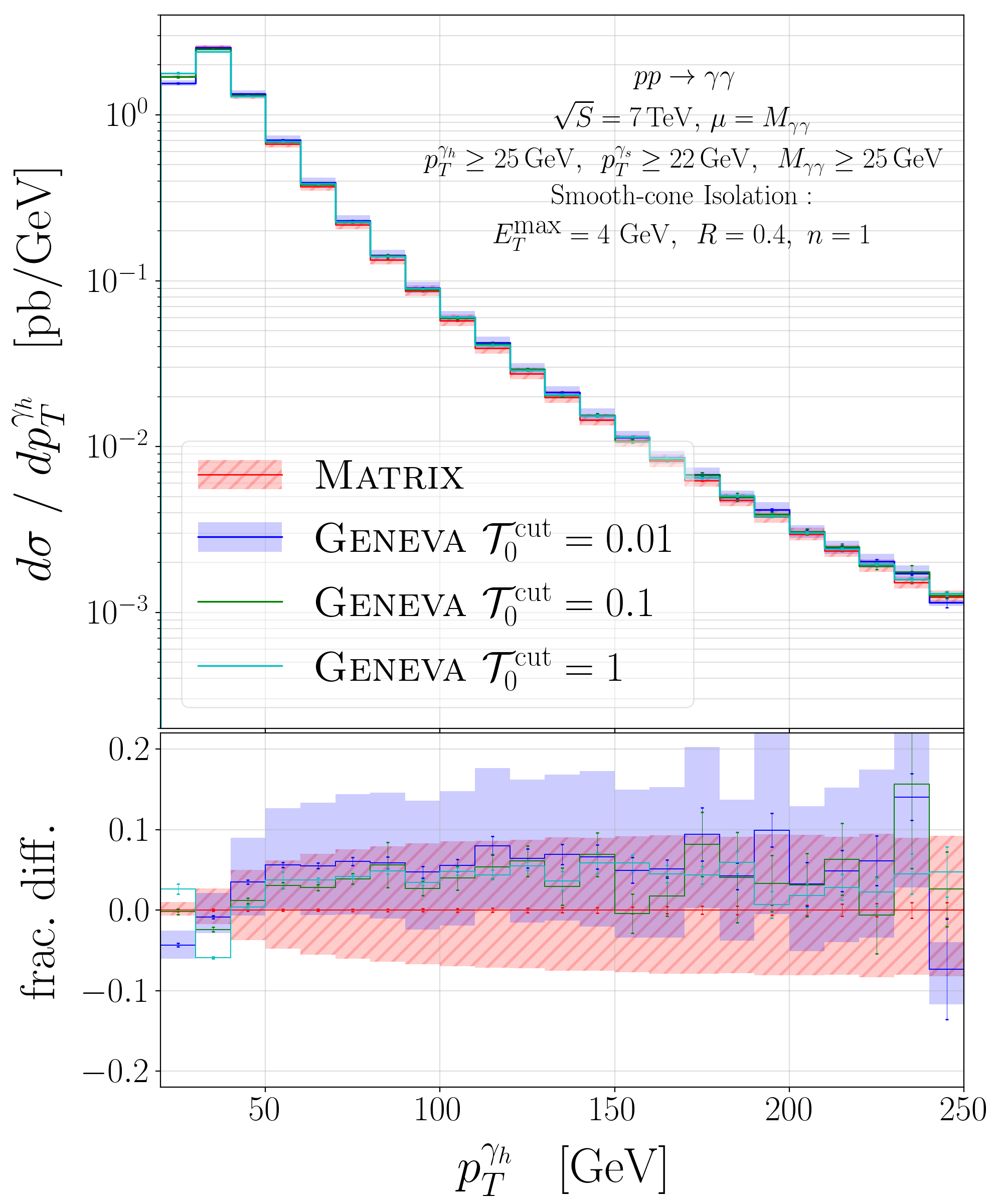} &\hspacebetweentwoplots&
\includegraphics[width=\rescaletwoplots]{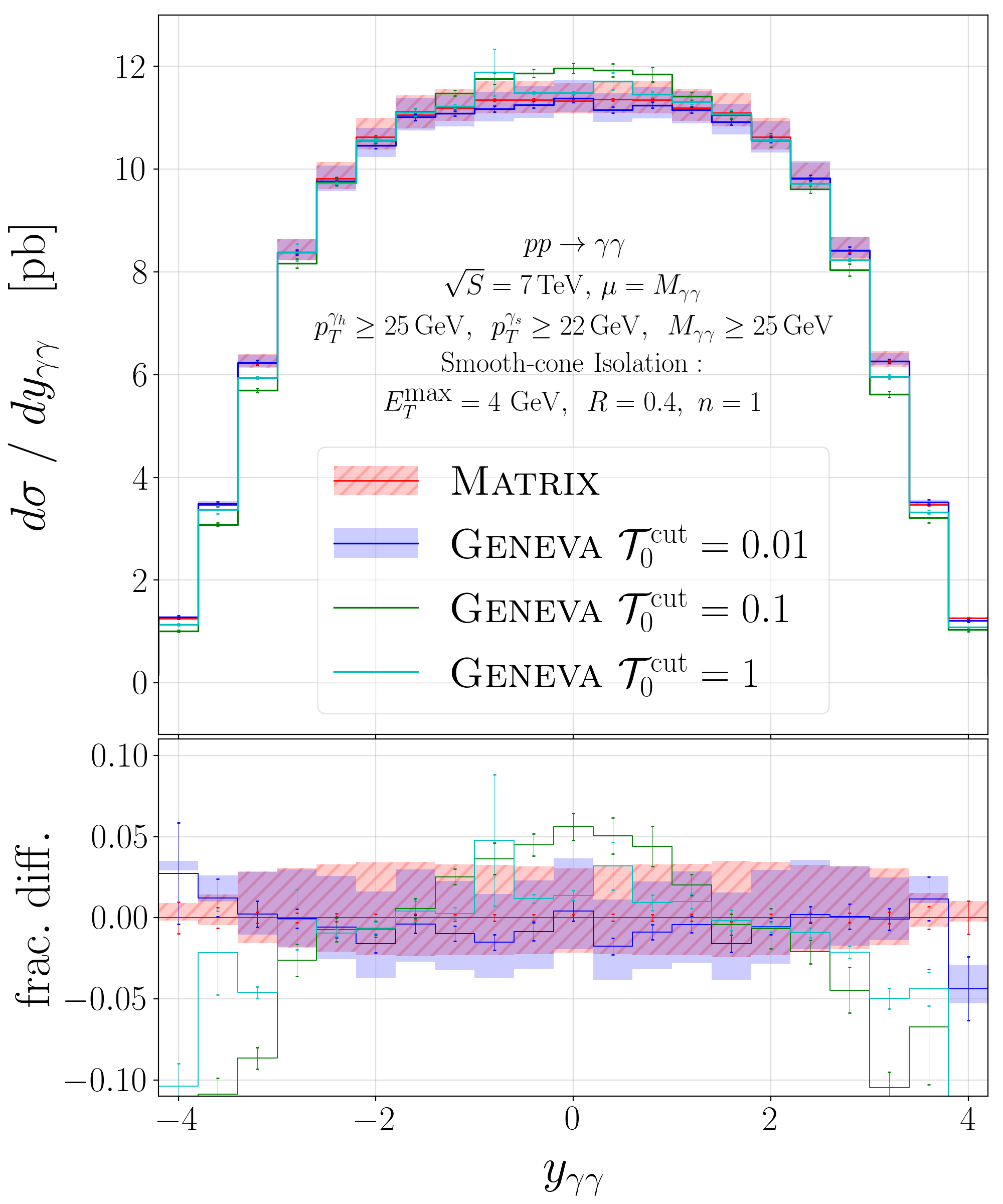} \\[\vspacebetweentwoplots]
\includegraphics[width=\rescaletwoplots]{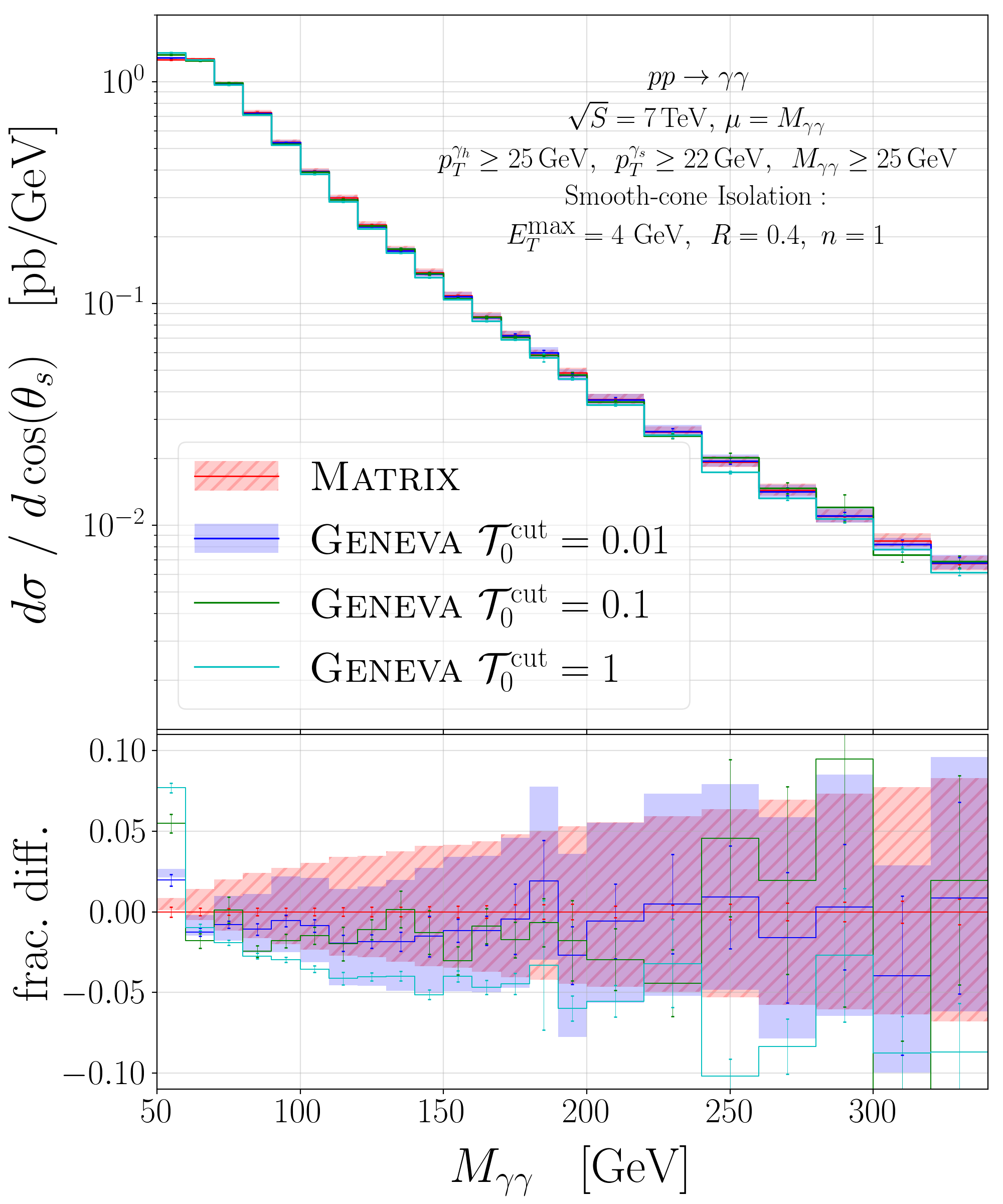} &\hspacebetweentwoplots&
\includegraphics[width=\rescaletwoplots]{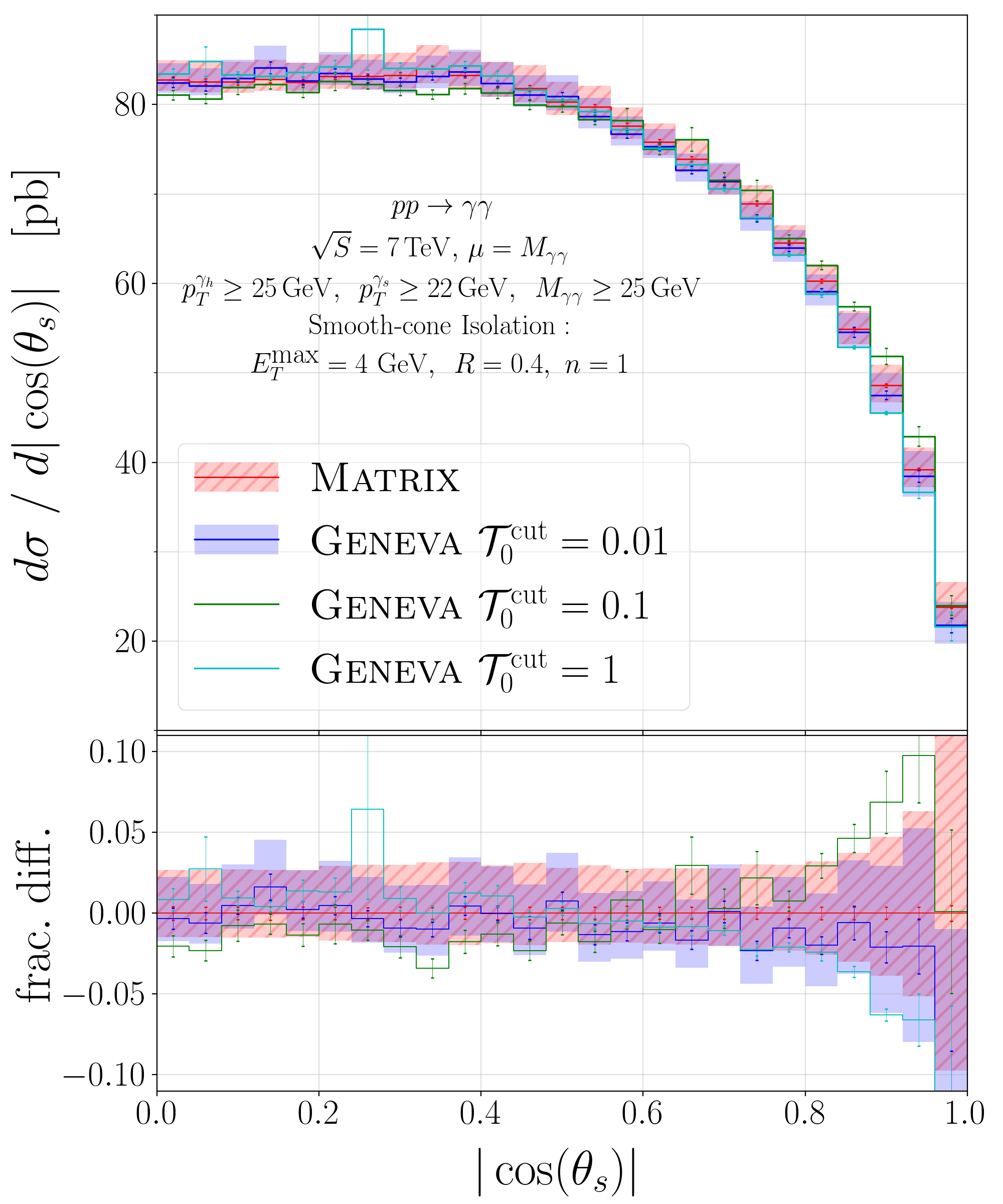} \\
\end{tabular}
\end{center}
\spaceabovefigurecaption
\caption{Comparison between \matrix and \geneva for different values of $\Tau^\cut_0$. We show the transverse momentum of the hardest photon (top left), rapidity of the diphoton system (top right), invariant mass of the diphoton system (bottom left) and the cosine of the photon scattering angle (bottom right).
\label{fig:gvavsmatrixqqbar}
}
\spacebelowfigurecaption
\end{figure}
%
\subsection{NNLO validation}
\label{subsec:genevavalidation}

After reweighting the $\Phi_0$ events for the central scale choice as
well as for its variations, we compare the inclusive (\ie not probing additional radiation) distributions
 obtained with \geneva to the independent NNLO
results obtained with \matrix~\cite{Grazzini:2017mhc}. This check
is nontrivial since the complete dependence on the $\Phi_0$ kinematics
is not, in general, captured by the reweighting.

In \fig{gvavsmatrixqqbar} we show the transverse momentum of the
hardest photon, the rapidity and the invariant mass of the diphoton
system, and the absolute value of the cosine of the photon scattering
angle in the frame of the LO partonic collision, defined as
\begin{align}
|\cos \theta_s| = \tanh\bigg(\frac{|\Delta y_{\gamma \gamma}|}{2}\bigg)\, ,
\end{align}
where $\Delta y_{\gamma \gamma}$ is the diphoton rapidity separation.
After comparing three different choices of
$\Tau^{\mathrm{cut}}_0=\{0.01,0.1,1\}\,\, \mathrm{GeV}$ in \geneva
we conclude that the best agreement with the NNLO predictions for
these inclusive distributions is obtained for
$\Tau^{\mathrm{cut}}_0=0.01$~GeV, as expected from the study of the
missing power corrections (see \subsec{powercorr}). We
therefore set $\Tau^\cut_0$ to this default value for all of our
predictions.

As mentioned above, the \geneva predictions, despite being NNLO
accurate, are not exactly equivalent to those of a NNLO calculation:
indeed, they differ by power-suppressed terms as a consequence of the
projective map which is used to define the $\Phi_0$ events and by
higher-order resummation effects that are not completely removed even
after the inclusion of the additional terms in
\eq{term}. Nevertheless, the agreement between \geneva and \matrix for
the distributions in \fig{gvavsmatrixqqbar} is good, within the
uncertainty bands of the two calculations (representing the 3-point
$\mu_r$ and $\mu_f$ variations).

\subsection{Resumming the 1-jet/2-jet separation at LL}
\label{subsec:genevaTau1res}

In order to provide an event generator which is as flexible as
possible, and thus also able to provide exclusive predictions for
higher-multiplicity bins, we proceed with the separation of the
inclusive $1$-jet cross section into an exclusive $1$-jet cross
section and an inclusive $2$-jet cross section.  We can achieve this
separation by using the $\Tau_1$ resolution variable.  This introduces
a new scale which in principle requires a simultaneous resummation of
all the different ratios between the scales $\Tau_0$, $\Tau_1$ and
$Q$, in all the possible kinematic regions. A fully satisfactory
treatment of this kind is still lacking, but, in the region $\Tau_1
\ll \Tau_0$, we can take the simpler approach explained next. We
concentrate first on the $\Tau_1$ resummation and start by separating
\begin{align}
\frac{\dsigMC_{1}}{\df\Phi_{1}} (\Tau_0 > \Tau_0^\cut; \Tau_{1}^\cut)
=&\; \frac{\df\sigma_1^{\rm LL}}{\df\Phi_{1}}(\Tau_0 > \Tau_0^\cut; \Tau_{1}^\cut)\, \nn \\
&\hspace*{2em}+ \frac{\df\sigma_1^{\mathrm{match}}}{\df \Phi_{1}}(\Tau_0 > \Tau_0^\cut; \Tau_{1}^\cut)
\,,  \label{eq:1master}
\\
\frac{\dsigMC_{\geq 2}}{\df\Phi_{2}} (\Tau_0 > \Tau_0^\cut, \Tau_{1}>\Tau_{1}^\cut)
=&\; \frac{\df\sigma^{\rm LL}_{\geq 2}}{\df\Phi_2}\big(\Tau_0 > \Tau_0^\cut , \Tau_{1} > \Tau_{1}^\cut \big)
\, \nn \\
&\hspace*{2em}+\frac{\df \sigma^{\mathrm{match}}_{\geq 2}}{\df \Phi_2}(\Tau_0 > \Tau_0^\cut, \Tau_1 > \Tau_1^\cut)
\,,
\label{eq:2master}
\end{align}
where the terms $\df\sigma_1^{\rm LL}$ and $\df\sigma^{\rm LL}_{\geq
2}$ contain the LL resummation of the $\Tau_1^\cut$ and $\Tau_1$
dependencies respectively.  The $\df\sigma_1^{\mathrm{match}}$ and $\df
\sigma^{\mathrm{match}}_{\geq 2}$ terms contain the matching
corrections to the required FO accuracy.  At this point we can
implement a unitary approach such that the resummed contributions take
the form
\begin{align} \label{eq:Tau1resum}
\frac{\df\sigma_1^{\rm LL}}{\df\Phi_{1}}(\Tau_0 > \Tau_0^\cut; \Tau_{1}^\cut) =&\; \frac{\df\sigma_{\geq 1}^C}{\df\Phi_1} \, U_1(\Phi_1, \Tau_1^\cut)\, \thetaPSiso (\Phi_1) \, \theta(\Tau_0 > \Tau_0^\cut)
\,,\\[1ex]
\label{eq:Tau1resum2}
\frac{\df\sigma^{\rm LL}_{\geq 2}}{\df\Phi_2}(\Tau_0 > \Tau_0^\cut, \Tau_1 > \Tau_1^\cut) =&\;
\frac{\df\sigma_{\geq 1}^C}{\df\Phi_1}\, U_1'(\widetilde{\Phi}_1, \Tau_1)\, \theta(\Tau_0 > \Tau_0^\cut) \Big\vert_{\widetilde{\Phi}_1 = \Phi_1^\Tau(\Phi_2)} \, \nn \\
&\hspace*{3em} \times \thetaPSiso (\Phi_2)\, \thetaProj(\widetilde{\Phi}_1)\,  \cP(\Phi_2) \, \theta\left(\Tau_1 > \Tau_1^\cut\right)\,.
\end{align}
The evolution function\footnote{The explicit expressions for the
  evolution function $U_1(\Phi_1, \Tau_1^\cut)$ up to NLL accuracy can
  be found in sec.~2 of Ref.~\cite{Alioli:2019qzz}.} (or Sudakov
factor) $U_1(\Phi_1, \Tau_1^\cut)$ resums the $\Tau_1^\cut$ dependence
at LL accuracy in the region $\Tau_1^\cut \ll \Tau_0$, while
$U_1'(\Phi_1, \Tau_1)$ is the derivative of the evolution function
with respect to $\Tau_1^\cut$.  The latter resums the differential
$\Tau_1$ dependence and is evaluated at the projected configuration
$\widetilde{\Phi}_1 = \Phi_1^\Tau(\Phi_2)$.  The function
$\cP(\Phi_2)$ is a normalised splitting probability which is defined
similarly to $\cP (\Phi_1)$ in \eq{Pnorm}.  The quantity
$\df\sigma_{\geq 1}^C/\df\Phi_1$, which appears both in
\eqs{Tau1resum}{Tau1resum2}, is the inclusive 1-jet cross section in
the singular $\Tau_1\to 0$ limit. Its NLO$_1$ expansion is given by
\begin{align} \label{eq:NLO1singular}
\biggl[\frac{\df\sigma^{C}_{\geq 1}}{\df\Phi_{1}}\biggr]_{\NLO_1}
=&\; \bigg(B_1(\Phi_1) + V_1(\Phi_1)  + \int\! \frac{\df\Phi_2}{\df \Phi^C_1} \,C_2(\Phi_2)\bigg)\, \thetaPSiso (\Phi_1) \, \nn \\
\equiv &\; (B_1 + V_1^C)(\Phi_1)  \thetaPSiso (\Phi_1) \,,
\end{align}
where $C_2(\Phi_2)$ reproduces the point-wise singular behaviour of
$B_2(\Phi_2)$ and acts as a local subtraction at
NLO~\cite{Frixione:1995ms} with its own projection
$\df\Phi_2/\df\Phi^C_1
\equiv \df\Phi_2\,\delta[\Phi_1 - \Phi_1^C(\Phi_2)]$.  After requiring
that $\dsigMC_{1}$ and $\dsigMC_{\geq 2}$ are accurate to NLO$_1$ and
LO$_2$ respectively, the matching corrections are expressed as
\begin{align}
\label{eq:1match}
\frac{\df\sigma_1^{\mathrm{match}}}{\df\Phi_{1}}(\Tau_0>\Tau_0^\cut;\Tau_1^\cut) =&\; \frac{\df\sigma_1^{{\rm NLO_1}}}{\df\Phi_{1}}(\Tau_0>\Tau_0^\cut;\Tau_1^\cut)\, \nn \\
&\hspace*{2em}- \biggl[\frac{\df\sigma_1^{\rm LL}}{\df\Phi_{1}}(\Tau_0>\Tau_0^\cut;\Tau_1^\cut) \biggr]_{\rm NLO_1}
\,,
\\[1ex]
\label{eq:2match}
\frac{\df\sigma_{\ge 2}^{\mathrm{match}}}{\df\Phi_{2}}(\Tau_0 > \Tau_0^\cut,\Tau_1 > \Tau_1^\cut) =&\; \frac{\df\sigma_{\ge 2}^{{\rm LO_2}}}{\df\Phi_{2}}(\Tau_0 > \Tau_0^\cut,\Tau_1 > \Tau_1^\cut) \, \nn \\
&\hspace*{2em}- \biggl[\frac{\df\sigma_{\geq 2}^{\rm LL}}{\df\Phi_2} \left(\Tau_0 > \Tau_0^\cut, \Tau_1 > \Tau_1^\cut\right)  \biggr]_{\rm LO_2} \,.
\end{align}
After inserting \eq{Tau1resum}, \eq{Tau1resum2} and
\eq{NLO1singular}  in the above equations and taking into account
the appropriate phase space restrictions we find
\begin{align}
\label{eq:Phi1Match}
\frac{\df\sigma_1^{\mathrm{match}}}{\df\Phi_1}(\Tau_0 > \Tau_0^\cut; \Tau_1^\cut)
= & \int\ \biggl[\frac{\df\Phi_{2}}{\df\Phi^\Tau_1}\,B_{2}(\Phi_2)\, \thetaPSiso(\Phi_2)\, \theta\left(\Tau_0(\Phi_2) > \Tau_0^\cut\right)\, \theta(\Tau_{1} < \Tau_1^\cut)\nn \\
& - \frac{\df\Phi_2}{\df \Phi_1^C}\, C_{2}(\Phi_{2}) \thetaPSiso (\Phi_1) \, \theta(\Tau_0 > \Tau_0^\cut) \biggr] \nn \\
&- B_1(\Phi_1)\, U_1^\one(\Phi_1, \Tau_1^\cut)\, \thetaPSiso (\Phi_1) \theta(\Tau_0 > \Tau_0^\cut) \, ,\\[-5ex]\nn
\end{align}
\begin{align}
\label{eq:Phi2Match}
\frac{\df\sigma_{\geq 2}^{\mathrm{match}}}{\df\Phi_2}(\Tau_0 > \Tau_0^\cut, \Tau_1 > \Tau_1^\cut)
= &\; \thetaPSiso(\Phi_2) \, \theta\left(\Tau_0(\Phi_2) > \Tau_0^\cut\right)\, \big[ B_2(\Phi_2)\, \theta(\Tau_{1}>\Tau^{\mathrm{cut}}_{1})\, \nn \\
&- B_1(\Phi_1^\Tau)\,U_1^{\one\prime}(\widetilde{\Phi}_1, \Tau_1)\,\cP(\Phi_2)\,\thetaProj(\widetilde{\Phi}_1)\, \theta(\Tau_1 > \Tau_1^\cut)
\big]\, \,.
\end{align}
In the above expressions $U^{(1)}_1(\Phi_1,\Tau^\cut_1)$ and $U^{(1)\,
\prime}_1(\Phi_1,\Tau_1)$ indicate the $\mathcal{O}(\alpha_s)$
expansions of the evolution function $U_1(\Phi_1, \Tau_1^\cut)$ and of
its derivative $U_1'(\Phi_1, \Tau_1)$ respectively. Since the $\Tau_1$
resummation is carried out to LL accuracy, the matching corrections
still contain subleading single-logarithmic terms.

So far we have presented a NLO$_1$+LL$_{\Tau_1}$ matched result, but
we still need to incorporate the $\Tau_0$ resummation
that we discussed in the previous sections. We achieve this by
requiring that the integral of the NLO$_1$+LL$_{\Tau_1}$ result
reproduces the $\Tau_0$-resummed result for the inclusive 1-jet MC
cross section $\dsigMC_{\geq 1}$
\begin{align} \label{eq:MC1plusMC2}
\frac{\dsigMC_{\geq 1}}{\df\Phi_1}(\Tau_0 > \Tau_0^\cut)
=&\; \frac{\dsigMC_{1}}{\df\Phi_{1}} (\Tau_0 > \Tau_0^\cut; \Tau_{1}^\cut)
+ \int\!\frac{\df\Phi_{2}}{\df\Phi^\Tau_{1}}\, \frac{\dsigMC_{\ge 2}}{\df\Phi_{2}} (\Tau_0 > \Tau_0^\cut, \Tau_{1} > \Tau_{1}^\cut) \, .
\end{align}
The next step for the process at hand is nontrivial and requires some detailed explanation. In the
case of Drell--Yan or $VH$ production, once this point was reached in
the derivation one could simply proceed by summing the two
contributions in the equation above. In this manner one could
obtain a result which was independent of the evolution function, of
its derivative and of the respective FO expansions, by
exploiting the unitarity condition
\begin{align}
\label{eq:unitarity}
U_1(\Phi_1, \Tau_1^\cut) + \int \! \frac{\df \Phi_2}{\df \Phi^\Tau_1}\, U_1'(\Phi_1, \Tau_1)\, \cP(\Phi_2)\,
\theta(\Tau_1 > \Tau_1^\cut) = 1  \,.
\end{align}
Unfortunately, in the case of diphoton production, this step is
complicated by the presence of additional phase space restrictions on
$\Phi_2$, due to the application of process-defining cuts, isolation
cuts and projection cuts.
Assuming that \eq{unitarity} holds to to a good approximation despite the presence of all  these cuts, we obtain
\begin{align}
\label{eq:MC1plusMC2approx}
\frac{\dsigMC_{\geq 1}}{\df\Phi_1}(\Tau_0 > \Tau_0^\cut) =&\; \frac{\df\sigma_{\geq 1}^C}{\df\Phi_1}\, \theta(\Tau_0 > \Tau_0^\cut)
+ \int \! \bigg[
\frac{\df \Phi_2}{\df \Phi^\Tau_1} \, B_2(\Phi_2)\,\thetaPSiso (\Phi_2) \,\theta\left(\Tau_0(\Phi_2) > \Tau_0^\cut\right) \, \nn \\
&- \frac{\df\Phi_2}{\df\Phi_1^C}\, C_2(\Phi_2)\,\thetaPSiso (\Phi_1) \,\theta(\Tau_0 > \Tau_0^\cut) \bigg] + \ldots\,,
\end{align}
where the dots represent the remaining unitarity violating terms.  We
verified that, when implementing the $\Tau_1$ resummation at LL, these
terms are indeed numerically small  both for the total cross
section and at the differential level for the $\Tau_0$ distribution.
Therefore, we neglect them in the
following.\footnote{ A possible general solution to this problem would
  be to enforce the identity in \eq{unitarity} and define
  $U_1(\Phi_1, \Tau_1^\cut)$ to be the function which fulfils
  \eq{unitarity} even in the presence of phase space cuts on the
  $\Phi_2$ integration. $U_1(\Phi_1, \Tau_1^\cut)$ could be computed
  numerically as
  \begin{align} U_1(\Phi_1, \Tau_1^\cut) \equiv 1- \int \! \frac{\df
    \Phi_2^\Tau}{\df \Phi_1}\, U_1'(\Phi_1, \Tau_1)\, \cP(\Phi_2)\,
    \theta(\Tau_1 > \Tau_1^\cut)\, \Theta^{\mathrm{PS}}(\Phi_2) \nn\,,
  \end{align} where $\Theta^{\mathrm{PS}}(\Phi_2)$ indicates a set of
  phase space cuts. However, such a choice would have the
    drawback that the introduction of cuts would make the $\Tau_1$ resummation
accuracy no longer clearly specified.}   By comparing the expression
on the r.h.s.\ of \eq{MC1plusMC2approx} with the r.h.s.\ of
\eq{sigma>=1} we obtain the result for $\df\sigma_{\geq 1}^C$\,,
\begin{align}
\label{eq:sigmaC>=1}
\frac{\df\sigma_{\geq 1}^C}{\df\Phi_{1}}(\Tau_0 > \Tau_0^\cut)
=&\; \Bigg\{\frac{\df\sigma^{\rm NNLL'}}{\df\Phi_{0}\df \Tau_0} -  \biggl[\frac{\df\sigma^{\rm NNLL'}}{\df\Phi_0 \df \Tau_0} \biggr]_{\rm NLO_1}\Bigg\}\, \cP(\Phi_1)\, \theta(\Tau_0 > \Tau_0^\cut)\,  \thetaProj(\widetilde{\Phi}_0) \, \nn \\
&+(B_1+V^C_1)(\Phi_1) \, \theta (\Tau_0(\Phi_1)> \Tau^{\mathrm{cut}}_0) \, .
\end{align}
Finally, we obtain the complete formulae for the exclusive 1-jet and
the inclusive 2-jet cross sections
as implemented in the \geneva code:
\begin{align}
\label{eq:1belowtau0}
\frac{\dsigMC_{1}}{\df\Phi_{1}} (\Tau_0 \le \Tau_0^\cut; \Tau_{1}^\cut)
=&\; \thetaPSiso (\Phi_1)  \, B_1\, (\Phi_1)\,\theta(\Tau_0<\Tau^{\mathrm{cut}}_0)\,  \big[ \overline{\Theta}_{\mathrm{iso}}^{\mathrm{proj}}(\widetilde{\Phi}_0)  + \overline{\Theta}^{\mathrm{FKS}}_{\mathrm{map}}(\Phi_1)  \big] \,,\\[1ex]
\label{eq:1masterfull}
\frac{\dsigMC_{1}}{\df\Phi_{1}} (\Tau_0 > \Tau_0^\cut; \Tau_{1}^\cut)
=&\; \thetaPSiso (\Phi_1)\Bigg\{\Bigg[ \frac{\df\sigma^{\rm NNLL'}}{\df\Phi_0\df\Tau_0}-\frac{\df\sigma^{\rm NNLL'}}{\df\Phi_0\df\Tau_0}\bigg\vert_{\NLO_1}\,\Bigg]\, \cP(\Phi_1)\, \thetaProj(\widetilde{\Phi}_0) \, \nn \\
&\hspace*{5em}+ \big[B_1 + V_1^C\big](\Phi_1)  \Bigg\} \times  U_1(\Phi_1, \Tau_1^\cut)\, \theta(\Tau_0 > \Tau_0^\cut)
\nn \\
&\hspace*{-6em}+\int\ \bigg[\frac{\df\Phi_{2}}{\df\Phi^\Tau_1}\,B_{2}(\Phi_2)\, \thetaPSiso (\Phi_2)\,\thetaProj (\widetilde{\Phi}_1)\, \theta\left(\Tau_0(\Phi_2) > \Tau_0^\cut\right)\,\theta(\Tau_{1} < \Tau_1^\cut)\, \nn \\
&\hspace*{-1em}- \frac{\df\Phi_2}{\df \Phi^C_1}\, C_{2}(\Phi_{2})\,\thetaPSiso (\Phi_1) \, \theta(\Tau_0 > \Tau_0^\cut) \bigg] \, \nn \\
&\hspace*{-6em}- B_1(\Phi_1)\, U_1^\one(\Phi_1, \Tau_1^\cut)\,\thetaPSiso (\Phi_1)\, \theta(\Tau_0 > \Tau_0^\cut)
\,, \\[-5ex] \nn
\end{align}
\begin{align}
\label{eq:2belowtau1}
\frac{\dsigMC_{\geq 2}}{\df\Phi_{2}}  (\Tau_0 > \Tau_0^\cut, \Tau_{1} \le \Tau_{1}^\cut)
=&\;  B_2(\Phi_2)\, \thetaPSiso (\Phi_2)\, \Big[\overline{\Theta}_{\mathrm{map}}^\Tau(\Phi_2)+\thetaBarProj(\widetilde{\Phi}_1)\Big]\,\nn \\
&\times \, \theta(\Tau_1 < \Tau_1^\cut)\, \theta\left(\Tau_0(\Phi_2) > \Tau_0^\cut\right) \,,\\[1ex]
\label{eq:2masterfull}
\frac{\dsigMC_{\geq 2}}{\df\Phi_{2}} (\Tau_0 > \Tau_0^\cut, \Tau_{1}>\Tau_{1}^\cut)
=&\; \thetaPSiso (\Phi_2)\, \Bigg\{ \bigg[ \frac{\df\sigma^{\rm NNLL'}}{\df\Phi_0\df\Tau_0} - \frac{\df\sigma^{\rm NNLL'}}{\df\Phi_0\df\Tau_0}\bigg|_{\NLO_1}\bigg]\, \cP(\widetilde{\Phi}_1)\, \thetaProj(\widetilde{\Phi}_0)\, \nn \\
&\hspace*{-11em} + (B_1 + V_1^C)(\widetilde{\Phi}_1)\Bigg\} \,  U_1'(\widetilde{\Phi}_1, \Tau_1)\, \theta(\Tau_0 > \Tau_0^\cut) \Big\vert_{\widetilde{\Phi}_1 = \Phi_1^\Tau(\Phi_2)} \!\! \,\thetaProj(\widetilde{\Phi}_1) \cP(\Phi_2) \, \theta(\Tau_1 > \Tau_1^\cut)
\nn \\
&\hspace*{-12em}+ \thetaPSiso (\Phi_2)\, \Big\{ B_2(\Phi_2)\, \theta(\Tau_{1}>\Tau^{\mathrm{cut}}_{1})
\, \nn \\
&\hspace*{-11em}- B_1(\Phi_1^\Tau)\,U_1^{\one\prime}\big(\widetilde{\Phi}_1, \Tau_1\big)\,\cP(\Phi_2)\,\thetaProj(\widetilde{\Phi}_1)\, \Theta(\Tau_1 > \Tau_1^\cut)
\Big\}\, \theta\left(\Tau_0(\Phi_2) > \Tau_0^\cut\right)\,.
\end{align}
In order to simplify the notation, in the above equations
we use the same symbol for two different projections. In
\eqs{1belowtau0}{1masterfull} the symbol $\widetilde{\Phi}_0$ refers to
the single projection $\Phi_1\to\widetilde{\Phi}_0$ using the FKS
mapping, while in \eq{2masterfull} the same symbol refers to the phase
space point obtained after a double projection $\Phi_2\to
\widetilde{\Phi}_1\to \widetilde{\Phi}_0$, where the first projection
is evaluated with the $\Tau_0$ preserving map and the second with the
FKS map.  The nonsingular contributions below $\Tau^\cut_0$ and below
$\Tau^\cut_1$ are also explicitly written in \eq{1belowtau0}
for the exclusive 1-jet MC cross section and in
\eq{2belowtau1} for the 2-jet inclusive MC cross section
respectively.

\subsection{Parton shower and hadronisation}
\label{subsec:genevashower}

The \geneva interface to the parton shower has been extensively discussed in
sec.~3 of Ref.~\cite{Alioli:2015toa}; here we summarise only the most
relevant features. The shower makes the calculation fully
differential at higher multiplicities by adding extra radiation to the
exclusive 0- and 1-jet cross sections and also further jets to the inclusive
2-jet cross section.
The extra emissions are added in a
recursive and unitary way. However, if additional analysis cuts are
applied after the shower, large differences are usually expected also
in distributions that are inclusive over the radiation.

For 0-jet events, the purpose of the shower is to
restore the emissions which were integrated over when
constructing the exclusive 0-jet cross section. In particular, the
shower supplements events with additional emissions
below the cut which are required to satisfy the
constraint $\Tau_0(\Phi_N)< \Tau^\cut_0$. In practice we allow for a
small spillover at the level of 5$\%$.

The showering of the 1- and 2-jet events requires a dedicated
treatment to avoid significantly altering the $\Tau_0$ spectrum
calculated at the partonic level. The $\Phi_2$ points generated after
the first emission must satisfy the restriction
$\Tau_1(\Phi_2)< \Tau^\cut_1$ as well as the projectability condition
onto $\Phi_1$ using the $\Tau_0$-preserving map
$\Tau_0(\Phi_2)=\Tau_0(\Phi_1)$ presented in \sec{Genevapart1}. In
order to fulfil these constraints, we carry out
the first emission in \geneva using the LL Sudakov factor
described in
\subsec{genevaTau1res}.  The subsequent emissions are controlled by
the shower and need only satisfy the
constraint $\Tau_1(\Phi_N)<\Tau^\cut_1$.  In addition, we multiply the
entire 1-jet cross section by a second Sudakov factor
$U_1(\Tau^\cut_1,\Lambda_1)$ as described in
Ref.~\cite{Alioli:2019qzz}:
\begin{align}
\frac{\mathrm{d} \sigma_{1}^{\mathrm{MC}}}{\df\Phi_{1}} (\Tau_0 > \Tau_0^\cut, \Tau_1^\cut, \Lambda_1) =&\; \frac{\dsigMC_{1}}{\df\Phi_{1}} (\Tau_0 > \Tau_0^\cut, \Tau_1^\cut) \, U_1(\Tau_1^\cut, \Lambda_1)\, ,\\
\frac{\mathrm{d} \sigma_{\geq 2}^{\mathrm{MC}}}{\df\Phi_{2}} (\Tau_0 > \Tau_0^\cut, \Tau_1^\cut,\Tau_{1}>\Lambda_1) =&\; \frac{\dsigMC_{\geq 2}}{\df\Phi_{2}} (\Tau_0 > \Tau_0^\cut, \Tau_{1}>\Tau_{1}^\cut) \nn \\
&+ \frac{\df}{\df \Tau_1} \, \frac{\mathrm{d} \sigma^{\mathrm{MC}}_{1}}{\df\Phi_{1}} (\Tau_0 > \Tau_0^\cut, \Tau_1^\cut, \Tau_1) \,\nn\\
&\quad \times\cP(\Phi_2) \,  \theta( \Lambda_1<\Tau_1<\Tau_1^{\max})
\,.
\end{align}
The parameter $\Lambda_1$ determines the ultimate 1-jet resolution cutoff which
we set to be much smaller than the \geneva cutoff parameter
$\Tau^\cut_1 = 1$~GeV. For the process at hand we use $\Lambda_1=
10^{-4}$~GeV with the consequence that the 1-jet cross section is
extremely suppressed and accounts for about $0.1\%$ of the total cross
section. The restrictions on the shower for such a small contribution to the
cross section can then be ignored. Regarding the showering of the
partonic $\Phi_2$ events, it was shown in Ref.~\cite{Alioli:2015toa} that
the first emission of the shower acting on these events affects the
$\Tau_0$ distribution at order $\as^3/\Tau_0$.  From the
above discussion it then follows that the showered
events originate either from $\df \sigma^{\mathrm{MC}}_0$ or $\df
\sigma^{\mathrm{MC}}_{\ge2}$.

In the following we compare the partonic, showered and hadronised
results for a selected set of distributions. We use the same inputs as
above with the only difference that, for these comparisons, the
fixed-order scale is set to $\mu_\FO=M_{\gamma\gamma}^T$.  We use the
parton shower program \pythiaEight~\cite{Sjostrand:2014zea} interfaced
to \geneva. At the hadronisation level, we switch off the hadron
decays in order to keep the analysis simple and avoid contributions
from secondary photons. For the same reason we also avoid including
multi-parton interactions (MPI) and a QED shower.

\begin{figure}[tp]
\begin{center}
\begin{tabular}{ccccc}
\includegraphics[width=\rescalethreeplots]{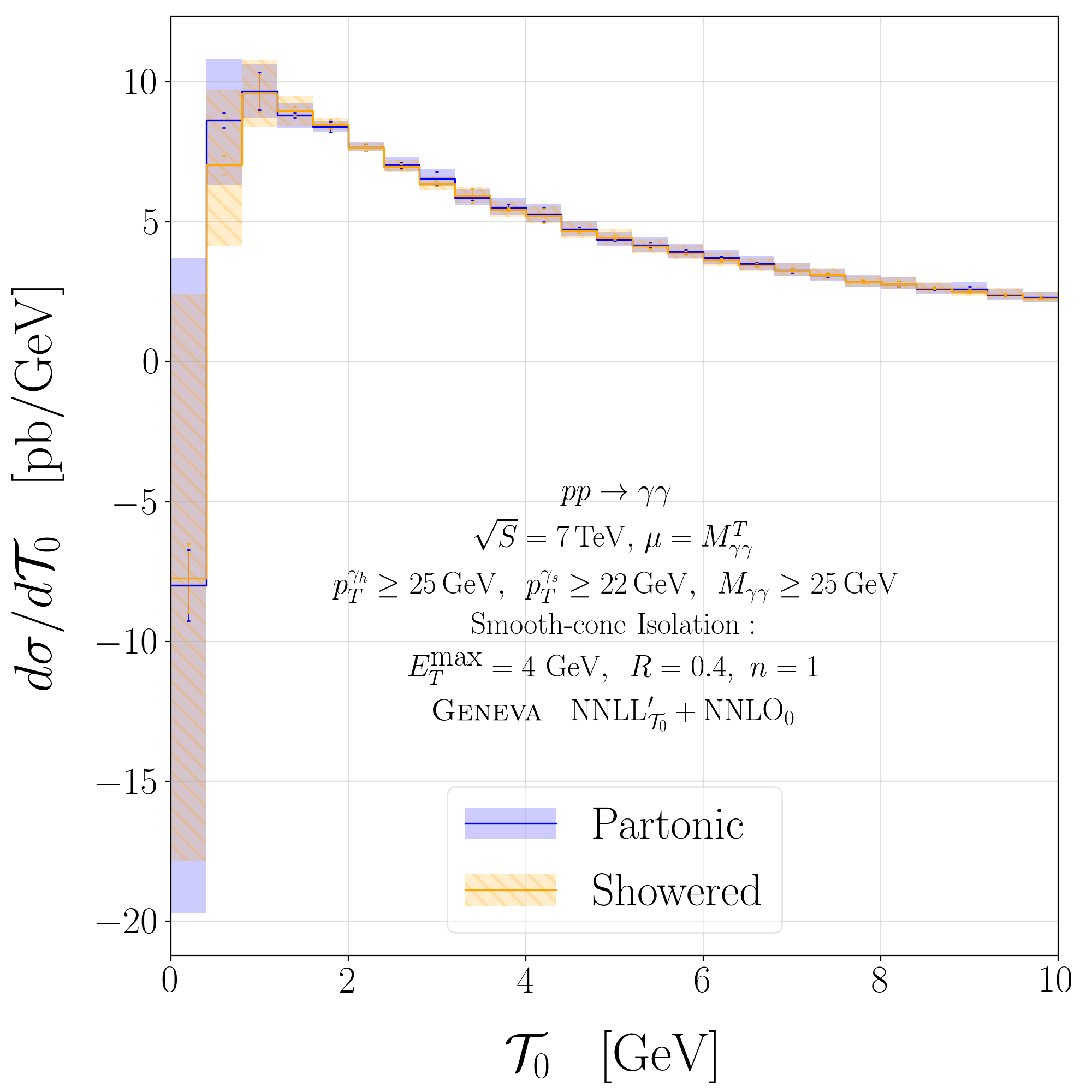} &\hspacebetweenthreeplots
\includegraphics[width=\rescalethreeplots]{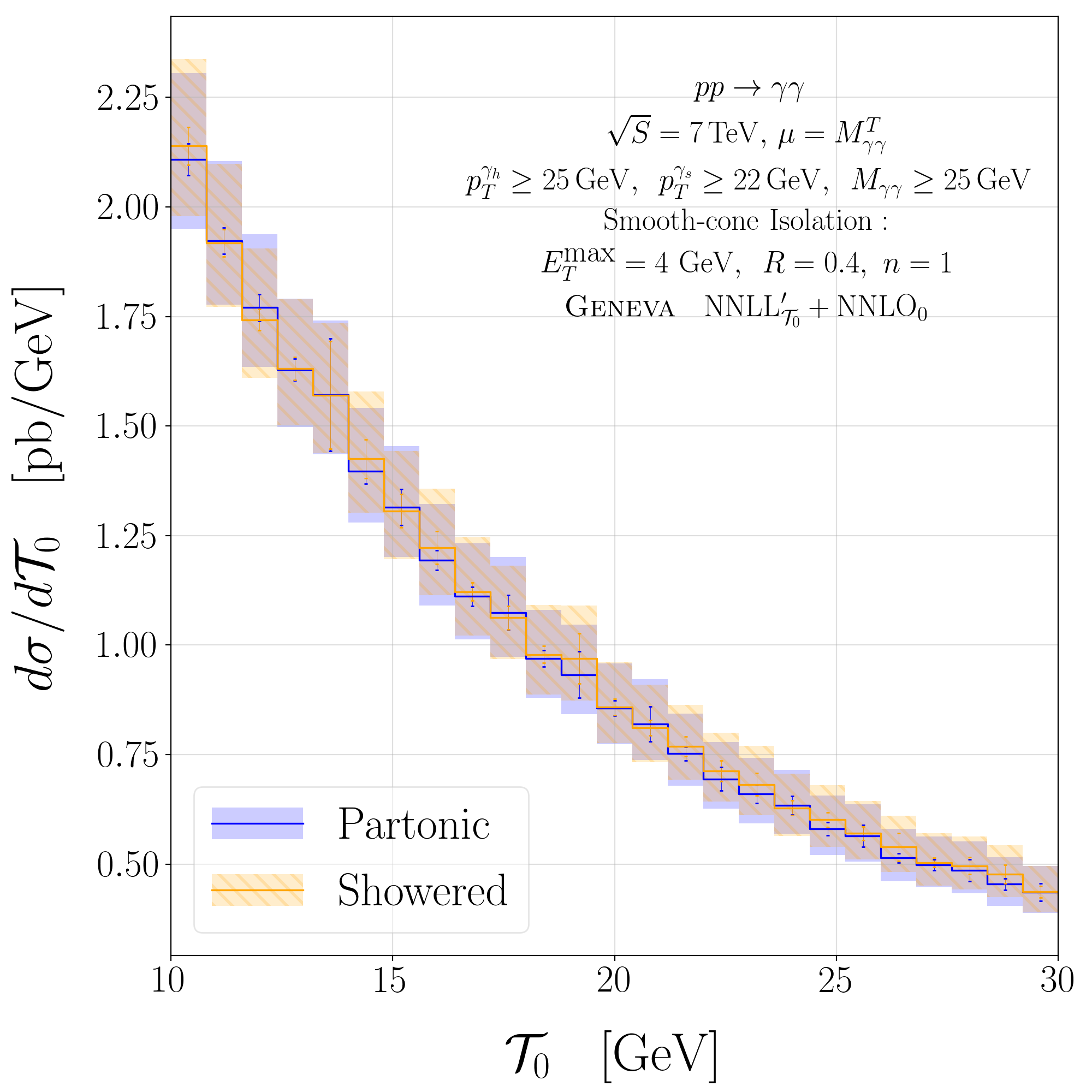} &\hspacebetweenthreeplots
\includegraphics[width=\rescalethreeplots]{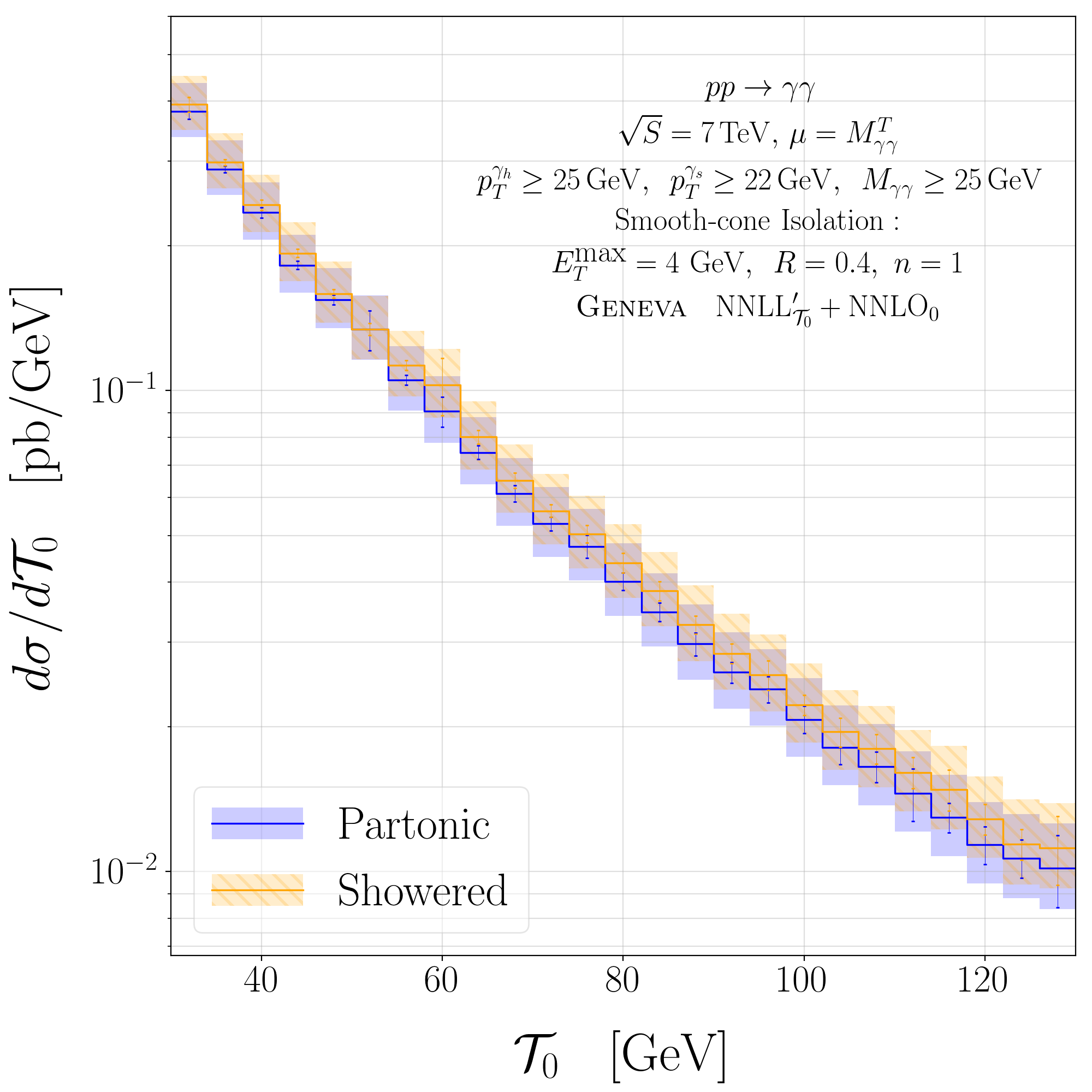} \\[\vspacebetweenthreeplots]
\includegraphics[width=\rescalethreeplots]{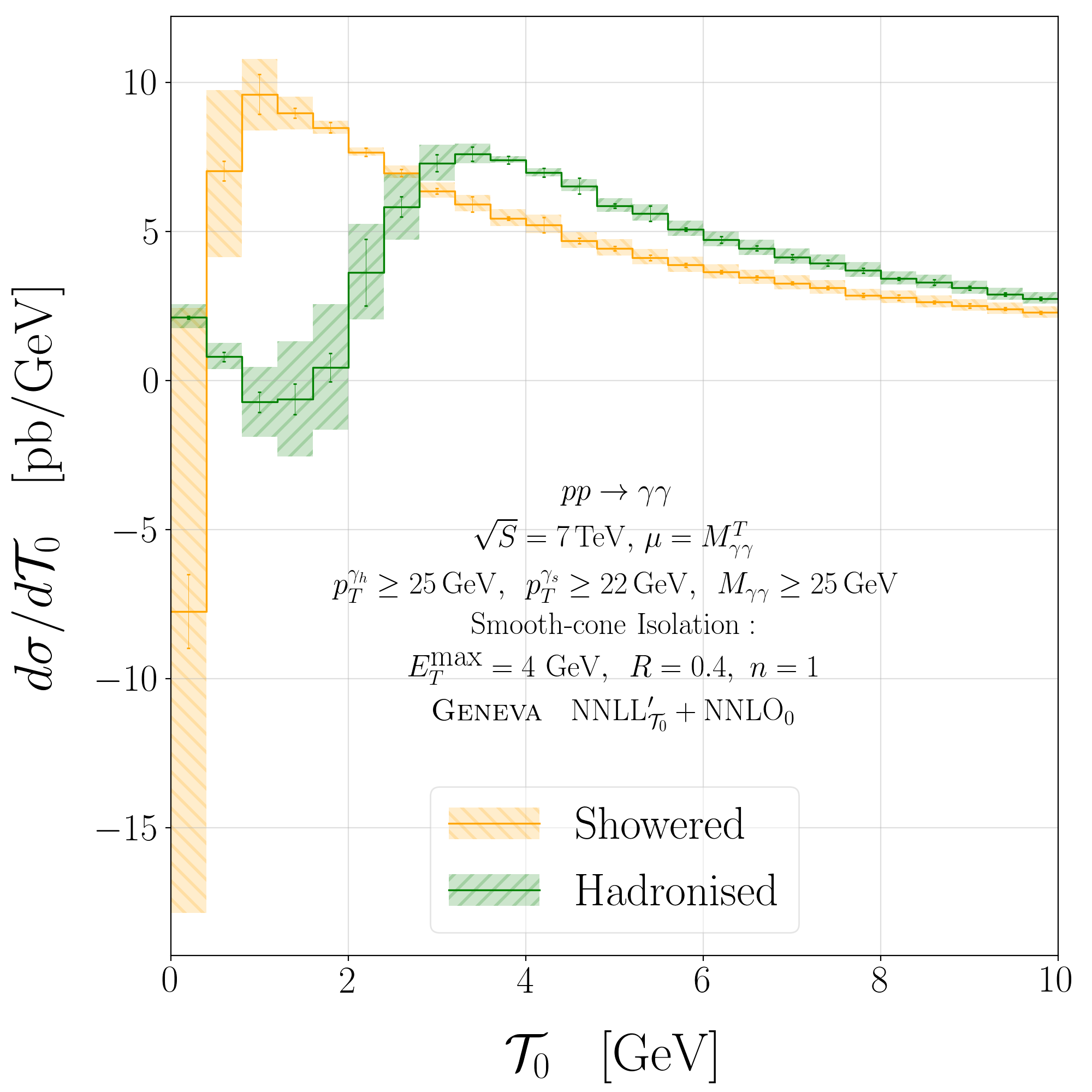} &\hspacebetweenthreeplots
\includegraphics[width=\rescalethreeplots]{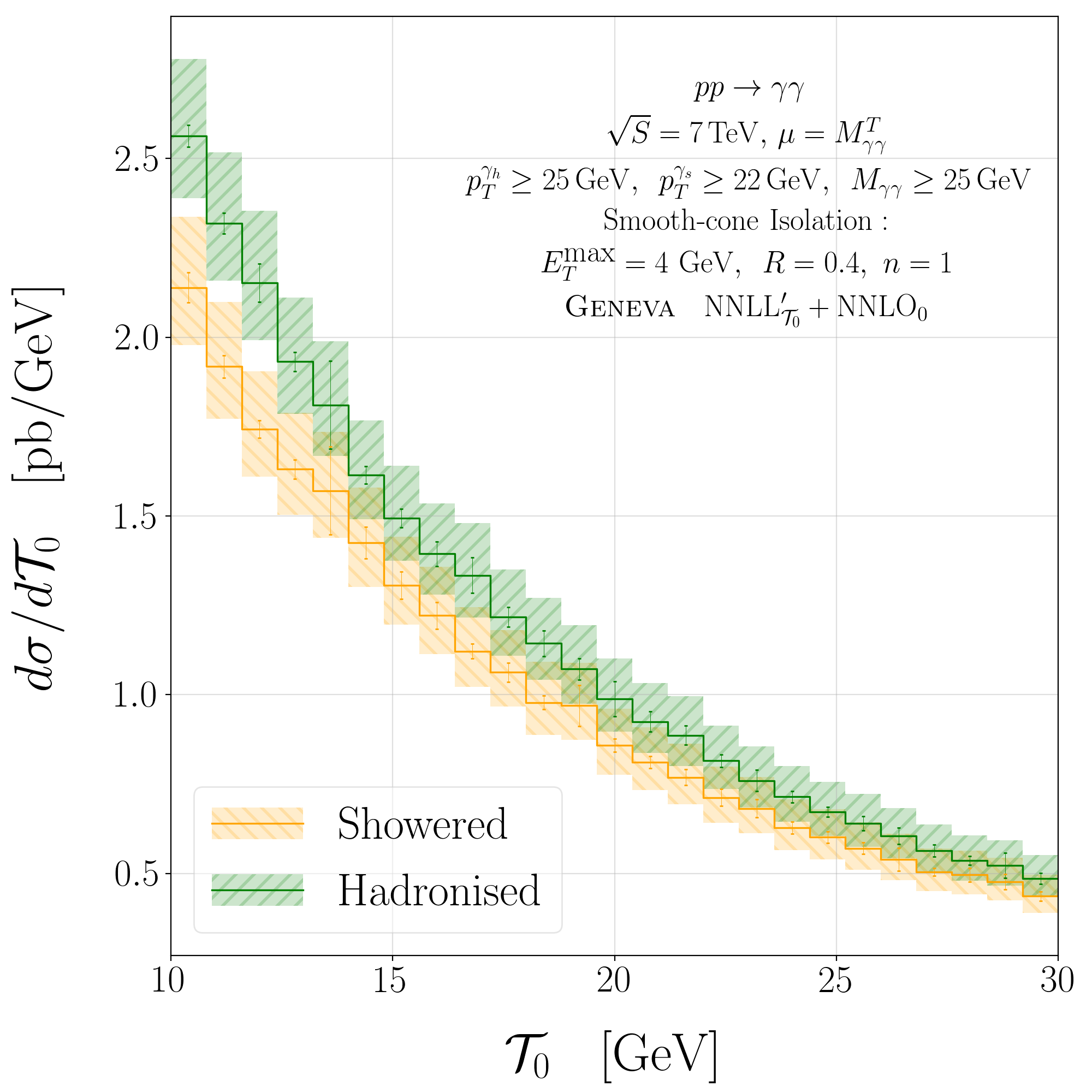} &\hspacebetweenthreeplots
\includegraphics[width=\rescalethreeplots]{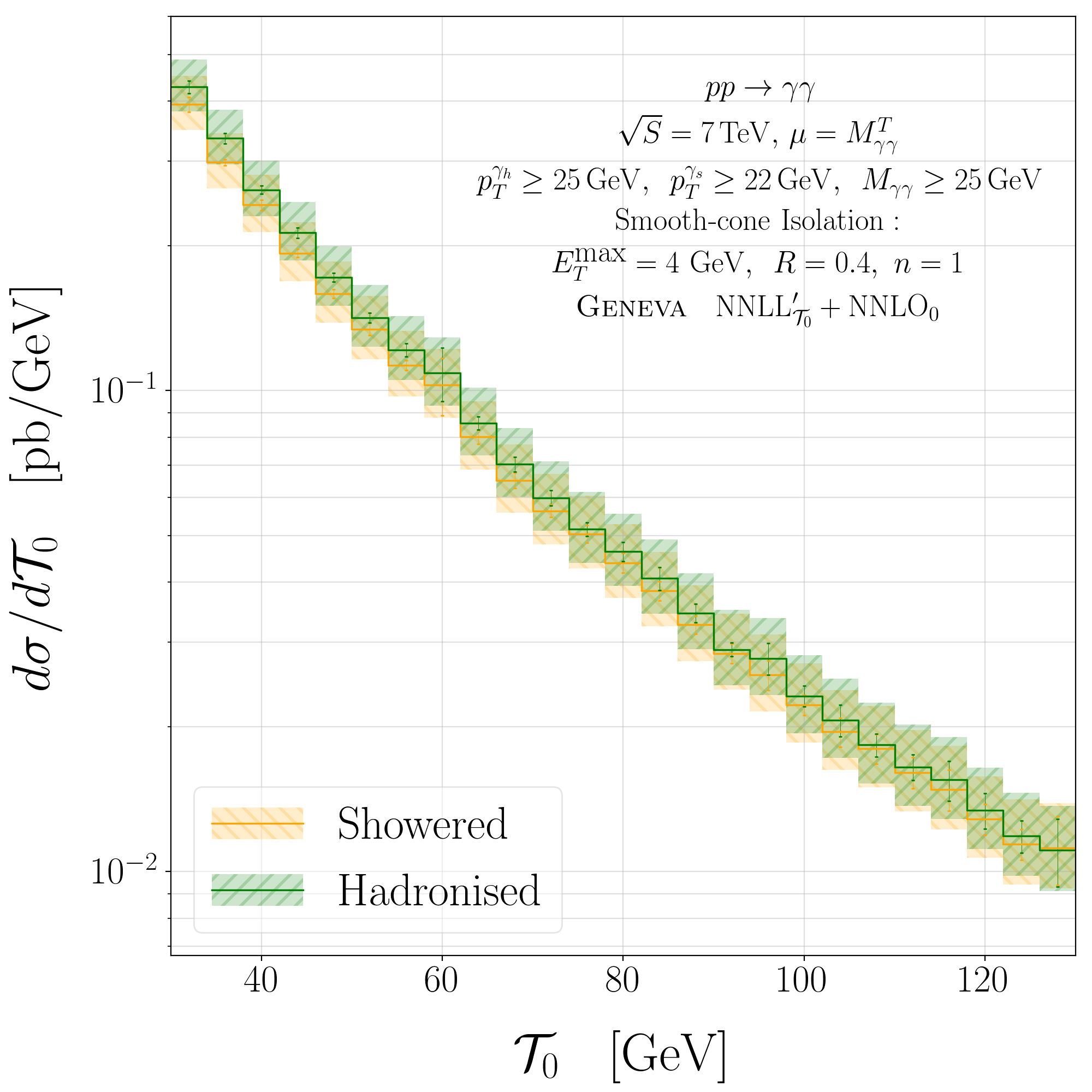}
\end{tabular}
\end{center}
\spaceabovefigurecaption
\caption{Comparison of $\Tau_0$ spectra between the partonic NNLO$_0$+NNLL$^\prime$ and the showered results, after interfacing to \pythiaEight, before the inclusion of non-perturbative effects (above). Comparison between the showered and hadronised $\Tau_0$ spectra (below). The peak (left), transition (centre) and tail (right) regions are shown.
\label{fig:Tau0showerqqbarNOancuts}
}
\spacebelowfigurecaption
\end{figure}
\begin{figure}[tp]
\begin{center}
\begin{tabular}{ccc}
\includegraphics[width=\rescaletwoplots]{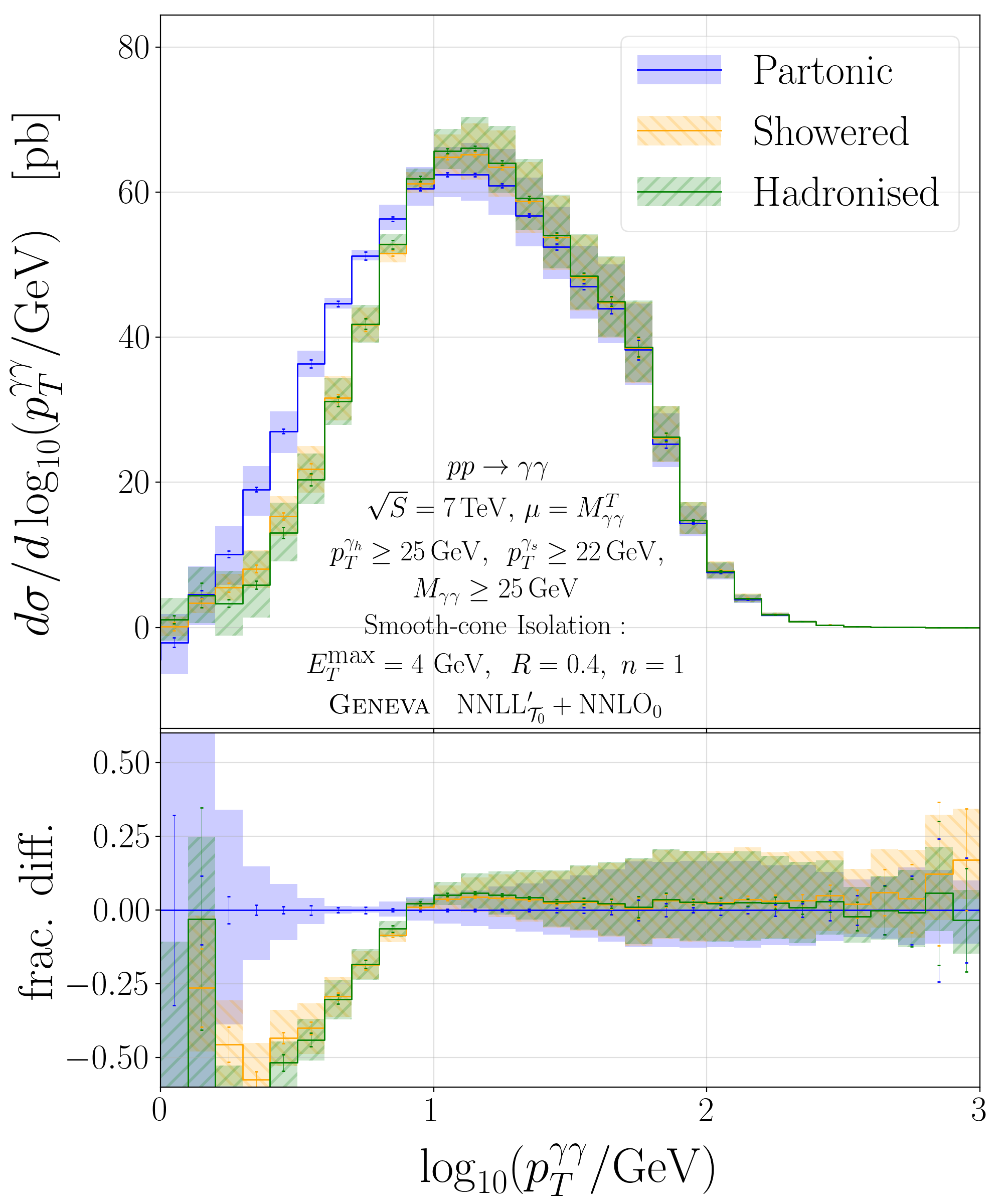} &\hspacebetweentwoplots&
\includegraphics[width=\rescaletwoplots]{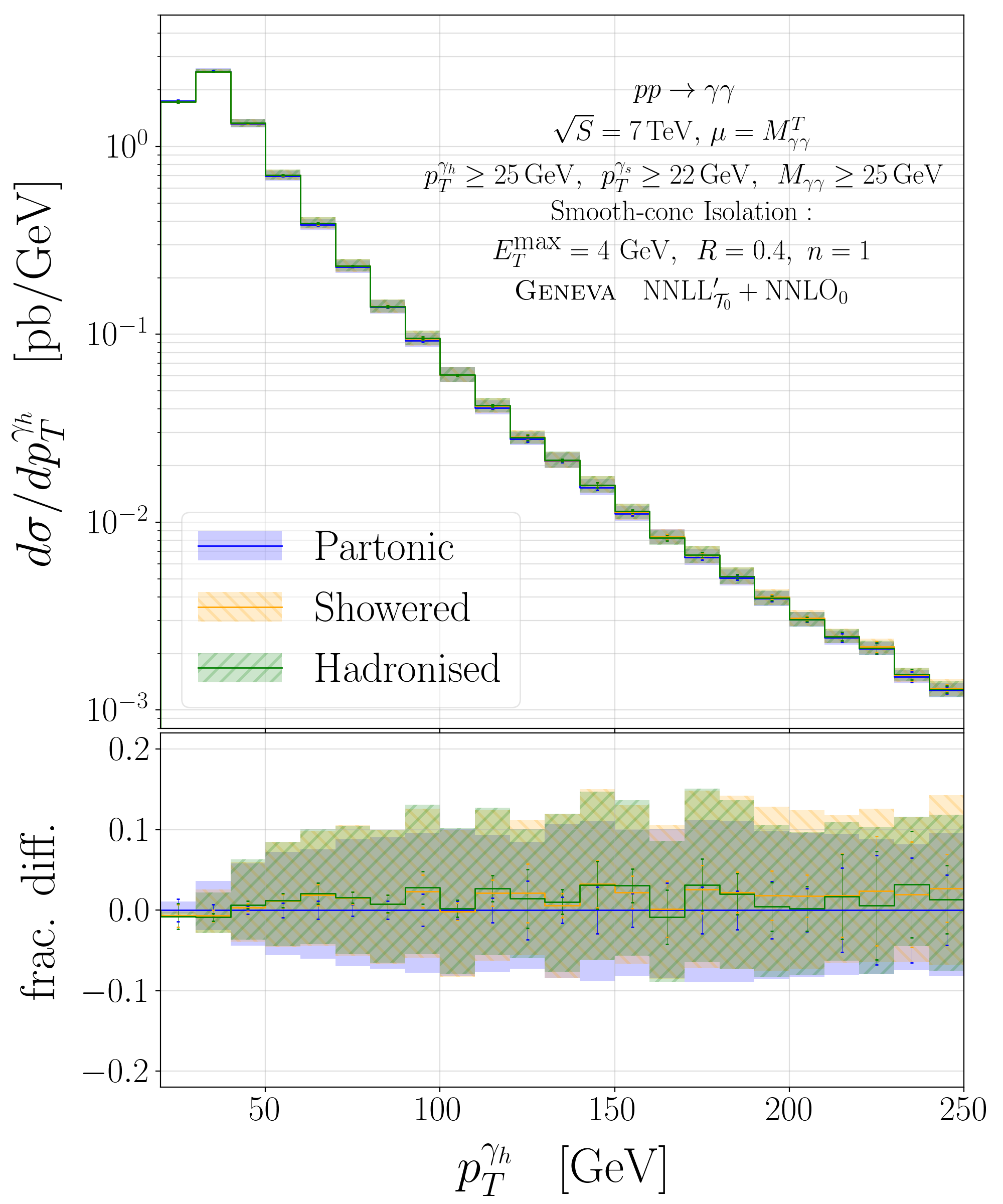} \\[\vspacebetweentwoplots]
\includegraphics[width=\rescaletwoplots]{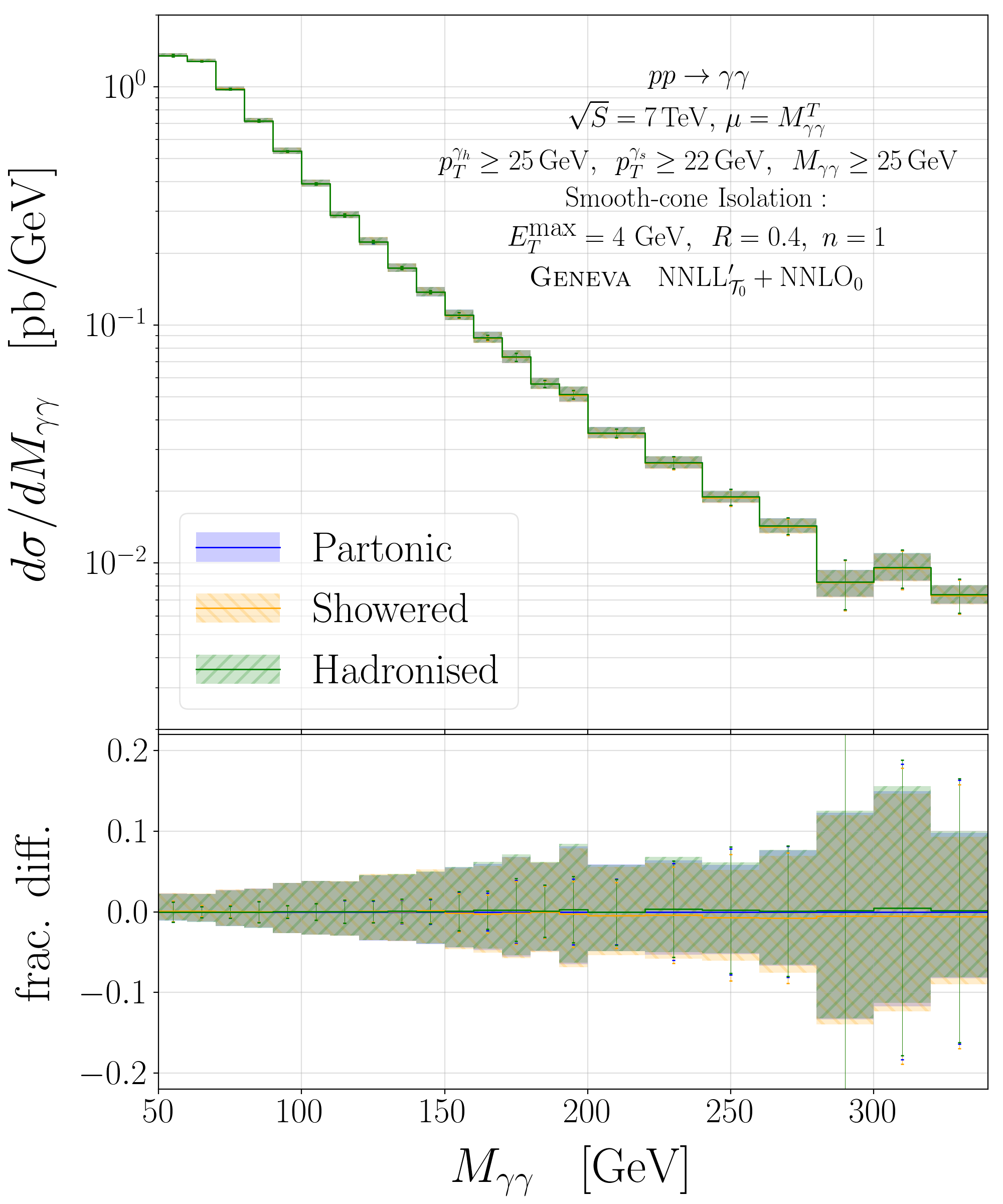} &\hspacebetweentwoplots&
\includegraphics[width=\rescaletwoplots]{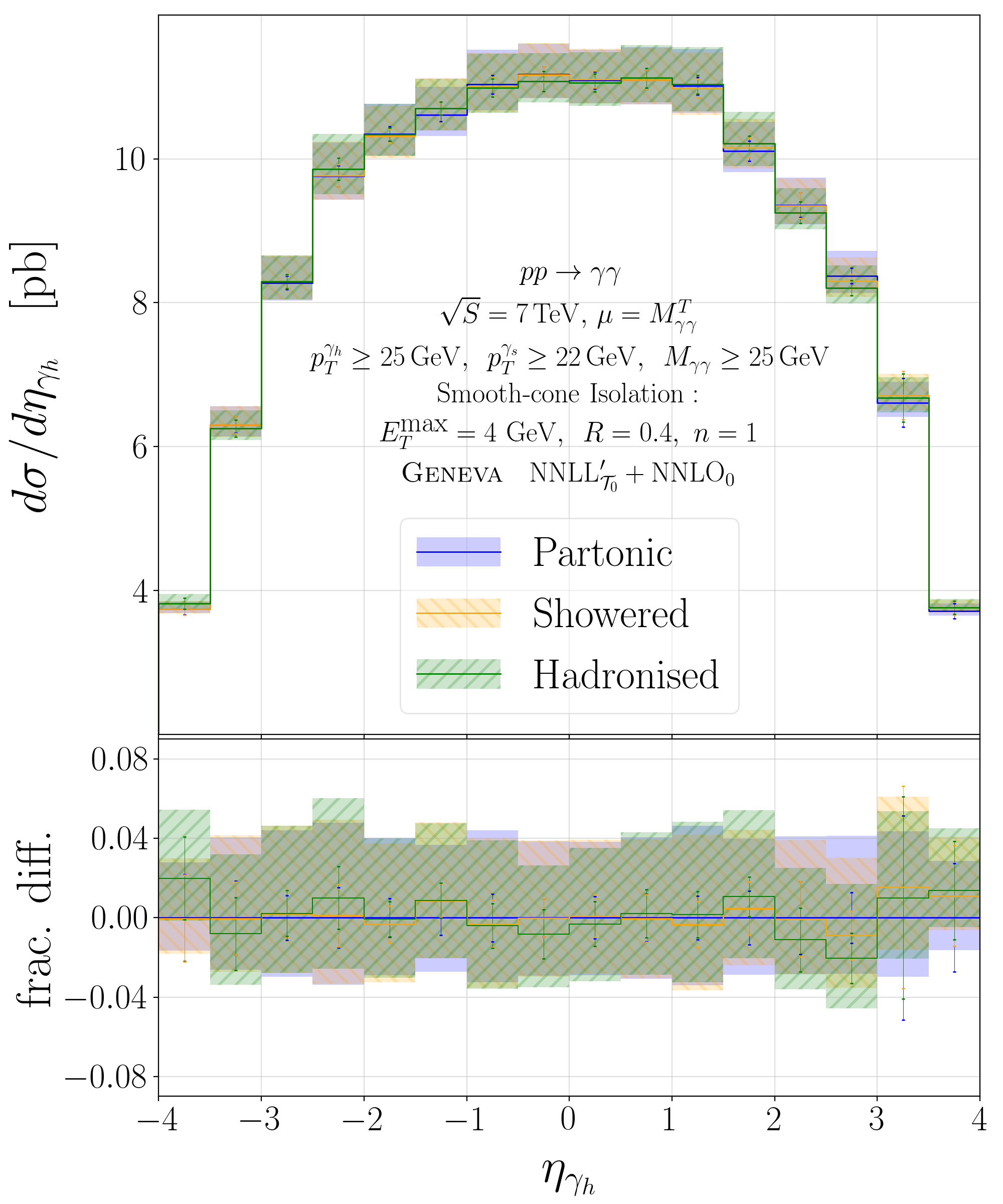}
\end{tabular}
\end{center}
\spaceabovefigurecaption
\caption{ Comparisons of the partonic, showered and hadronised results for a selected set of distributions.
\label{fig:showerqqbarNOancuts}
}
\spacebelowfigurecaption
\end{figure}
\begin{figure}[tp]
\begin{center}
\begin{tabular}{ccc}
\includegraphics[width=\rescaletwoplots]{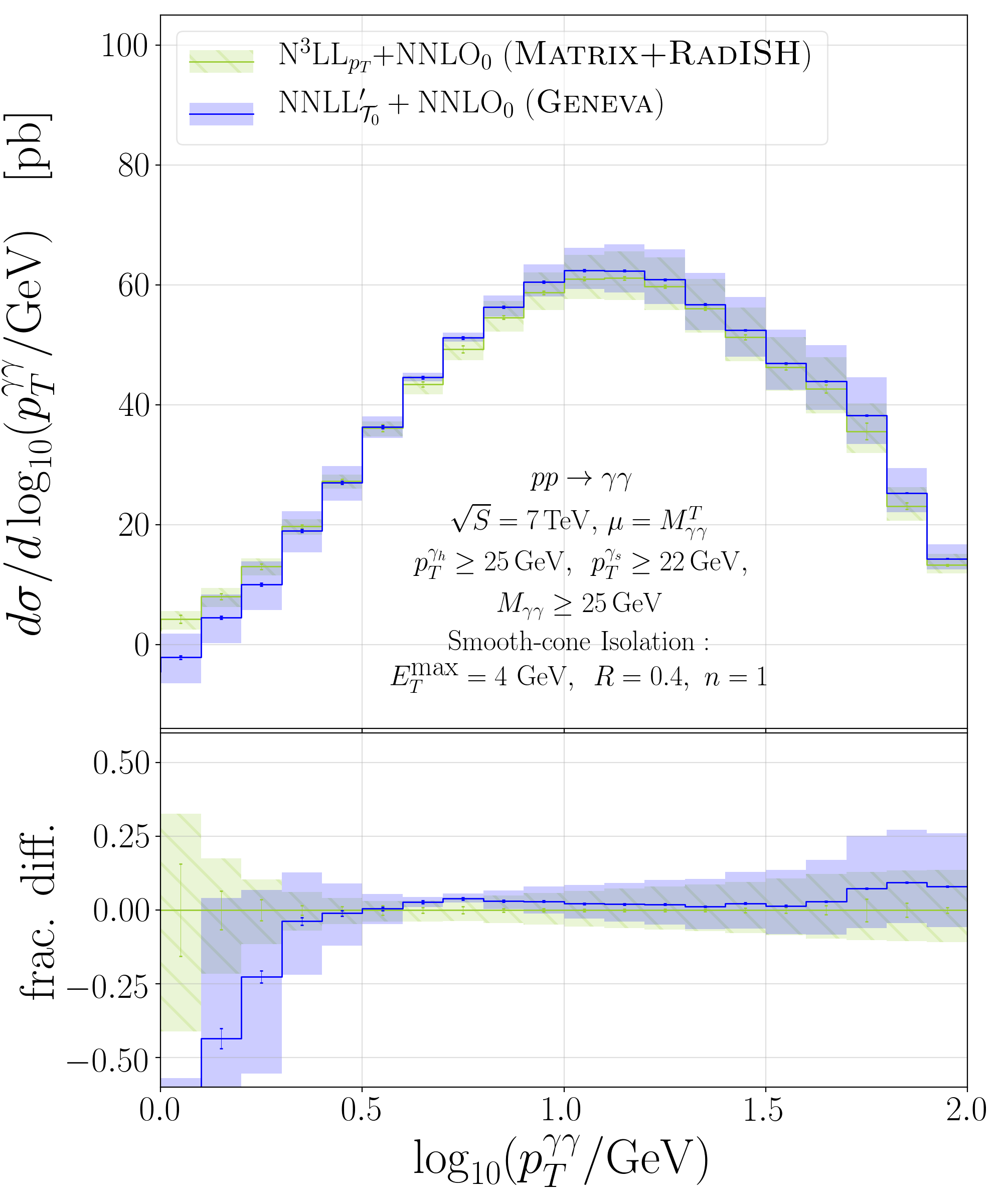} &\hspacebetweentwoplots&
\includegraphics[width=\rescaletwoplots]{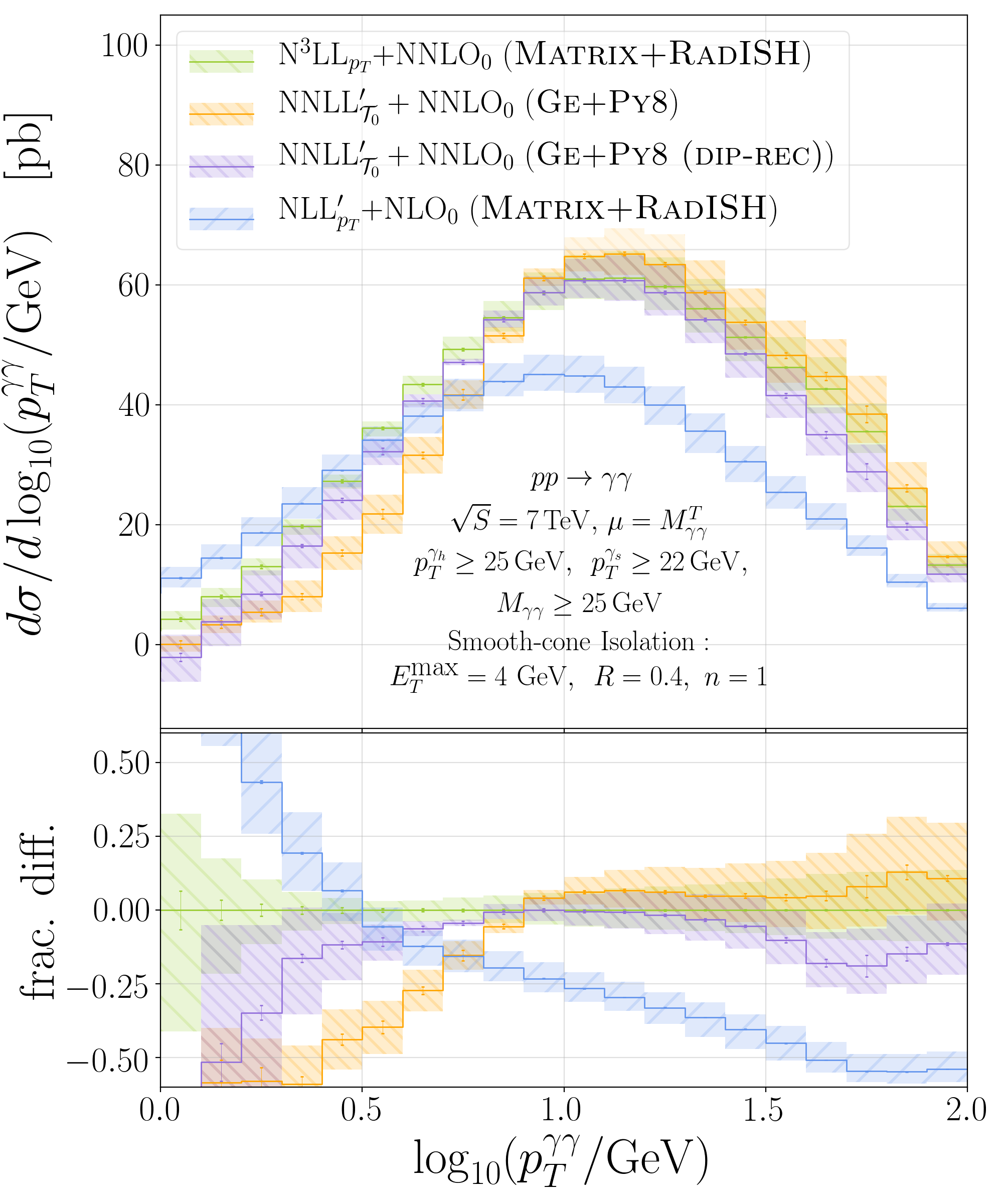}
\end{tabular}
\end{center}
\spaceabovefigurecaption
\caption{Comparison with \textsc{Matrix+RadISH} for the $p^{\gamma \gamma}_T$ distribution at different resummation accuracies. \geneva results before showering are shown on the left panel, after showering but before hadronisation on the right panel.
\label{fig:cmpradish}
}
\spacebelowfigurecaption
\end{figure}

We present in the upper panel of \fig{Tau0showerqqbarNOancuts} the
comparison between the partonic and showered results for the $\Tau_0$
distribution, showing that the $\Tau_0$ distribution is not modified
by the shower above $\Tau_0^\cut$. Below $\Tau^\cut_0$, instead, the
shape of the $\Tau_0$ distribution is determined entirely by the
shower; the effects are hardly visible since the cutoff is set to a
very small value ($\Tau^\cut_0=10^{-2}$~GeV).

We study the impact of hadronisation, which provides the nonperturbative effects, by comparing, in the bottom
panel of \fig{Tau0showerqqbarNOancuts},
the showered and hadronised $\Tau_0$ distributions. As expected, we notice a large
difference between the two results only in the peak region, since the
$\Tau_0$ observable is very sensitive to additional low-energy
hadronic emissions. At larger values of $\Tau_0$ these corrections
are instead suppressed as $\mathcal{O}(\Lambda_{\mathrm{QCD}}/Q)$ and
their effects are lessened.

We further study effects due to the parton shower and hadronisation in
\fig{showerqqbarNOancuts}, where we compare the partonic, showered and
hadronised results for the transverse momentum of the photon pair, the
transverse momentum of the hardest photon, the invariant mass of the
photon pair and the pseudorapidity of the hardest photon.  We first
observe that for all the inclusive distributions, the NNLO accuracy is
maintained at a very precise numerical level after both the showering
and the hadronisation processes.

Moreover, although the transverse momentum of the photon pair, or any
other exclusive observable, formally have the same logarithmic
accuracy as the shower, we expect that it could also benefit from the
high resummation accuracy of the $\Tau_0$ distribution. We observe
that the distribution is significantly modified after the shower only
in the region below $10$~GeV, while for larger values of
$p^{\gamma \gamma}_T$, the higher-order partonic result is practically
recovered.  In order to quantify the quality of our predictions for
this observable we can compare with the direct resummation of
$p^{\gamma \gamma}_T$, which is performed in the
\textsc{Matrix+RadISH} interface~\cite{Kallweit:2020gva} up to
N$^3$LL$_{\pt}$+NNLO$_0$ accuracy.

In the left panel of \fig{cmpradish} we show such a comparison at the
partonic level, \ie before the shower, observing a very good
agreement.~\footnote{We compare against results at N$^3$LL because the
public version of \textsc{Matrix+RadISH} does not presently allow for
NNLL$^\prime$ accuracy. In order to have a like-for-like comparison
with \geneva results we have also selected an additive scheme for the
matching of the resummation to the fixed-order.} In the right panel of
the same figure, we compare our results after showering but before
hadronisation against the \textsc{Matrix+RadISH} results at both
N$^3$LL$_{\pt}$+NNLO$_0$ and NLL$^\prime_{\pt}$+NLO$_0$
accuracy. We include results with two different schemes
for the shower recoil: the default shower recoil of \pythiaEight and a
second more local scheme in which the spectator parton absorbs the
recoil of the initial-final dipole, preserving the transverse momentum
of colourless particles. This second recoil scheme is labeled
{\mbox{DIP-REC}} in the figures. Even after adding the shower
effects, in particular when using the new recoil scheme, the Geneva
results are in better agreement with those with higher logarithmic
accuracy.

\begin{figure}[tp]
\begin{center}
\begin{tabular}{ccc}
\includegraphics[width=\rescaletwoplots]{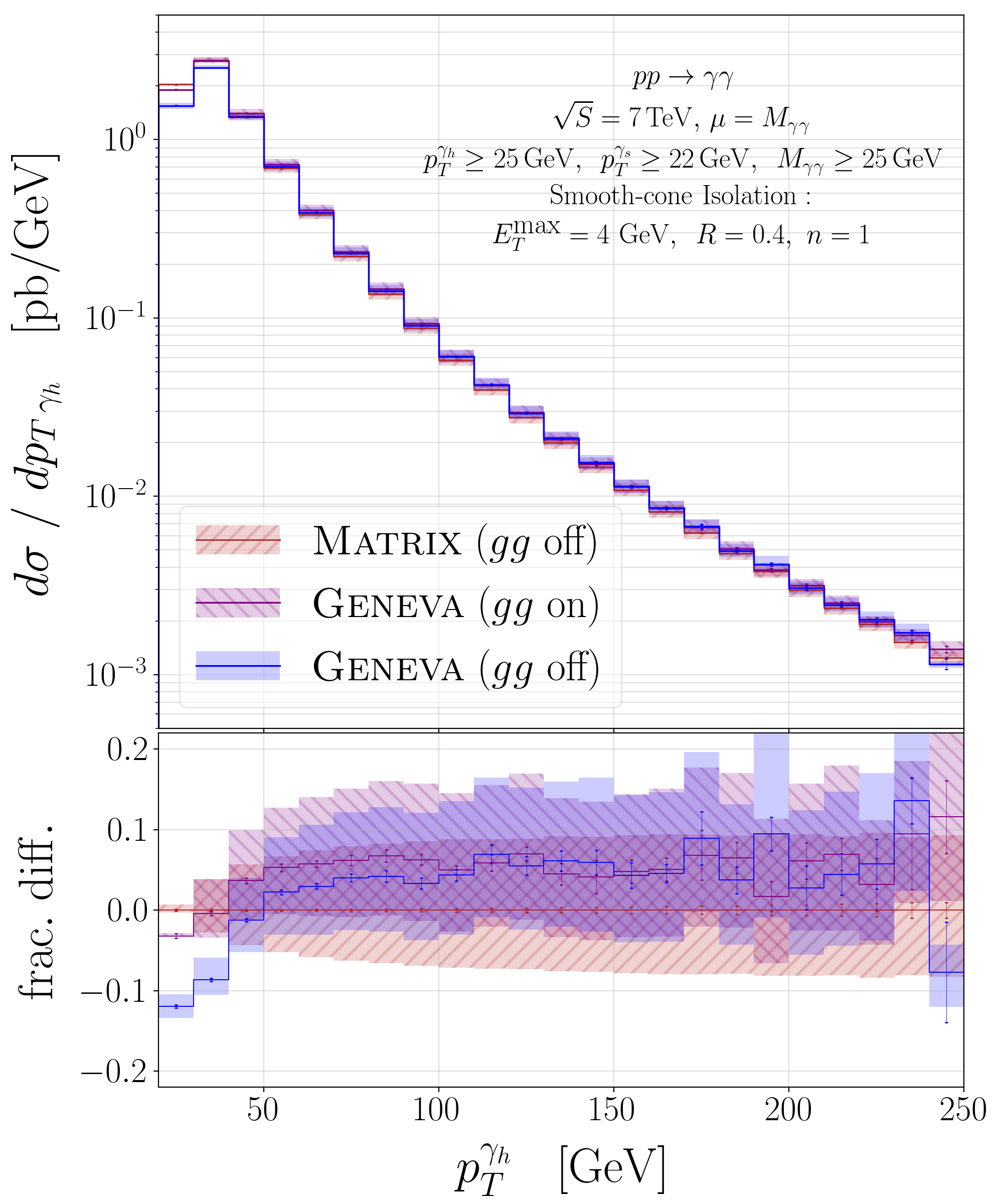} &\hspacebetweentwoplots&
\includegraphics[width=\rescaletwoplots]{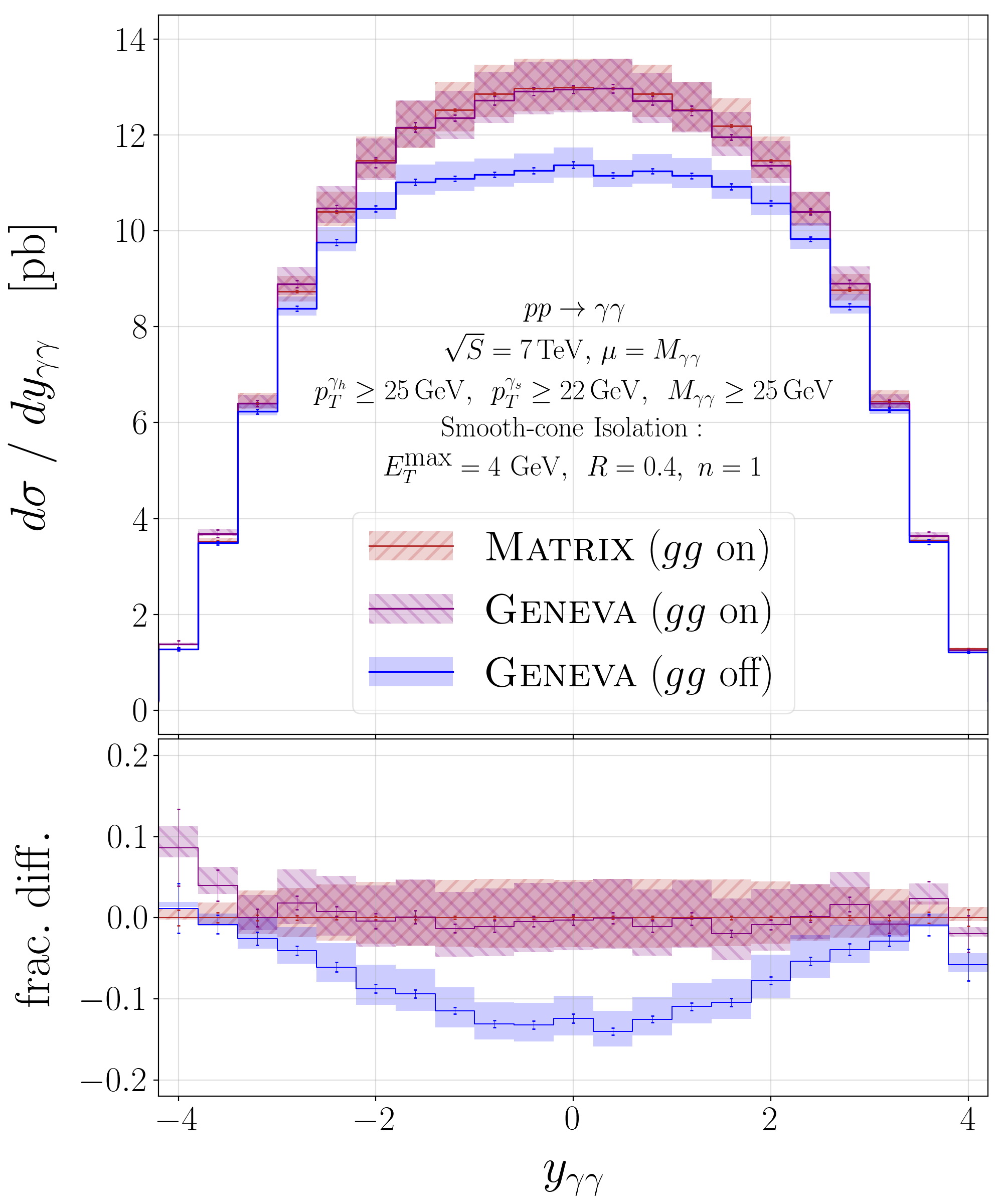} \\[\vspacebetweentwoplots]
\includegraphics[width=\rescaletwoplots]{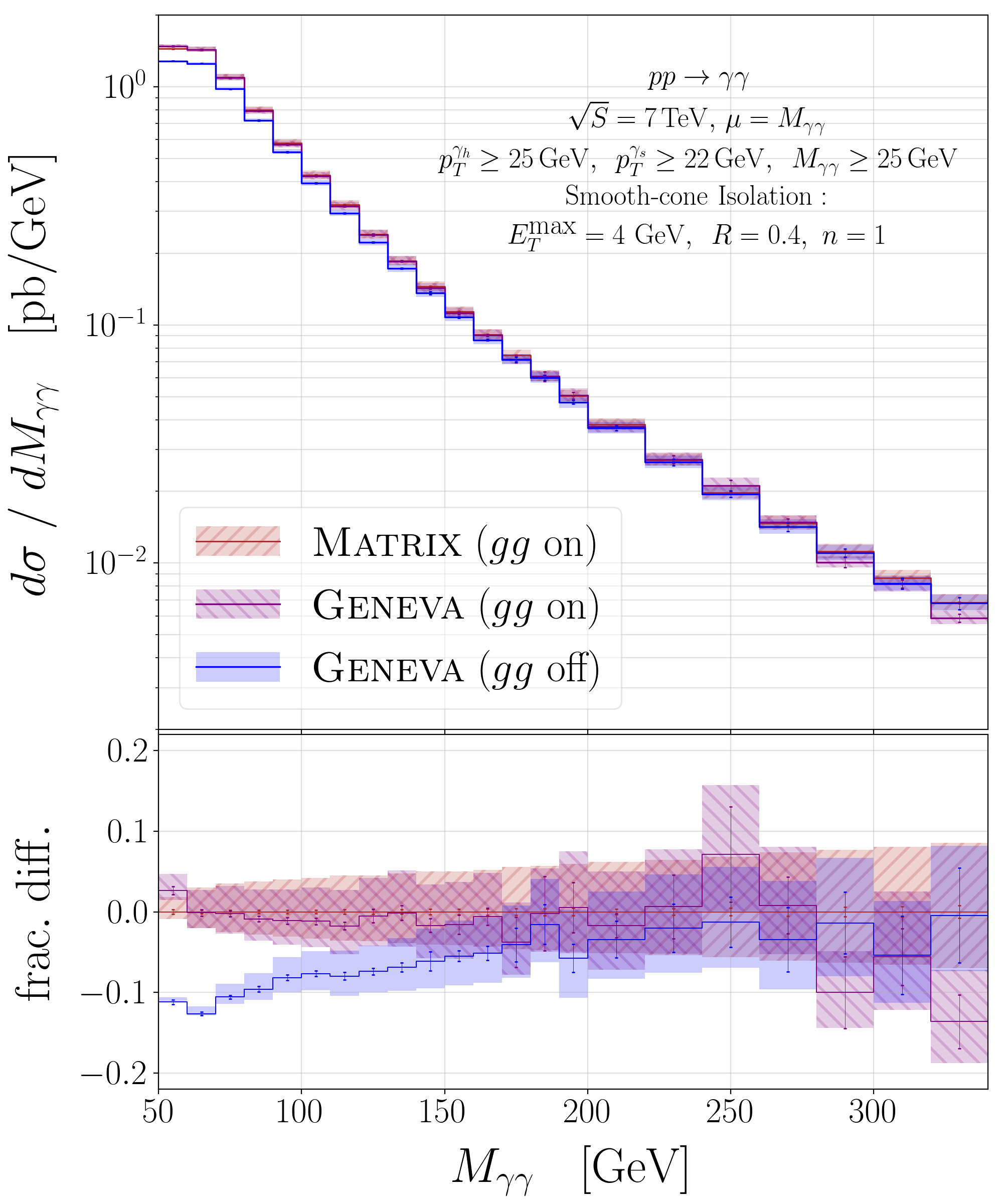} &\hspacebetweentwoplots&
\includegraphics[width=\rescaletwoplots]{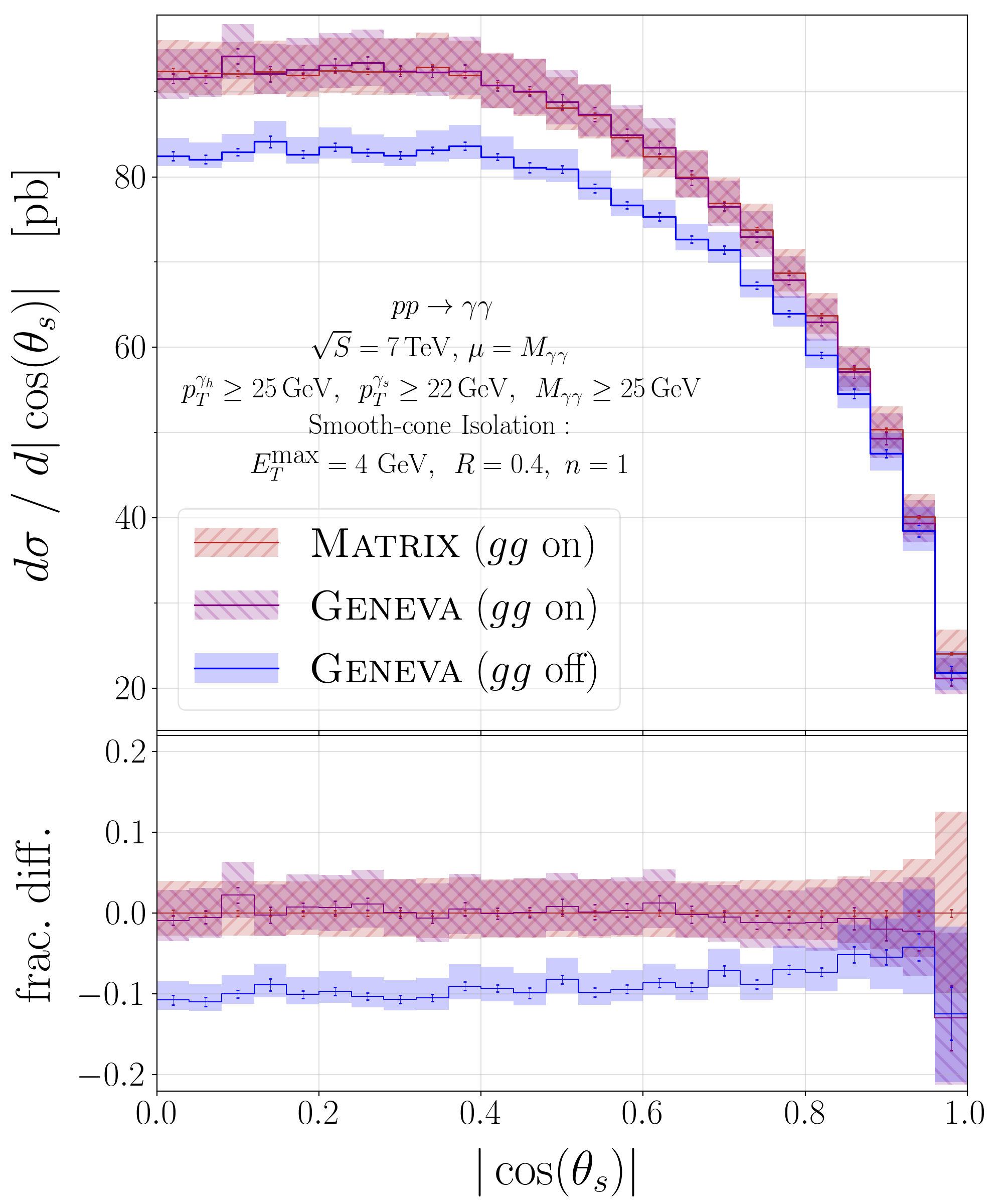} \\
\end{tabular}
\end{center}
\spaceabovefigurecaption
\caption{Comparisons between \geneva and \matrix after the inclusion of the $gg$ channel contribution. We also show the \geneva results before the inclusion of the $gg$ channel.
\label{fig:gvavsmatrixgg}
}
\spacebelowfigurecaption
\end{figure}
\begin{figure}[tp]
\begin{center}
\begin{tabular}{ccc}
\includegraphics[width=\rescaletwoplots]{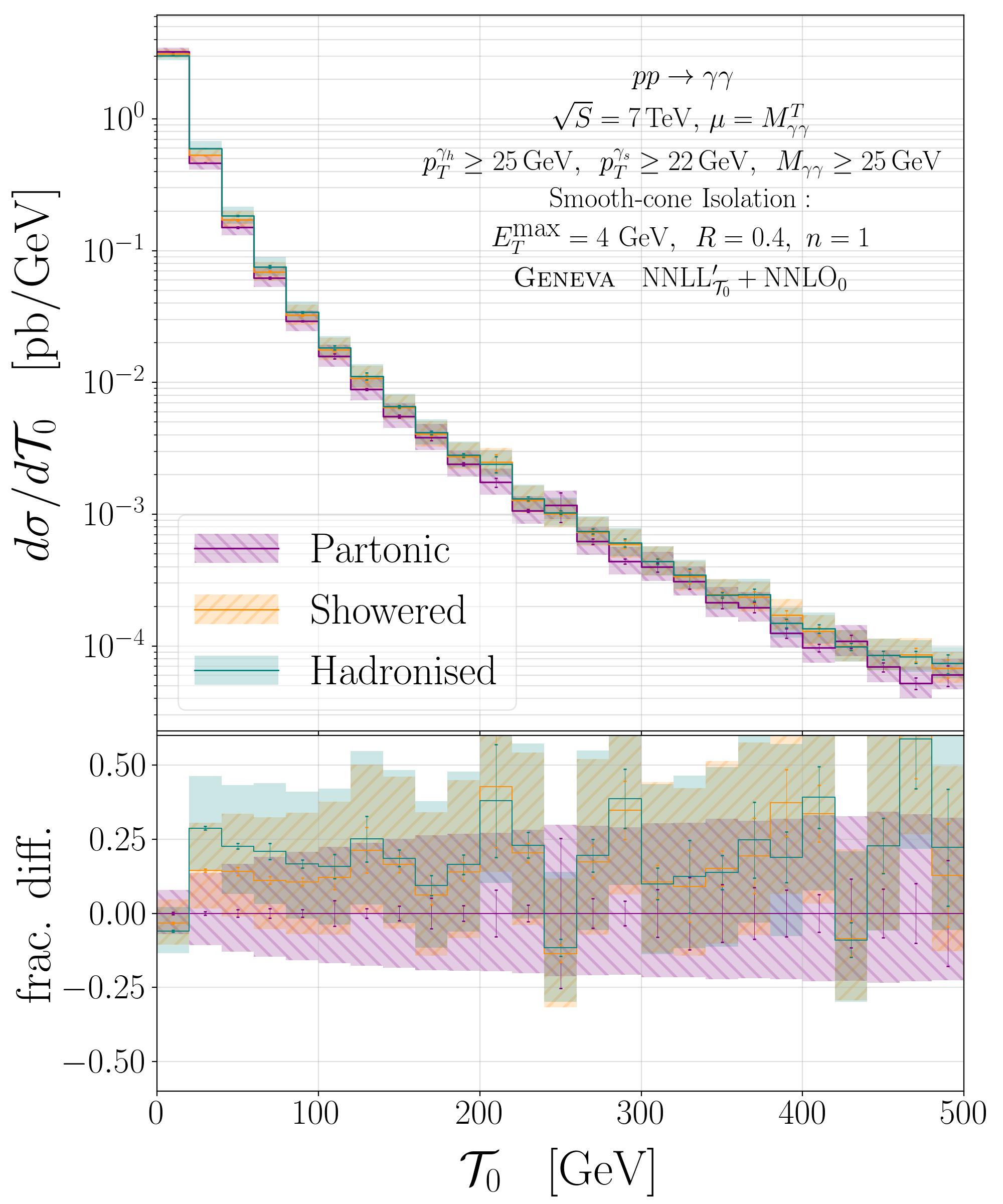} &\hspacebetweentwoplots&
\includegraphics[width=\rescaletwoplots]{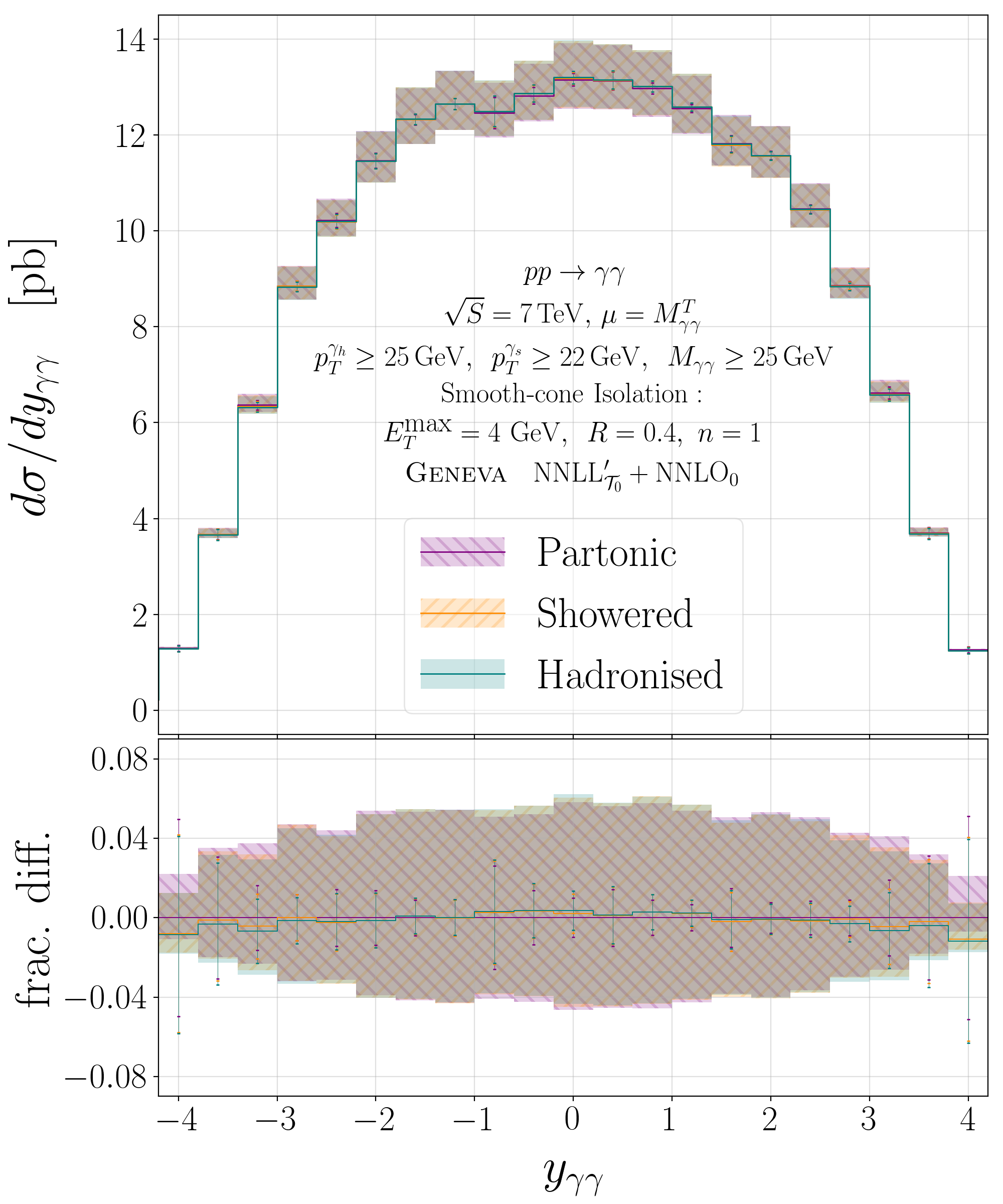} \\[\vspacebetweentwoplots]
\includegraphics[width=\rescaletwoplots]{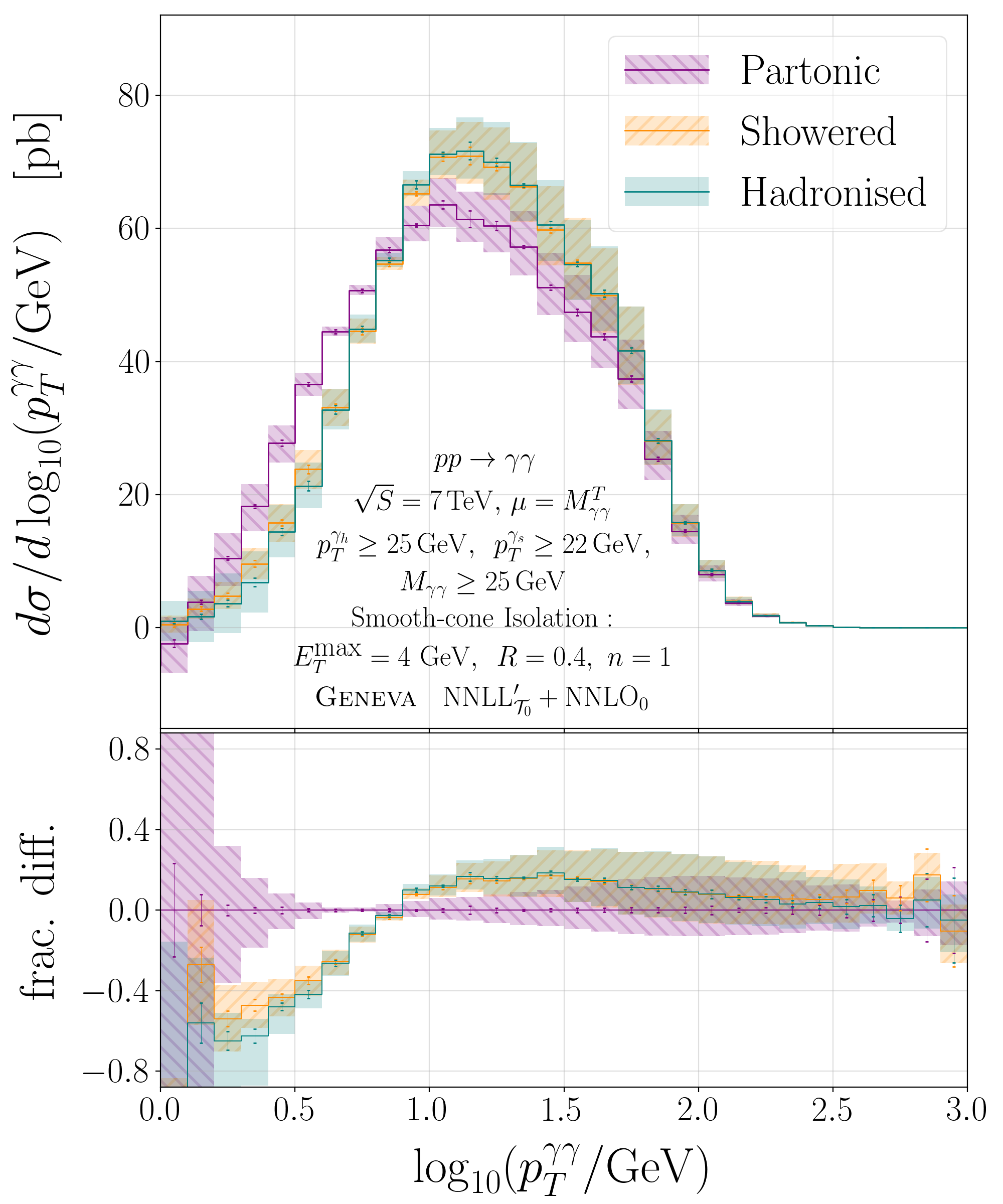} &\hspacebetweentwoplots&
\includegraphics[width=\rescaletwoplots]{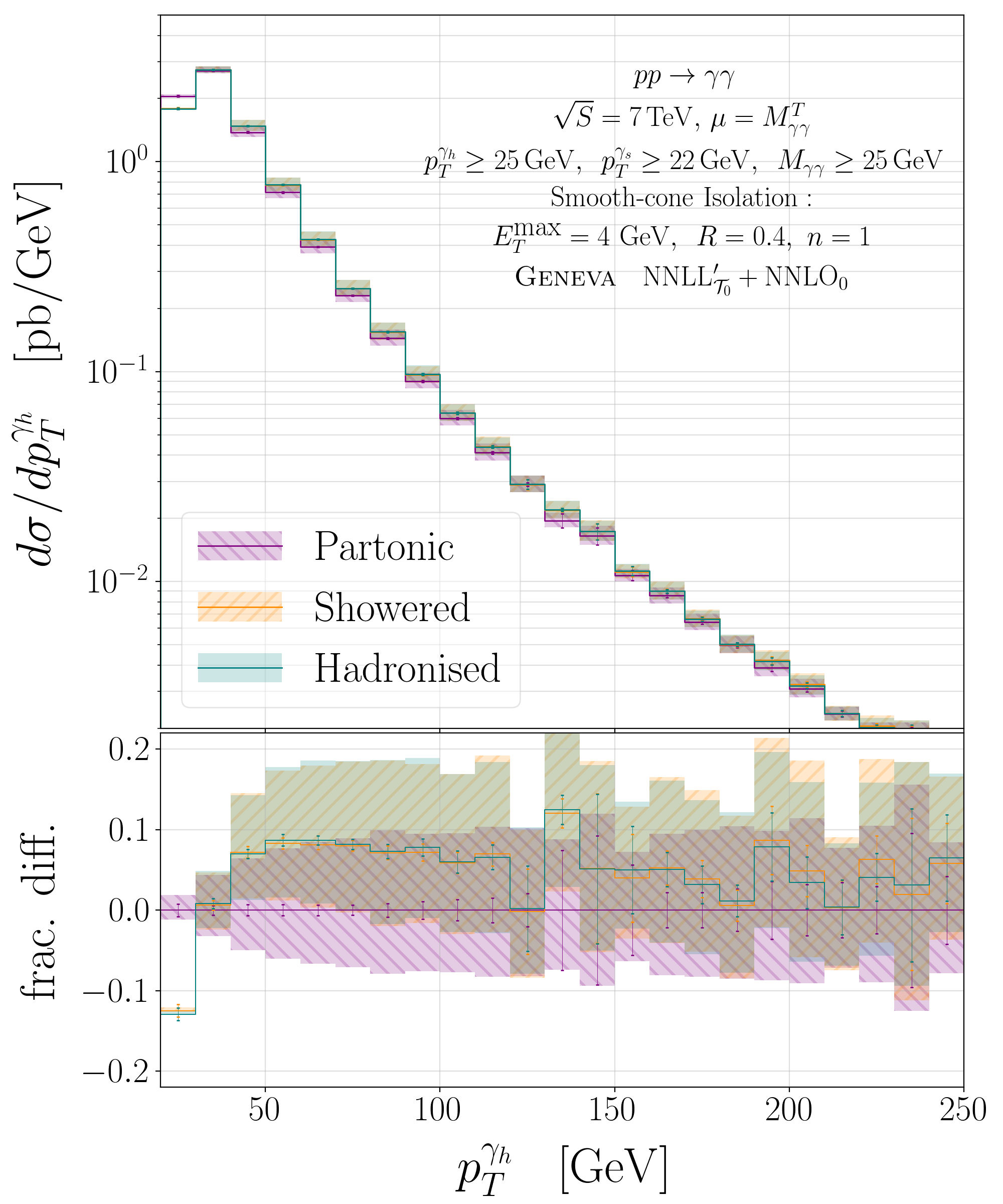} \\
\end{tabular}
\end{center}
\spaceabovefigurecaption
\caption{Comparison between the partonic, showered and hadronised spectra after the inclusion of the $gg$ channel contribution.
\label{fig:showergginclNOancuts}
}
\spacebelowfigurecaption
\end{figure}
%

\subsection{Inclusion of the $gg$ channel contribution}
\label{subsec:genevagg}

The effects of including the $gg$ channel contribution are quite large
both for the total cross section (in the $6$--$10\%$ range) and the
differential distributions. This is a consequence of the relative size
of the gluon parton distributions at the LHC.

In \fig{gvavsmatrixgg}
we compare the results of \geneva with \matrix after the inclusion of
the $gg$ channel contribution for the same set of inclusive
distributions presented in \fig{gvavsmatrixqqbar}. As shown in the
plots, we find very good agreement between the two calculations.  We
also show the effect of including the $gg$ channel contributions by
comparing to the \geneva results before its inclusion. Due to the
numerical relevance of this channel, its NLO QCD corrections have been
the subject of dedicated
studies~\cite{Bern:2002jx,Maltoni:2018zvp}. However, since these terms
are formally of higher order (N$^3$LO) with respect to the $q\bar{q}$
channel contribution, we neglect them in our calculation.

When showering events in the gluon fusion channel, we set the starting
scale of the shower to be equal to the highest scale present in the
process, which is the partonic centre-of-mass energy. The reason for
doing so is that we do not presently resum these contributions,
whose resummation accuracy is then entirely given by the shower.  A
dedicated higher-accuracy resummation of this channel is of course
possible but is left to future investigation.

In \fig{showergginclNOancuts} we show the comparison between the
partonic, showered and hadronised results after the inclusion of this
channel for the $\Tau_0$ distribution, the rapidity of the diphoton
system, the transverse momentum of the photon pair and the transverse
momentum of the hardest photon.  We observe somewhat larger effects after the
inclusion of the shower compared to the case of the $q\bar{q}$ channel
alone, especially for the $\Tau_0$, $p^{\gamma \gamma}_T$ and
$p^{\gamma_h}_T$ distributions. The $y_{\gamma \gamma}$ distribution
is instead left untouched by the shower.  These are
most likely due to the high scale at which we start the showering
process. A similar behaviour was also observed for the $VH$ production
process in \geneva after including the $gg$ channel contribution, as
well as in the \powheg and \mcatnlo implementations of similar
processes~\cite{Alioli:2016xab,Heinrich:2017kxx}.

\begin{figure}[tp]
\begin{center}
\begin{tabular}{ccc}
\includegraphics[width=\rescaletwoplots]{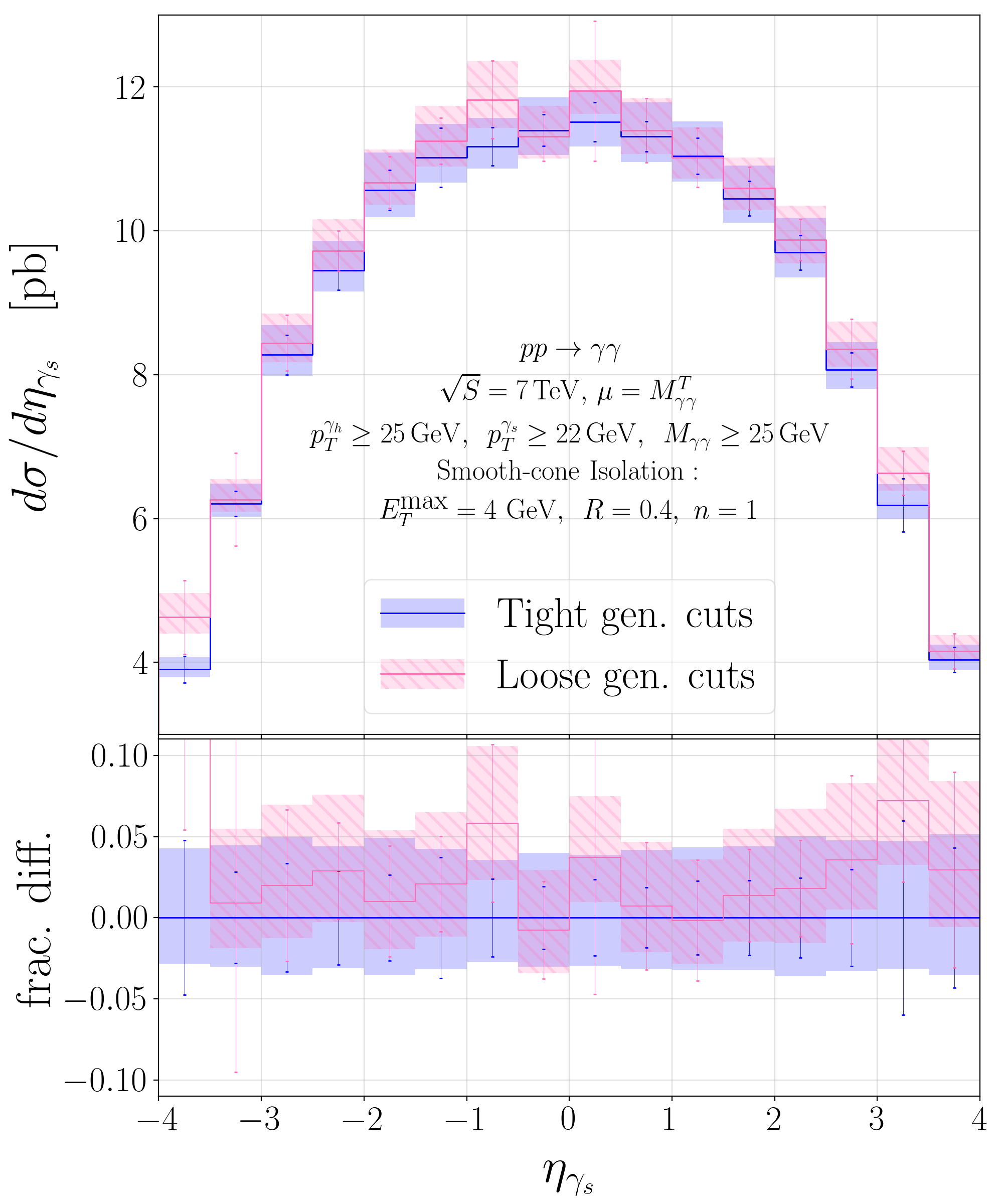} &\hspacebetweentwoplots&
\includegraphics[width=\rescaletwoplots]{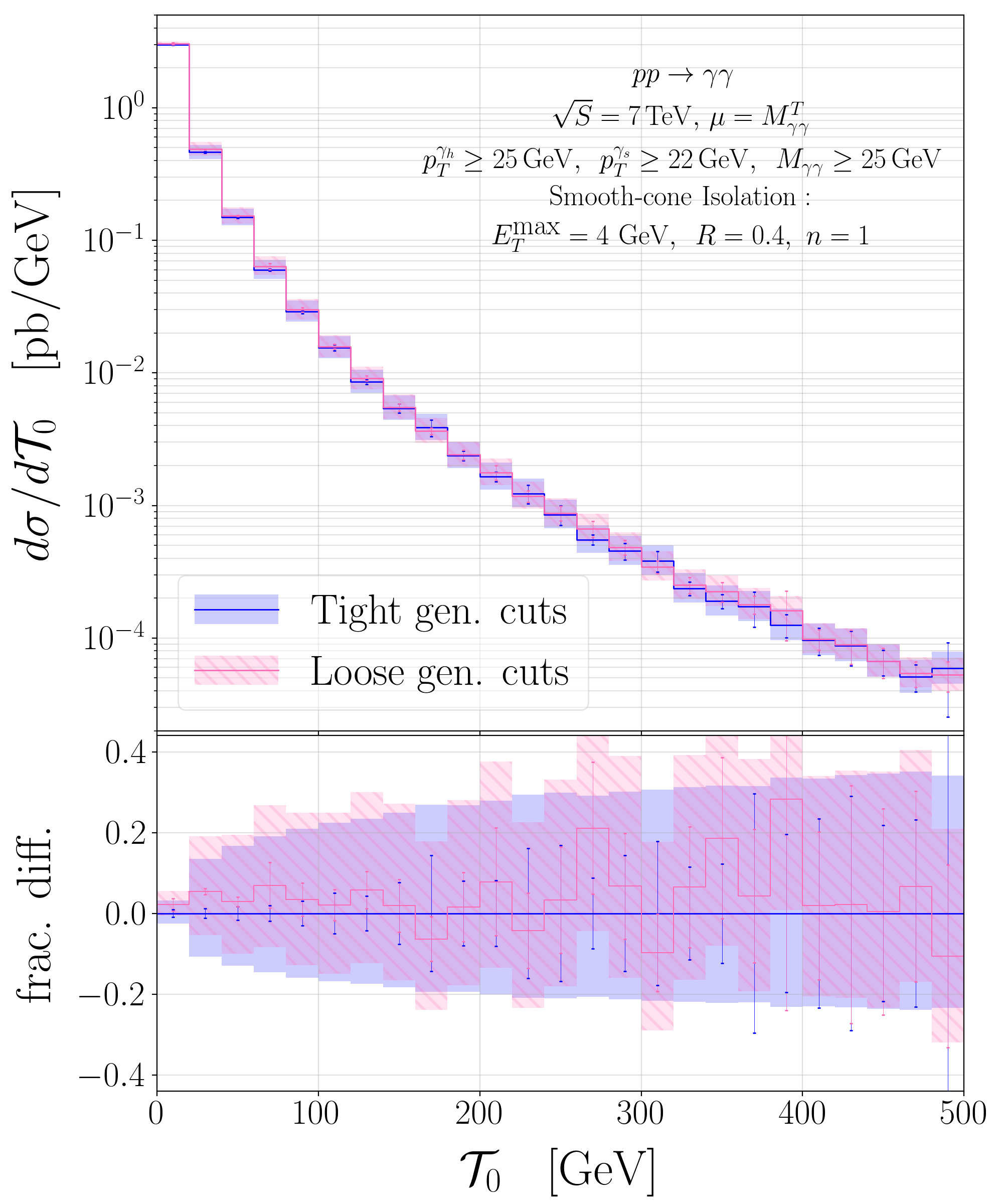}
\end{tabular}
\end{center}
\spaceabovefigurecaption
\caption{ Comparison between two different sets of generation cuts
  (see text for additional details) for the pseudorapidity of the
  softer photon (left) and the $\Tau_0$ distribution (right).
\label{fig:cmpgencuts}
}
\spacebelowfigurecaption
\end{figure}
\begin{figure}[tp]
\begin{center}
\begin{tabular}{ccc}
\includegraphics[width=\rescaletwoplots]{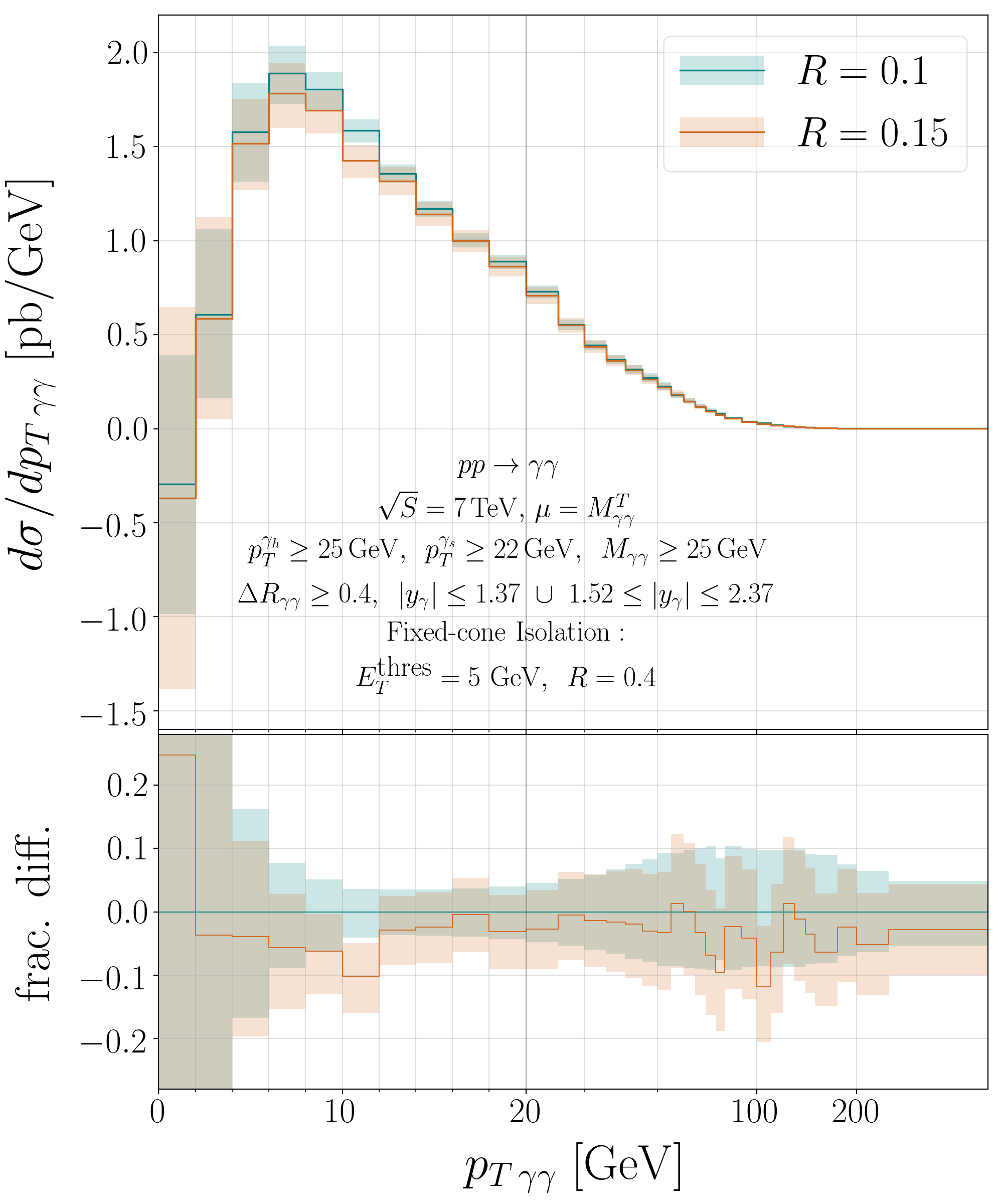} &\hspacebetweentwoplots&
\includegraphics[width=\rescaletwoplots]{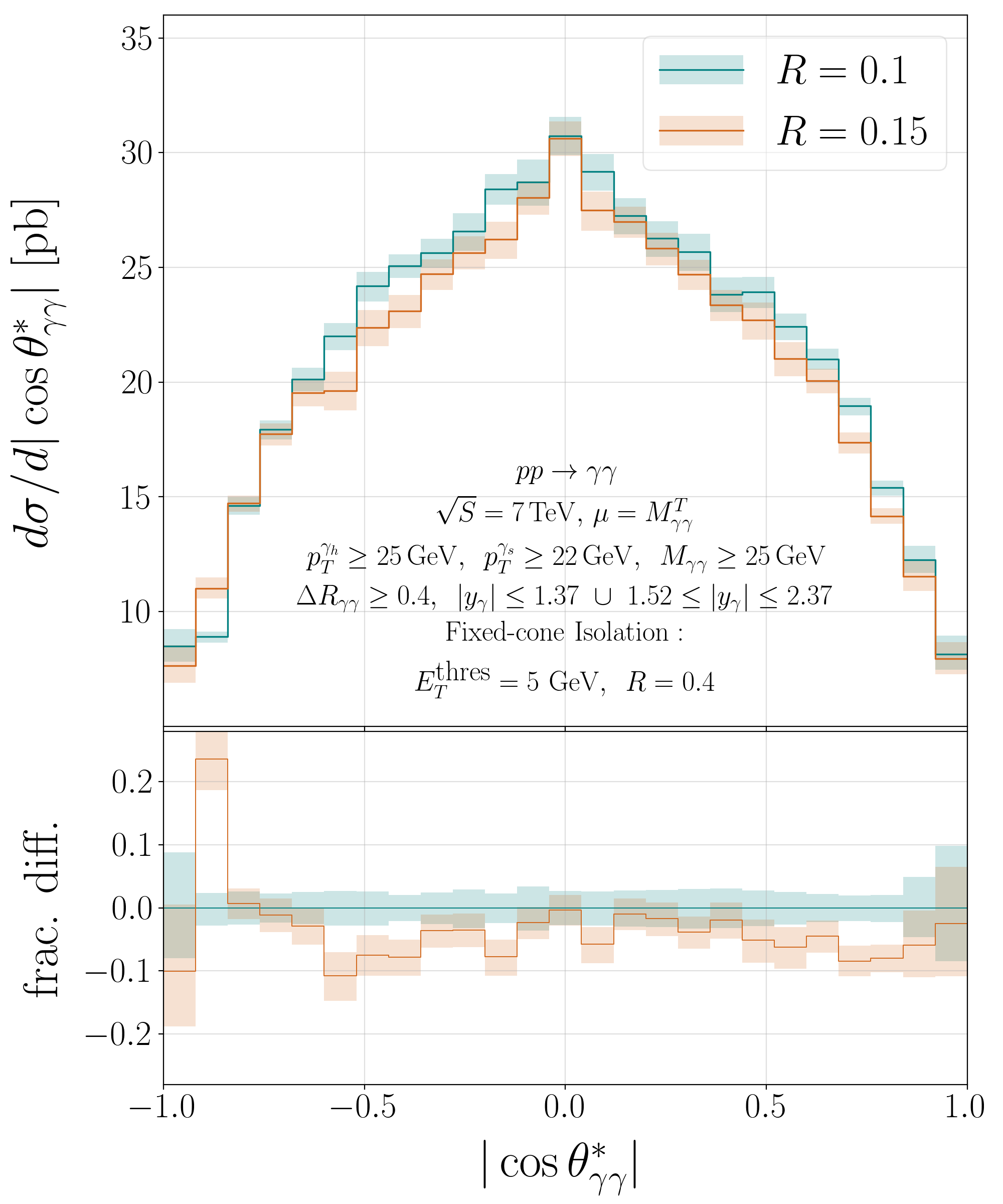}
\end{tabular}
\end{center}
\spaceabovefigurecaption
\caption{Comparison between \geneva+\;\pythiaEight results after
  applying two different generation cuts and the same analysis
  cuts. The theoretical predictions have been produced by applying the
  \rivet analysis {\tt {ATLAS\_2012\_I1199269}} to the hadronised
  events. We show the transverse momentum of the photon pair (left)
  and the cosine of the photon angle in the Collins--Soper frame
  (right).
\label{fig:isodep}
}
\spacebelowfigurecaption
\end{figure}
%

\subsection{Event generation and analysis cuts}
\label{subsec:genevaanalysis}

In this subsection we study the effects of applying process-defining
and isolation cuts at the generation and analysis levels, both before
and after shower and hadronisation. At the generation level, we are
forced to use a smooth-cone isolation procedure in order to generate
well-defined, IR-finite events, without fragmentation
contributions. At the analysis level, however, when one is interested
in comparing with data, a fixed-cone isolation algorithm is
needed. For these reasons, in \sec{results} we will apply a hybrid
isolation procedure, \ie first imposing a very loose smooth-cone
isolation cut at the generation level followed by a tighter fixed-cone
isolation at the analysis level.

In order to check the consistency of this approach, we must first
quantify the dependence of the results at the various levels of the
analysis from the cuts imposed at generation. We separate this
investigation into two parts: in the first, at parton level, we
examine the power-suppressed isolation effects due to the phase-space
projections below the jet resolution cutoffs; in the second, after the
shower, we study the effect of the random momenta reshuffling due to
recoil and hadronisation.

For the first part, we use the set of ``tight" cuts introduced in
\eqs{ptcuts}{isocuts}, which we report here for convenience
\begin{align}\label{eq:tightcuts}
p^{\gamma_h}_T \ge 25\,\, \mathrm{GeV}, \quad p^{\gamma_s}_T &\ge 22\,\, \mathrm{GeV},\quad M_{\gamma \gamma} \ge 25\,\, \mathrm{GeV}\, ,\nn \\
E^{\mathrm{max}}_T = 4 \,\, \mathrm{GeV}, \quad &R_{\mathrm{iso}}  = 0.4 ,\quad \mathrm{and} \quad n = 1 \, ,
\end{align}
and the second set of ``loose" cuts given by
\begin{align}\label{eq:loosecuts}
p^{\gamma_h}_T \ge 18\,\, \mathrm{GeV}, \quad p^{\gamma_s}_T &\ge 15\,\, \mathrm{GeV},\quad M_{\gamma \gamma} \ge 25\,\, \mathrm{GeV}\, ,\nn \\
E^{\mathrm{max}}_T = 4 \,\, \mathrm{GeV}, \quad &R_{\mathrm{iso}}  = 0.1 ,\quad \mathrm{and} \quad n = 1 \, .
\end{align}
We first generate the events by applying the set of loose cuts in
\eq{loosecuts} and, as a second step, we analyse them by applying the
tighter cuts of \eq{tightcuts} before showering.  We compare these
predictions to the results obtained by directly applying the set of
tight cuts at generation level.

This is shown in \fig{cmpgencuts} for
the pseudorapidity of the softer photon and the $\Tau_0$ distribution,
where we show the results of the calculation directly carried out with
tight generation cuts together with that where we apply loose
generation cuts (as in \eq{loosecuts}) and tighter cuts at the
analysis level. The two predictions are in good agreement and this
gives us confidence that our results are not strongly dependent on the
generation cuts applied.

For the second part, one should expect that power-suppressed effects
connected with the recoil after any emission could modify the momenta
of the final-state particles and, consequently, result in a different
rate of events passing the analysis cuts compared to those passing the
generation cuts. This effect is particularly severe after the shower,
since multiple emissions can greatly reshuffle the final-state
momenta. The same applies to the reshuffle used by SMC
programs to impose momentum conservation after hadronisation.

In order to quantify these effects we compare in \fig{isodep} results
obtained employing the loose generation cuts in \eq{loosecuts} with
the values $R_{\textrm{iso}}=0.1$ and $R_{\textrm{iso}}=0.15$ and
applying the ATLAS analysis cuts which are introduced later in
\eqs{atlascuts1}{atlascuts2} of \sec{results}. The figure shows
reasonable agreement between the two predictions for the transverse
momentum of the photon pair and the cosine of the photon angle in the
Collins--Soper frame, demonstrating that the size of these effects is
not large for variations of the isolation radius at generation
level. However, qualitatively we did find a stronger dependence of the
final results on the choice of the generation cuts on the photons'
transverse momenta.

\begin{figure}[tp]
\begin{center}
\begin{tabular}{ccc}
\includegraphics[width=\rescaletwoplots]{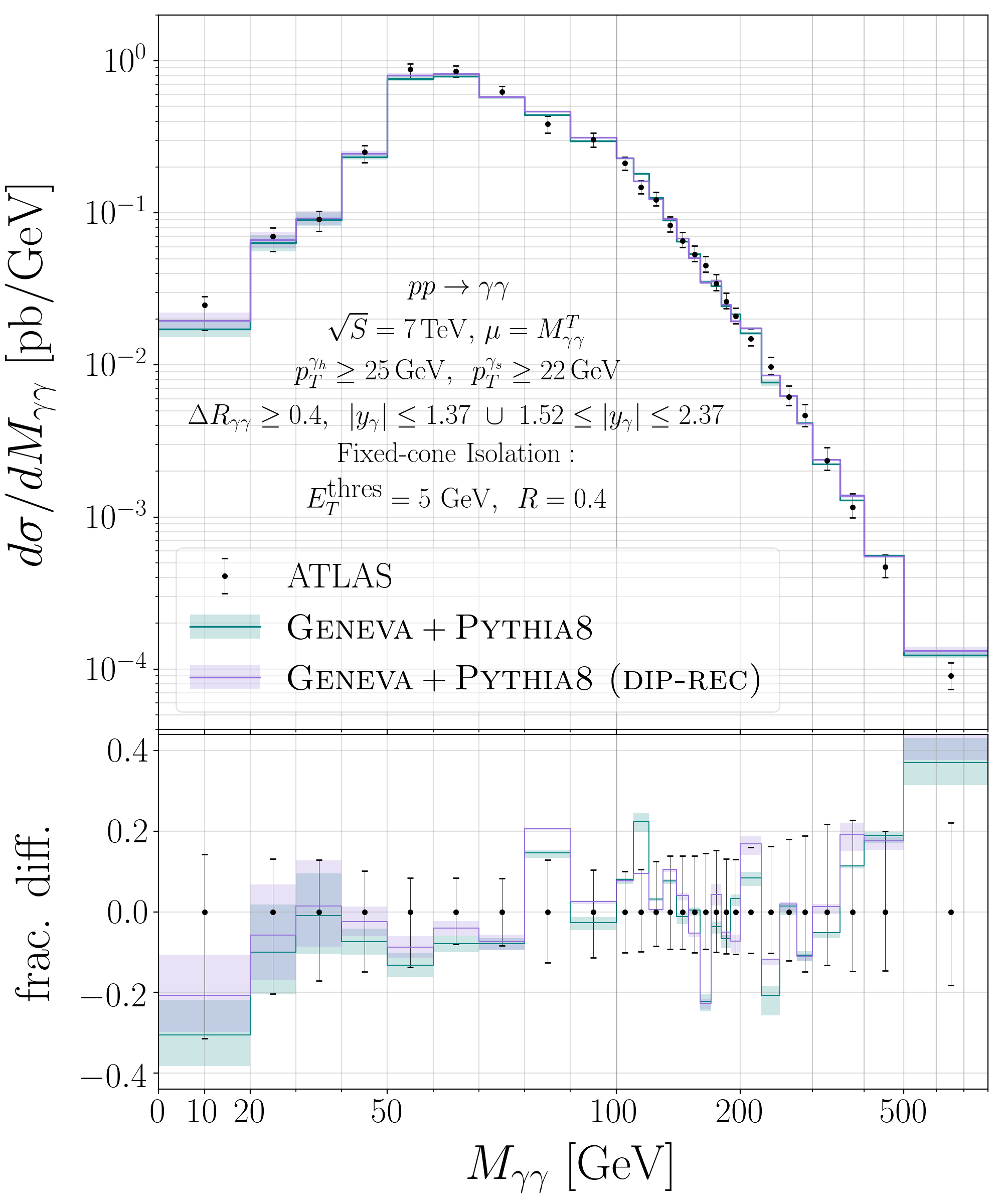} &\hspacebetweentwoplots&
\includegraphics[width=\rescaletwoplots]{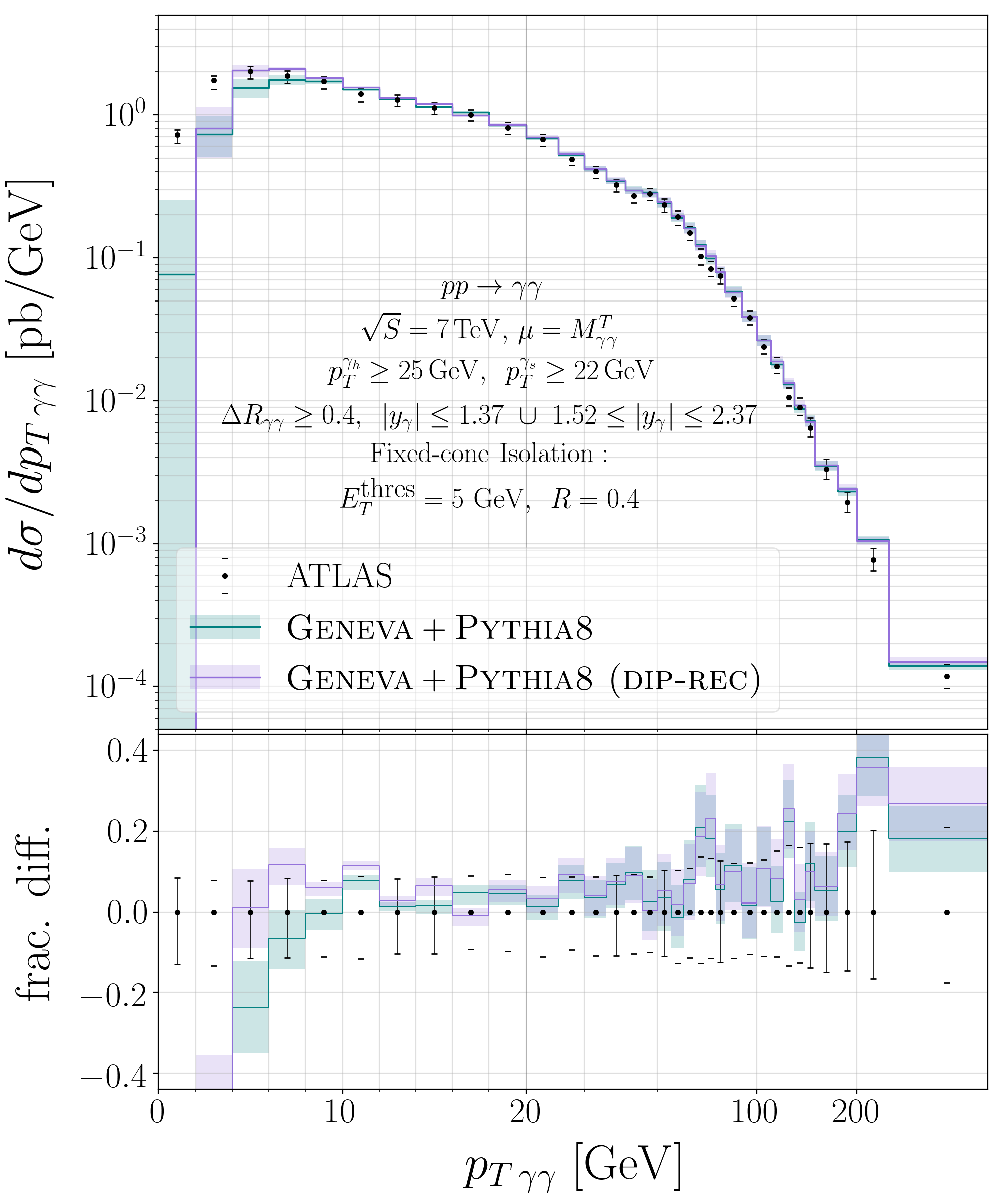} \\[\vspacebetweentwoplots]
\includegraphics[width=\rescaletwoplots]{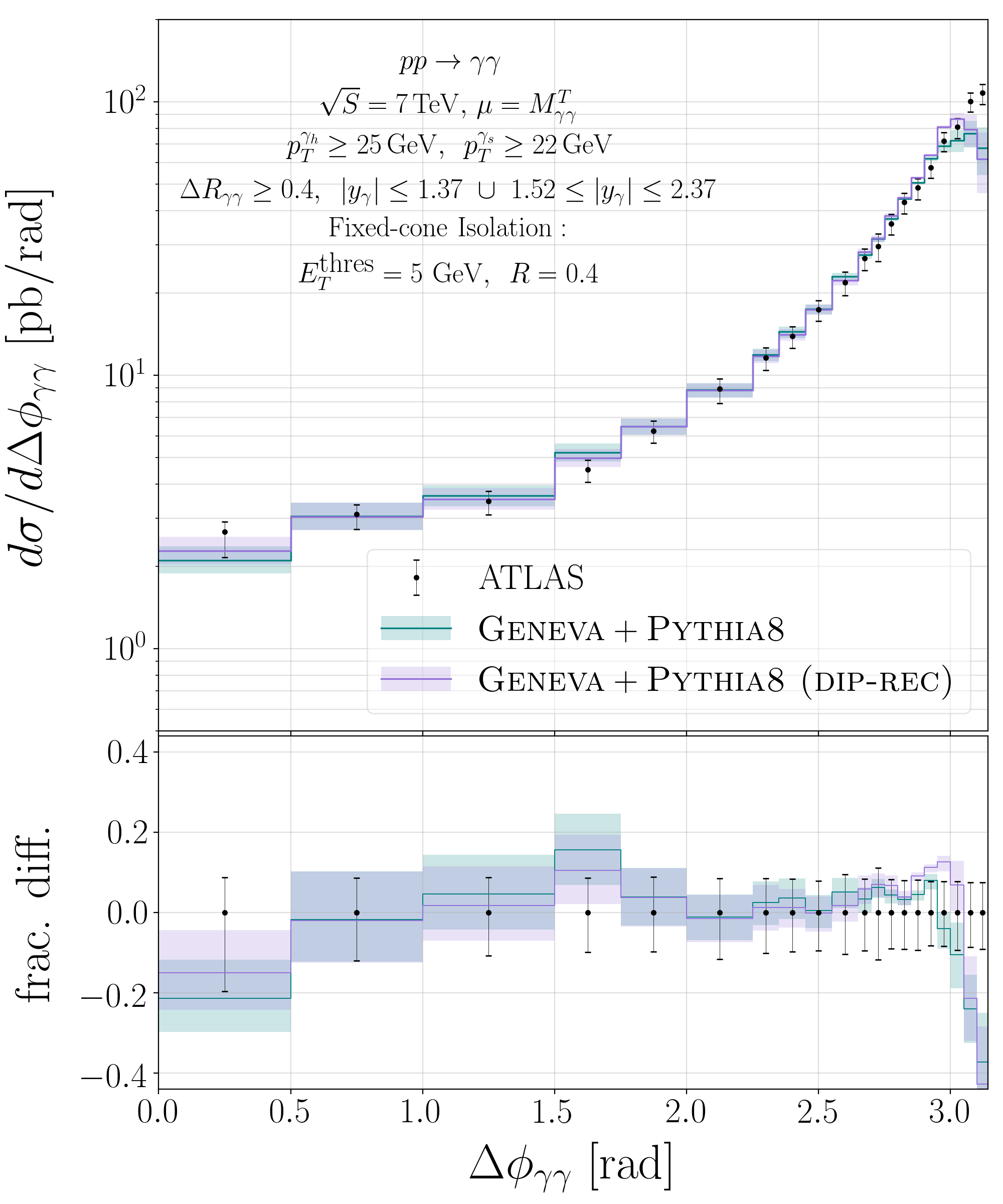} &\hspacebetweentwoplots&
\includegraphics[width=\rescaletwoplots]{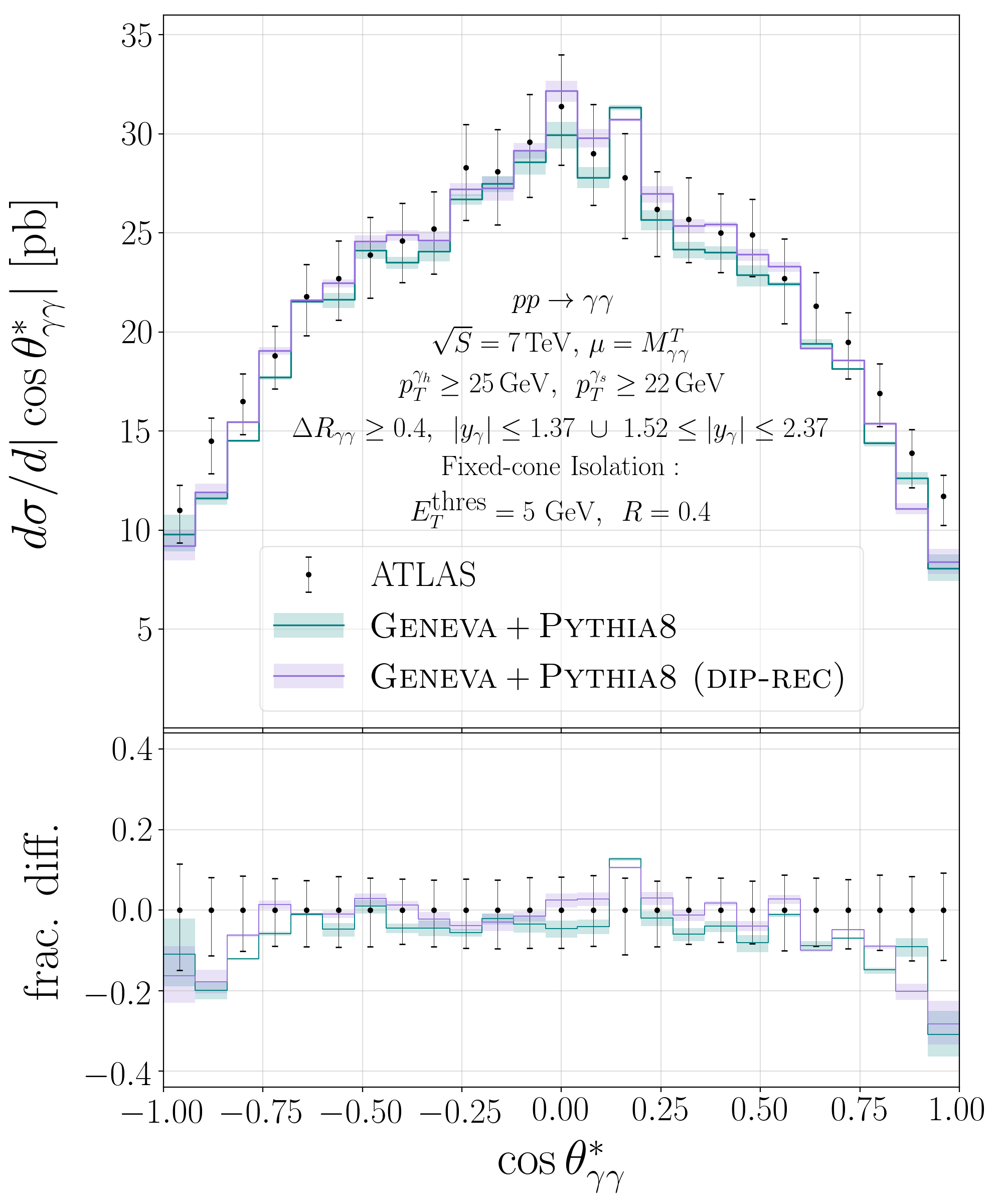} \\
\end{tabular}
\end{center}
\spaceabovefigurecaption
\caption{Comparison between \geneva+\;\pythiaEight and the 7 TeV data
  from ATLAS~\cite{Aad:2012tba}. The theoretical predictions have been
  produced by applying the \rivet analysis {\tt
    {ATLAS\_2012\_I1199269}} to the hadronised events.  We show the
  invariant mass of the photon pair (top left), the transverse
  momentum of the diphoton system (top right), the azimuthal-angle
  separation between the two photons (bottom left) and the cosine of
  the polar angle in the Collins--Soper frame of the diphoton system
  (bottom right).
\label{fig:rivetATLASMPIQEDV2}
}
\spacebelowfigurecaption
\end{figure}
\begin{figure}[tp]
\begin{center}
\begin{tabular}{ccc}
\includegraphics[width=\rescaletwoplots]{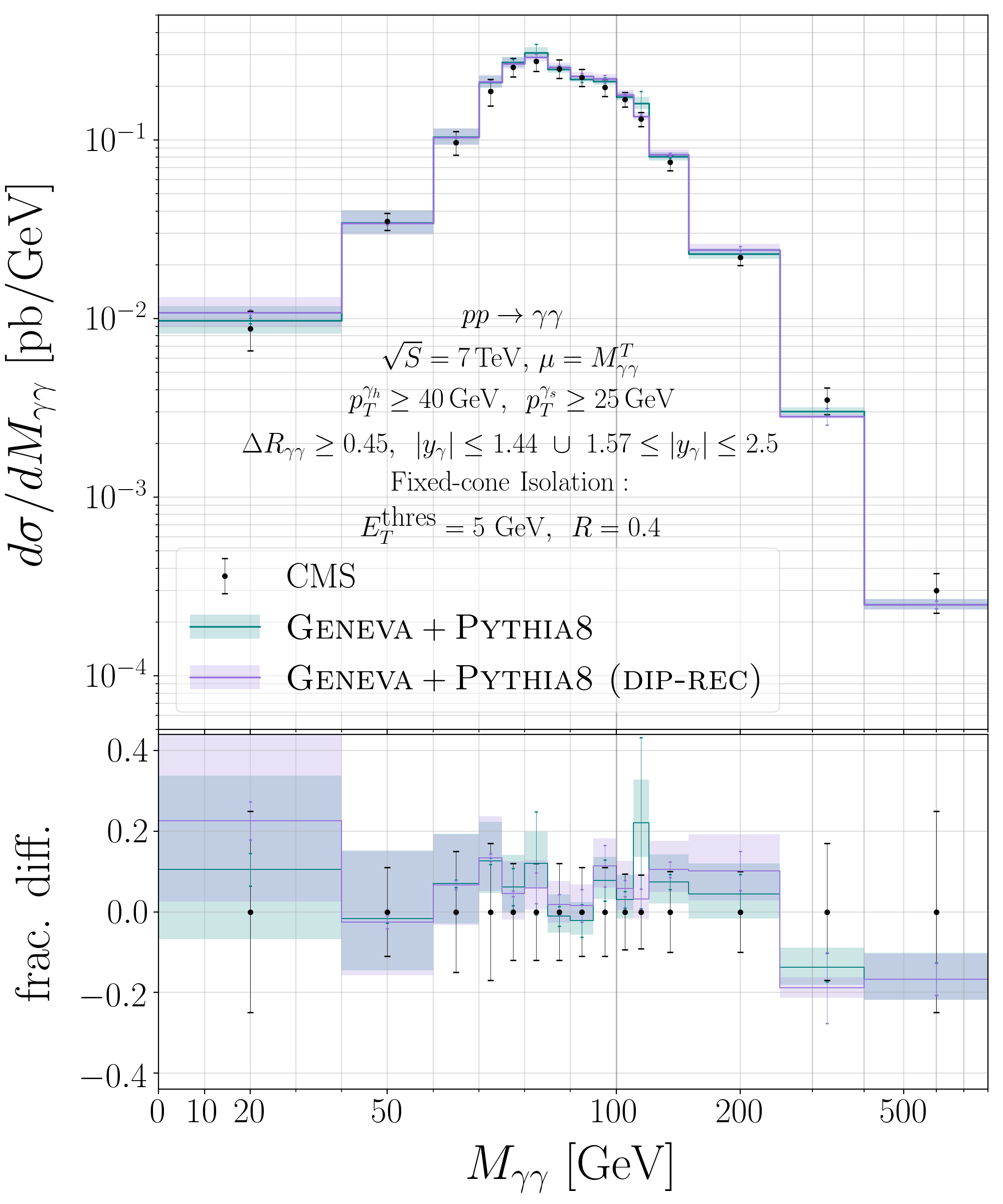} &\hspacebetweentwoplots&
\includegraphics[width=\rescaletwoplots]{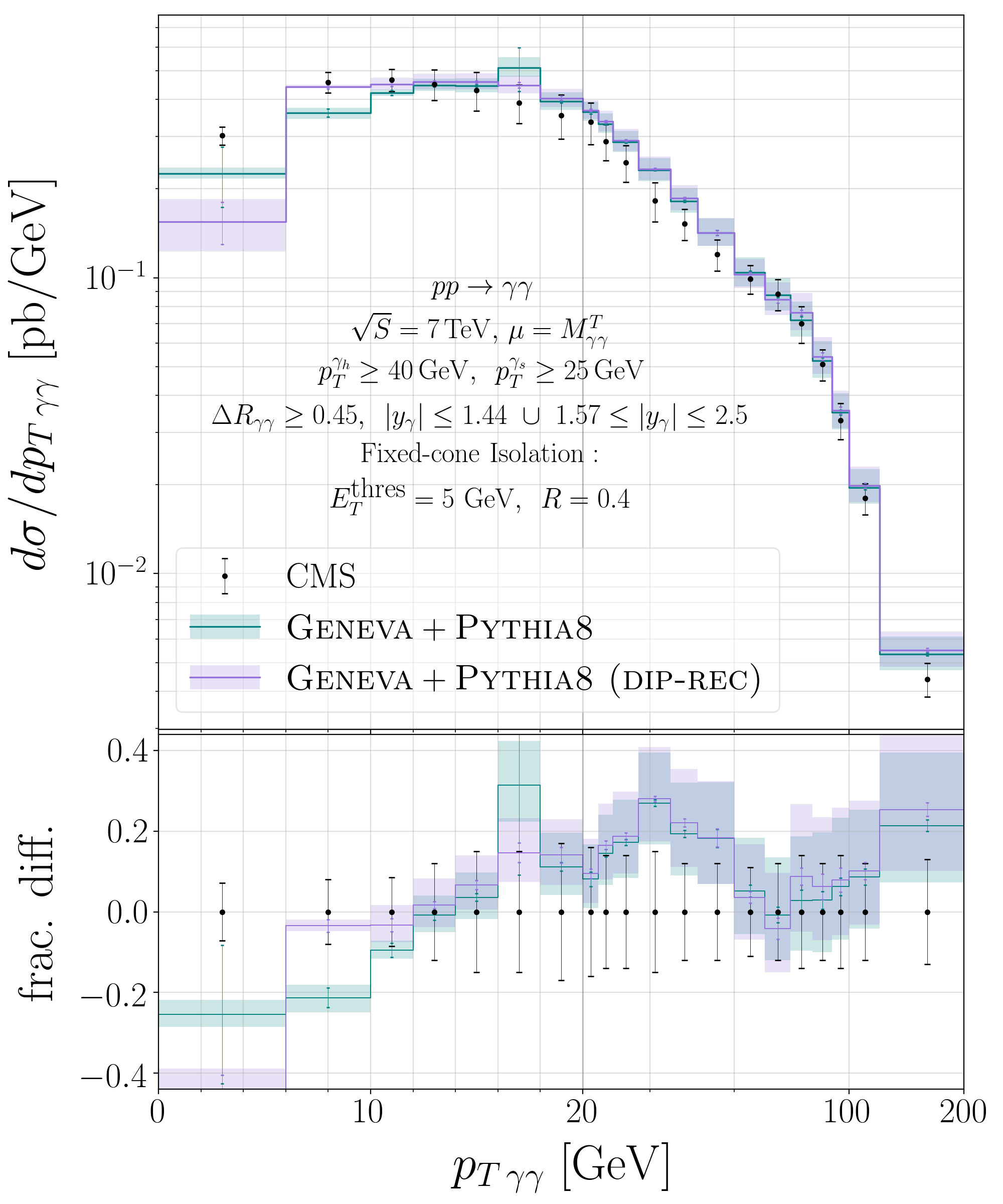} \\[\vspacebetweentwoplots]
\includegraphics[width=\rescaletwoplots]{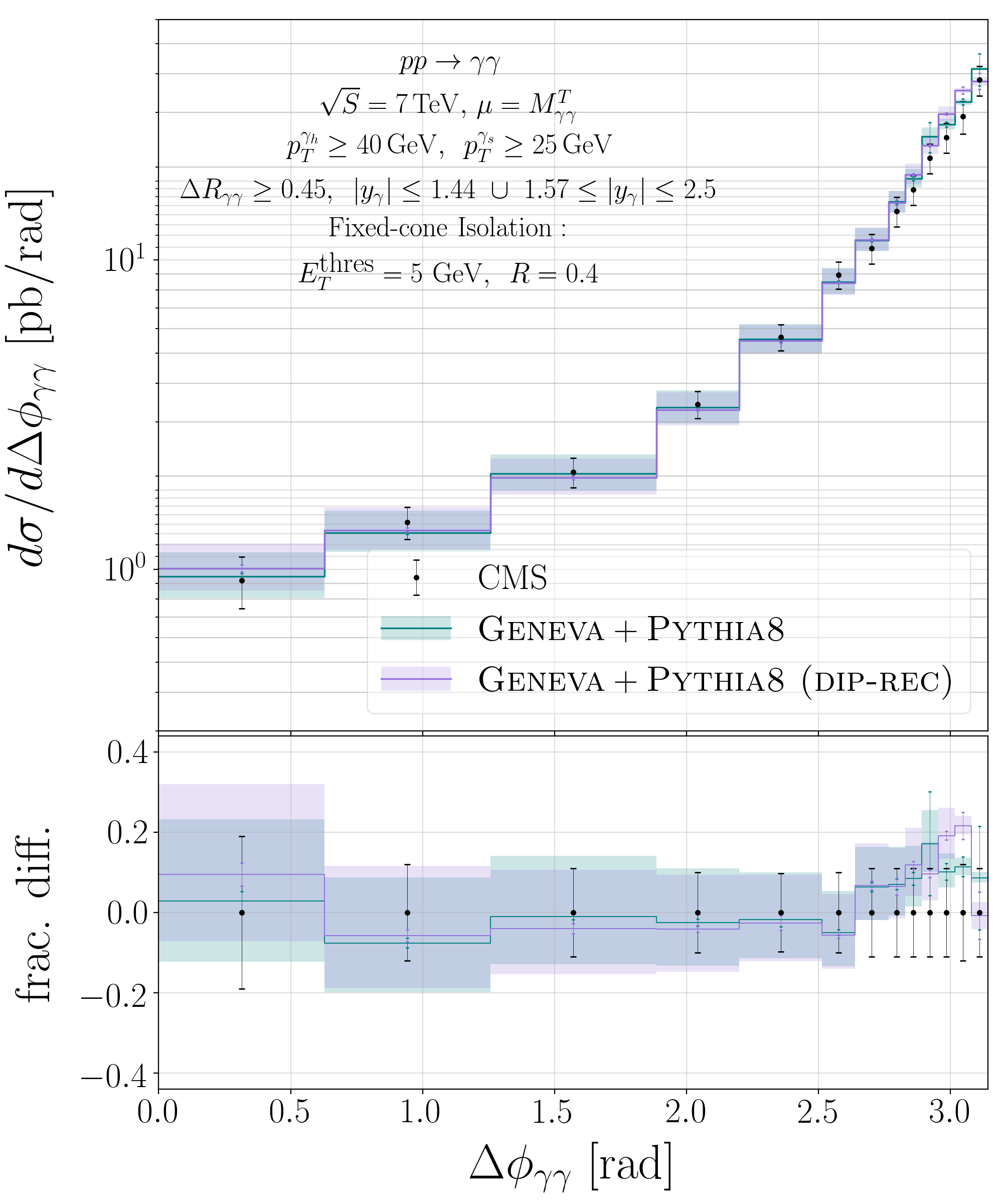} &\hspacebetweentwoplots&
\includegraphics[width=\rescaletwoplots]{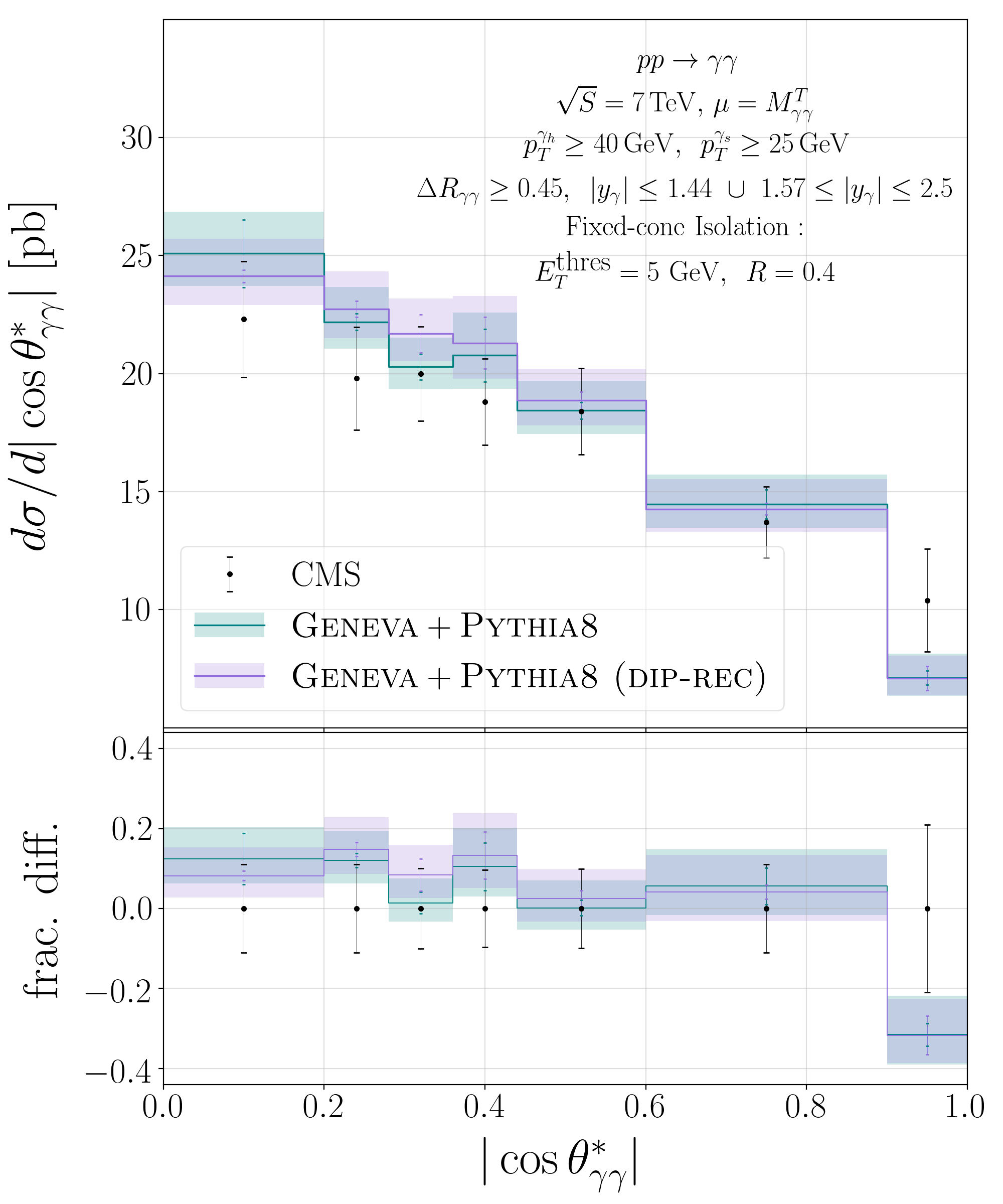} \\
\end{tabular}
\end{center}
\spaceabovefigurecaption
\caption{Comparison between \geneva+\;\pythiaEight and the 7 TeV data from CMS~\cite{Chatrchyan:2014fsa}. We show
invariant mass of the photon pair (top left), the
transverse momentum of the diphoton system  (top
right), the azimuthal-angle separation between the two photons (bottom left) and the cosine of the polar angle in the Collins--Soper frame of the diphoton
system (bottom right).
\label{fig:CMSMPIQED}
}
\spacebelowfigurecaption
\end{figure}
%

\section{Results and comparison to LHC data}
\label{sec:results}

In this section we compare our predictions against $7$~TeV LHC data
obtained from both ATLAS~\cite{Aad:2012tba} and
CMS~\cite{Chatrchyan:2014fsa}.  We employ the hybrid isolation
procedure, as detailed in \sec{PhotonIso} and
\subsec{genevaanalysis}. This means that we first generate partonic
events with looser smooth-cone isolation cuts, and only after the
shower and hadronisation procedures do we apply the tighter analysis
cuts and fixed-cone isolation algorithms which are used by the ATLAS
and CMS experiments.

For these particular comparisons, we generate events using the
\texttt{NNPDF31\_nnlo\_as\_0118} PDF set~\cite{Ball:2017nwa}.  We set
the FO scale to $\mu_\FO=M_{\gamma \gamma}^T$ and apply the following
process-defining cuts at generation level:
\begin{align}\label{eq:geninputsLHCcompare}
p^{\gamma_h}_T \ge 18\,\, \mathrm{GeV}, \quad p^{\gamma_s}_T &\ge 15\,\, \mathrm{GeV},\quad M_{\gamma \gamma} \ge 1\,\, \mathrm{GeV}\, ,\nn \\
E^{\mathrm{max}}_T = 4 \,\, \mathrm{GeV}, \quad &R_{\mathrm{iso}}  = 0.1 ,\quad \mathrm{and} \quad n = 1 \, .
\end{align}
Note that, in principle,  there is no need to
require a lower limit on the
invariant mass of the photon pair, but,  since
our hard function is evaluated at $\mu_H=M_{\gamma \gamma}$ in the
resummation region, we set this lower cutoff so that $\alpha_s(\mu_H)$
is not evaluated at scales which are too small.

We then shower and
hadronise the partonic events generated with the cuts in
\eq{geninputsLHCcompare}. To this end, we use \pythiaEight
and, in order to avoid any contamination by photons coming from
hadronic jets, we prevent the decay of hadrons. For this comparison we
also include the MPI and the QED shower effects. However, we do not
allow QED splittings of photons into quarks or leptons.\footnote{We
achieve this by changing the default \pythiaEight shower options to
{\tt HadronLevel:Decay = off} and \texttt{TimeShower:QEDshowerByGamma
= off}.} We obtained our predictions by using two
different recoil schemes for the shower: the default shower recoil
of \pythiaEight and a more local dipole recoil (\mbox{DIP-REC}).

In \fig{rivetATLASMPIQEDV2} we show the comparison between our
predictions and the ATLAS data at 7 TeV~\cite{Aad:2012tba} for the
invariant mass of the photon pair $M_{\gamma \gamma}$, the transverse
momentum of the diphoton system $p^{\gamma \gamma}_T$, the
azimuthal-angle separation between the two photons $\Delta
\phi_{\gamma \gamma}$ and the cosine of the polar angle
$\theta^*_{\gamma \gamma}$ in the Collins--Soper frame of the diphoton
system. The comparison is carried out by applying to the showered and
hadronised events the \rivet~\cite{Buckley:2010ar} analysis {\tt
  {ATLAS\_2012\_I1199269}}, which is provided by the ATLAS
collaboration. This requires the presence of two isolated photons by
means of a fixed-cone isolation criterion with parameters
\begin{align}
  E^{\mathrm{max}}_T = 4\,\,\mathrm{GeV}, \quad R_{\mathrm{iso}}=0.4\, ,
  \label{eq:atlascuts1}
\end{align}
and the set of cuts
\begin{align}
  p^{\gamma_h}_T \ge 25\,\, \mathrm{GeV}, \quad p^{\gamma_s}_T &\ge
  22\,\, \mathrm{GeV},\quad \Delta R_{\gamma \gamma} \geq 0.4\quad
  \mathrm{and}\quad |y_\gamma| \leq 1.37\ \cup \ 1.52 \leq |y_\gamma|
  \leq 2.37 \, ,
\label{eq:atlascuts2}
\end{align}
where $\Delta R_{\gamma \gamma} = \sqrt{\Delta {\eta_{\gamma
      \gamma}}^2 + \Delta {\phi_{\gamma \gamma}}^2}$ is the separation
between the photons.  Overall, we observe very good agreement between
the theoretical predictions and the ATLAS data.  For the invariant
mass distribution, in the region \mbox{$M_{\gamma
    \gamma}\geq350$}~GeV, above the $t\bar{t}$ production threshold,
the theoretical predictions seem to depart from data.  Here we expect
that the inclusion of EW corrections and of the two-loop diagrams with
a closed top-quark loop in the hard function will improve the
theoretical description.  We also observe a deviation in the extreme
region $\Delta \phi_{\gamma \gamma}\sim \pi$ of the $\Delta
\phi_{\gamma \gamma}$ distribution.

Next, we compare against the CMS data at 7
TeV~\cite{Chatrchyan:2014fsa}.  The CMS analysis uses a fixed-cone
isolation algorithm with parameters
\begin{align}
E^{\mathrm{max}}_T = 5\,\,\mathrm{GeV}, \quad R_{\mathrm{iso}}=0.4\, ,
\end{align}
and the set of cuts
\begin{align}
  p^{\gamma_h}_T \ge 40\,\, \mathrm{GeV}, \quad p^{\gamma_s}_T &\ge
  25\,\, \mathrm{GeV},\quad \Delta R_{\gamma \gamma} \geq 0.45\quad
  \mathrm{and}\quad |y_\gamma| \leq 1.44\ \cup \ 1.57 \leq |y_\gamma|
  \leq 2.5 \, .
\end{align}

The comparison between the predictions obtained with
\geneva+\;\pythiaEight and the CMS data is shown in \fig{CMSMPIQED}
for the same set of distributions presented for the ATLAS
case. Unfortunately, since no corresponding \rivet analysis is
available, we have implemented the aforementioned cuts in the \geneva
analyser. The $M_{\gamma \gamma}$ distribution is shown with a linear
scale on the abscissa up to $100$~GeV and a logarithmic scale
beyond. Similarly, the $p^{\gamma \gamma}_T$ distribution is shown
with a linear scale up to $20$~GeV and a logarithmic scale beyond.  We
observe a similarly good agreement for the inclusive distributions as
for ATLAS. Close to $\Delta \phi_{\gamma \gamma} \sim \pi$ the
photons' azimuthal separation shows an opposite trend compared to that
of ATLAS, but our predictions in this case always agree with CMS data
within experimental errors.

For both experiments, the $p_{T,\gamma\gamma}$ prediction also shows
some systematic trends, undershooting the data at the very low end of
the spectrum. However, the local dipole recoil scheme seems to perform significantly better than the default \pythiaEight, providing a good description of the data down to very low values of $p_{T,\gamma\gamma}$ for both ATLAS and CMS. We also stress that the theoretical uncertainties on our predictions for this observable are not completely exhaustive as they
do not yet include \eg the uncertainties related to the matching to
the shower or to the variations of the shower parameters.  Since this
discrepancy seems entirely caused by shower recoil and nonperturbative effects (\textit{cfr.}~\figs{showerqqbarNOancuts}{cmpradish}) we expect that
including shower uncertainties and hadronisation tuning will improve
the agreement with data.

\section{Conclusions}
\label{sec:conc}

We have presented the first calculation for the production of isolated
photon pairs at the LHC resummed in the $0$-jettiness resolution
variable to NNLL$^\prime$ accuracy and matched to the NNLO
calculation. This has been performed within the \geneva Monte Carlo
framework, allowing us to interface to the \pythiaEight parton shower
and hadronisation model.  Our work constitutes the first NNLO event
generator matched to a parton shower (NNLO+PS) for this process.

The implementation of photon pair production in an event generator is
complicated by the nontrivial process definition, which suffers
from QED singularities.  In order to solve this problem we
used a smooth-cone isolation algorithm in \geneva to remove such divergences.
We studied the dependence of our results on the parameters of the
isolation applied at generation level and
also investigated the
differences due to the recoil between the standard resummation approach and our
implementation in \geneva.
We have validated our
calculation using the \matrix program to NNLO accuracy and found good
agreement when using $\Tau^\cut_0=0.01$~GeV, which sufficiently reduces the size of
the neglected subleading power corrections.

We have further studied the effects of parton shower and hadronisation.
We first ensured that the $\Tau_0$ distribution is not
affected by the shower alone if no additional phase space cuts are
applied after showering. We have then quantified the size of the
nonperturbative effects provided by hadronisation at the low end of
the $\Tau_0$ spectrum, finding the expected modifications.

We have also found that the NNLO description of the inclusive distributions
is preserved by the shower and hadronisation procedures.  Larger
effects instead appear for more exclusive distributions. After
including the $gg$ channel at $\mathcal{O}(\alpha^2_s)$, we observed
larger shower effects, both for inclusive and exclusive distributions,
connected to the usage of a higher starting scale of the shower for
this contribution.

Finally, after including both MPI and QED effects in \pythiaEight, we
have used a hybrid isolation procedure to compare to LHC data at 7
TeV.  We find in general a good agreement with both ATLAS and CMS,
with minor tensions appearing for exclusive distributions such as
$p_{T,\gamma\gamma}$ and $\Delta \phi_{\gamma \gamma}$. These are reduced when using a more local shower recoil scheme which has a smaller impact on the colour singlet system. Therefore our predictions could possibly be ameliorated after the inclusion of theoretical
uncertainties connected to the matching to the shower or to the
variations of the shower and hadronisation parameters.

Possible directions for future work include the NLL$^\prime$
resummation of the gluon fusion channel contribution, which is
currently included only at LO+PS, the addition of electroweak
corrections, and of the top-quark mass effects in the two-loop hard
function.  Moreover, other interesting processes with a single photon
in the final state, such as $Z\gamma$ and $W\gamma$, could be also
implemented in the \geneva framework.

The code used for the simulations presented in this work is available
upon request from the authors and will be made public in a future
release of \geneva.
\vskip 0.3cm
\noindent{\bf Note added:} On the same day the present article appeared on \texttt{arXiv.org}, the matching of NNLO corrections to parton showers for another genuine $2 \to 2$ process, namely $Z\gamma$, was posted \cite{Lombardi:2020wju}.

\section*{Acknowledgements}
\label{sec:Acknowledgements}

We thank Nigel Glover for providing the two-loop squared amplitudes
from which we extracted the hard functions and Jonas Lindert for his help
with the usage of \openloops~2. We also thank Leandro Cieri for his comments on the manuscript and for discussions in the early stage of the project.
The work of SA, AB, SK,
RN, DN and LR is supported by the ERC Starting Grant REINVENT-714788.
SA and ML acknowledge funding from Fondazione Cariplo and Regione
Lombardia, grant 2017-2070.  The work of SA is also supported by MIUR
through the FARE grant R18ZRBEAFC.  We acknowledge the CINECA award
under the ISCRA initiative and the National Energy Research Scientific
Computing Center (NERSC), a U.S. Department of Energy Office of
Science User Facility operated under Contract No. DEAC02-05CH11231,
for the availability of the high performance computing resources
needed for this work.

\appendix

\section{Hard functions for $\mathbf{\gamma\gamma}$ production}
\label{app:hardfunction}
\def\aligneq{\;=\;}

The hard function is one of the ingredients of the SCET factorisation
formula in \eq{factorization}, and, in order to achieve
NNLL$^\prime$ accuracy, it is needed up to $\mathcal{O}(\alpha^2_s)$
in perturbation theory.
It can be calculated as the square of the hard interaction matching coefficients.
The relevant partonic process is given by
\begin{align}
	q(p_1) + \bar{q} (p_2) \rightarrow \gamma(p_3) +  \gamma(p_4) +X  \, ,
\end{align}
and the hard function can be extracted from the $2 \to 2$ matrix
elements, which were computed up to two-loop level in
Ref.~\cite{Anastasiou:2002zn}.  We introduce the following Mandelstam
invariants:
\begin{align}
	s \aligneq (p_1 + p_2)^2,\quad t \aligneq (p_2-p_3)^2, \quad u \aligneq (p_1-p_3)^2 \, .
\end{align}
Momentum conservation implies that they satisfy the relation $s+t+u =
0$, where $s>0$ and $t,u <0$. Hence only two of the invariants are
independent and it is convenient to express the results in terms of
$s$ and the dimensionless parameter $x = -t/s$.  The UV-renormalised
amplitudes have the following perturbative expansion\footnote{In the SCET literature it is customary to
	express the perturbative expansions of the hard matching coefficients
	of certain operators in powers of $\alpha_s/4\pi$, rather than $\alpha_s/2\pi$. This introduces a $2^L$ conversion factor for $L$-loop amplitudes.}
\begin{align} \label{eq:amplitudes}
	|\mathcal{M}_{q\bar{q}\gamma\gamma}(\epsilon,s,x) \rangle
        \aligneq&\; 4 \pi \alpha \bigg[
          |\mathcal{M}^{(0)}_{q\bar{q}\gamma\gamma} \rangle +
          \bigg(\frac{\alpha_s}{4 \pi}\bigg)
          |\mathcal{M}^{(1)}_{q\bar{q}\gamma\gamma} \rangle +
          \bigg(\frac{\alpha_s}{4 \pi}\bigg)^2
          |\mathcal{M}^{(2)}_{q\bar{q}\gamma\gamma} \rangle +
          \mathcal{O}(\alpha_s^3)\bigg]\, ,
\end{align}
where $\epsilon=(4-d)/2$ is the dimensional regulator. Note that the
perturbative coefficients on the r.h.s.\ depend also on the
renormalisation scale $\mu_r$ while the all-order amplitude on the
l.h.s.\ is independent of it.

After UV renormalisation the virtual amplitudes still
contain IR poles in the dimensional regulator $\epsilon$.\footnote{The
precise structure of the IR poles depends on the regularisation scheme
employed in the calculation. For the relation between different
regularisation schemes, see the analysis done in
Ref.~\cite{Broggio:2015dga}.} We subtract these poles in the
$\overline{\text{MS}}$ scheme by acting on the amplitudes with a
renormalisation factor ${\boldsymbol{Z}}$,
\begin{align}
	\label{eq:IRrenamp}
	| \mathcal{M}^{\text{ren}}_{q\bar{q}\gamma\gamma}(s,x,\mu)
        \rangle \aligneq \lim_{\epsilon \to 0} {\boldsymbol{Z}}^{-1}
        (\epsilon,s,x,\mu)
        |\mathcal{M}_{q\bar{q}\gamma\gamma}(\epsilon,s,x) \rangle\, ,
\end{align}
where the IR-finite amplitudes now depend on the
renormalisation scale $\mu$. The explicit form of the renormalisation
factor ${\boldsymbol{Z}}$ was recently determined up to
$\mathcal{O}(\alpha_s^4)$ in massless QCD in
Ref.~\cite{Becher:2019avh} by investigating the structure of the
associated anomalous dimension. For our specific computation we need
it only up to $\mathcal{O}(\alpha^2_s)$ in the case of a colourless
final state. We define its perturbative expansion as
\begin{align}
	{\boldsymbol{Z}}^{-1} (\epsilon,s,\mu) \aligneq 1 +
        \bigg(\frac{\alpha_s}{4 \pi}\bigg)
        \,{\boldsymbol{Z}}^{(1)}(\epsilon,s,\mu)+
        \bigg(\frac{\alpha_s}{4 \pi}\bigg)^2\,
             {\boldsymbol{Z}}^{(2)}(\epsilon,s,\mu) +
             \mathcal{O}(\alpha^3_s)\, .
\end{align}
Up to $\mathcal{O}(\alpha^2_s)$ we find
\begin{align}
	|\mathcal{M}^{(1),\text{ren}}_{q\bar{q}\gamma\gamma}(\mu)
	\rangle \aligneq&\; \lim_{\epsilon \to 0} \Big[
	|\mathcal{M}^{(1)}_{q\bar{q}\gamma\gamma}(\epsilon) \rangle +
	{\boldsymbol Z}^{(1)}(\epsilon,\mu)
	|\mathcal{M}^{(0)}_{q\bar{q}\gamma\gamma}(\epsilon) \rangle
	\Big]\,
	, \label{eq:ren1}\\ |\mathcal{M}^{(2),\text{ren}}_{q\bar{q}\gamma\gamma}(\mu)
	\rangle \aligneq&\; \lim_{\epsilon \to 0} \Big[
	|\mathcal{M}^{(2)}_{q\bar{q}\gamma\gamma}(\epsilon) \rangle +
	{\boldsymbol Z}^{(1)}(\epsilon,\mu)
	|\mathcal{M}^{(1),\text{ren}}_{q\bar{q}\gamma\gamma}(\epsilon)
	\rangle \, \nonumber\\ &\hspace*{2.5em}+ \big({\boldsymbol
	Z}^{(2)}(\epsilon,\mu) - \big({\boldsymbol
	Z}^{(1)}(\epsilon,\mu)\big)^2\big)
	|\mathcal{M}^{(0)}_{q\bar{q}\gamma\gamma}(\epsilon) \rangle
	\Big] \, \label{eq:ren2},
\end{align}
where for simplicity we dropped the common kinematic dependence on
$s,x$ from all terms in the above equations. The renormalised
amplitudes on the l.h.s.\ are free of IR poles. For the diphoton
production process the ${\boldsymbol Z}$ factors are defined in terms of
anomalous dimension coefficients
\begin{align}
	{\boldsymbol{Z}}^{(1)}(\epsilon, \mu) \aligneq& -
        \frac{\Gamma^\prime_0}{4 \epsilon^2} - \frac{{\bf \Gamma}_0}{2
          \epsilon}\, , \\ {\boldsymbol{Z}}^{(2)}(\epsilon, \mu)
        \aligneq&\; \frac{(\Gamma^\prime_0)^2}{32 \epsilon^4} +
        \frac{\Gamma^\prime_0}{8 \epsilon^3}\bigg({\bf
          \Gamma}_0+\frac{3}{2} \beta_0 \bigg) + \frac{{\bf
            \Gamma}_0}{8 \epsilon^2} ({\bf \Gamma}_0+2 \beta_0) -
        \frac{\Gamma^\prime_1}{16 \epsilon^2} - \frac{{\bf
            \Gamma}_1}{4 \epsilon} \, ,
\end{align}
where $\beta_0 = 11/3 C_A - 4/3 T_F n_f$, and for this process
\begin{align}
	\Gamma^\prime_i \aligneq& -2 C_F \gamma^{\text{cusp}}_i\, , & &i=0,1\, ,\\
	{\bf \Gamma}_i \aligneq& - C_F \gamma^{\text{cusp}}_i \ln\bigg(\frac{\mu^2}{-s}\bigg) +  2 \gamma_i^q\, , & &i=0,1\, .
\end{align}
The perturbative coefficients of the anomalous dimensions
$\gamma^{\text{cusp}}$ and $\gamma^q$ entering in the above equations
are given by
\begin{align}
	\gamma^{\textrm{cusp}}_0 \aligneq&\; 4\, ,
        \\ \gamma^{\textrm{cusp}}_1 \aligneq& \bigg( \frac{268}{9} -
        \frac{4 \pi^2}{3}\bigg) C_A - \frac{80}{9} T_F n_f\, ,
        \\ \gamma_0^q \aligneq& -3 C_F\, , \\ \gamma_1^q \aligneq&\;
        C^2_F \bigg(-\frac{3}{2} + 2\pi^2-24 \zeta_3 \bigg) +C_F C_A
        \bigg( -\frac{961}{54} -\frac{11 \pi^2}{6} +26 \zeta_3\bigg)
        \, \nonumber \\ &\; + C_F T_F n_f \bigg(\frac{130}{27} +
        \frac{2 \pi^2}{3} \bigg)\, .
\end{align}
Quite often the IR parts of analytic results reported in the literature
are expressed in terms of the Catani IR operators ${\boldsymbol
I}^{(1)}$ and ${\boldsymbol I}^{(2)}$. The amplitudes are
then expressed similarly to \eqs{ren1}{ren2} as
\begin{align}
	|\mathcal{M}^{(1),\text{fin}}_{q\bar{q}\gamma\gamma}(\mu)
	\rangle \aligneq&\; \lim_{\epsilon \to 0} \Big[
	|\mathcal{M}^{(1)}_{q\bar{q}\gamma\gamma}(\epsilon) \rangle -
	{\boldsymbol I}^{(1)}(\epsilon,\mu)
	|\mathcal{M}^{(0)}_{q\bar{q}\gamma\gamma}(\epsilon) \rangle
	\Big]\,
	, \label{eq:fin1}\\ |\mathcal{M}^{(2),\text{fin}}_{q\bar{q}\gamma\gamma}(\mu)
	\rangle \aligneq&\; \lim_{\epsilon \to 0} \Big[
	|\mathcal{M}^{(2)}_{q\bar{q}\gamma\gamma}(\epsilon) \rangle -
	{\boldsymbol I}^{(1)}(\epsilon,\mu)
	|\mathcal{M}^{(1),\text{fin}}_{q\bar{q}\gamma\gamma}(\epsilon)
	\rangle \, \nonumber\\ &\hspace*{2.5em}- \big({\boldsymbol
	I}^{(2)}(\epsilon,\mu) + \big({\boldsymbol
	I}^{(1)}(\epsilon,\mu)\big)^2\big)
	|\mathcal{M}^{(0)}_{q\bar{q}\gamma\gamma}(\epsilon) \rangle
	\Big] \, \label{eq:fin2},
\end{align}
where the difference with the IR-finite amplitudes in
\eqs{ren1}{ren2} amounts to finite terms in
$\epsilon$. The translation between the $\overline{\text{MS}}$ and
Catani subtraction scheme for IR poles can be computed explicitly and
is given by~\cite{Broggio:2014hoa}
\begin{align}
	|\mathcal{M}^{(1),\text{ren}}_{q\bar{q}\gamma\gamma}(\mu)
	\rangle \aligneq&\;
	|\mathcal{M}^{(1),\text{fin}}_{q\bar{q}\gamma\gamma}(\mu)
	\rangle + \lim_{\epsilon \to 0} \Big[{\boldsymbol
	I}^{(1)}(\epsilon,\mu) + {\boldsymbol Z}^{(1)}(\epsilon,\mu)
	\Big] |\mathcal{M}^{(0)}_{q\bar{q}\gamma\gamma} \rangle\,
	, \label{eq:ren1fin1}\\ |\mathcal{M}^{(2),\text{ren}}_{q\bar{q}\gamma\gamma}(\mu)
	\rangle \aligneq&\;
	|\mathcal{M}^{(2),\text{fin}}_{q\bar{q}\gamma\gamma}(\mu)
	\rangle + \lim_{\epsilon \to 0} \Big[{\boldsymbol
	I}^{(1)}(\epsilon,\mu) + {\boldsymbol Z}^{(1)}(\epsilon,\mu)
	\Big]
	|\mathcal{M}^{(1),\text{fin}}_{q\bar{q}\gamma\gamma}(\mu)
	\rangle\, \nonumber \\ &\hspace*{-0.5em}+ \lim_{\epsilon \to
	0} \Big[{\boldsymbol I}^{(2)}(\epsilon,\mu) +
	\big({\boldsymbol I}^{(1)}(\epsilon,\mu) + {\boldsymbol
	Z}^{(1)}(\epsilon,\mu)\big)\, {\boldsymbol
	I}^{(1)}(\epsilon,\mu) + {\boldsymbol Z}^{(2)}(\epsilon,\mu)
	\Big] |\mathcal{M}^{(0)}_{q\bar{q}\gamma\gamma}
	\rangle\, \label{eq:ren2fin2} .
\end{align}
We briefly comment on these equations.  It is important to notice
that in the above equations the pole content of
${\boldsymbol{Z}}^{(2)}$ and ${\boldsymbol{I}}^{(2)}$ is
different, but when all the other terms are taken into account, the
subtraction removes the same poles and the result is finite. While
the ${\boldsymbol{Z}}(\epsilon)$ factors are defined in the $\overline{\text{MS}}$
scheme and only contain poles, the ${\boldsymbol{I}}$ operators
need to be expanded to the correct order in $\epsilon$. For example,
the term between parentheses in the second term of the second line of
\eq{ren2fin2} is finite in $\epsilon$ but must be expanded to
$\mathcal{O}(\epsilon^2)$ since it multiplies ${\boldsymbol
I}^{(1)}$, which contains up to a double pole in $\epsilon$.

The squared amplitudes (summed over colours and spins) are defined as
\begin{align}
	\langle \mathcal{M}_{q\bar{q}\gamma\gamma} |
        \mathcal{M}_{q\bar{q}\gamma\gamma} \rangle \aligneq\sum
        |\mathcal{M}_{q\bar{q}\gamma\gamma}(\epsilon,s,x) |^2 \aligneq
        \mathcal{A}_{q\bar{q}\gamma\gamma} (\epsilon, s, x)\, ,
\end{align}
and are expanded perturbatively as follows:
\begin{align}
	\mathcal{A}_{q\bar{q}\gamma\gamma} (\epsilon, s, x)
        \aligneq&\; 16 \pi^2 \alpha^2 \bigg[
          \mathcal{A}^{\text{LO}}_{q\bar{q}\gamma\gamma}(\epsilon, s,
          x) + \bigg(\frac{\alpha_s}{4 \pi}\bigg)
          \mathcal{A}^{\text{NLO}}_{q\bar{q}\gamma\gamma}(\epsilon, s,
          x) \, \nonumber \\ &\hspace*{4.5em}+ \bigg(\frac{\alpha_s}{4
            \pi}\bigg)^2
          \mathcal{A}^{\text{NNLO}}_{q\bar{q}\gamma\gamma}(\epsilon,
          s, x) +\mathcal{O}(\alpha^3_s)\bigg]\, .
\end{align}
The $\epsilon$ dependence of the LO coefficient is left intentionally
since the higher-order $\epsilon$ terms are important for the
extraction of the higher-order hard function coefficients when the computation is
performed in the conventional dimensional regularisation (CDR) scheme.
Explicitly, the squared amplitudes are computed by interfering the perturbative
coefficients in \eq{amplitudes}. One obtains
\begin{align}
	\mathcal{A}^{\text{LO}}_{q\bar{q}\gamma\gamma}(\epsilon, s, x)
        \aligneq&\; \langle \mathcal{M}^{(0)}_{q\bar{q}\gamma\gamma}
        |\mathcal{M}^{(0)}_{q\bar{q}\gamma\gamma} \rangle \, ,
        \nonumber
        \\ \mathcal{A}^{\text{NLO}}_{q\bar{q}\gamma\gamma}(\epsilon,
        s, x) \aligneq&\; \langle
        \mathcal{M}^{(0)}_{q\bar{q}\gamma\gamma}
        |\mathcal{M}^{(1)}_{q\bar{q}\gamma\gamma} \rangle + \langle
        \mathcal{M}^{(1)}_{q\bar{q}\gamma\gamma}
        |\mathcal{M}^{(0)}_{q\bar{q}\gamma\gamma} \rangle \, ,
        \nonumber
        \\ \mathcal{A}^{\text{NNLO}}_{q\bar{q}\gamma\gamma}(\epsilon,
        s, x) \aligneq&\;
        \langle\mathcal{M}^{(1)}_{q\bar{q}\gamma\gamma}
        |\mathcal{M}^{(1)}_{q\bar{q}\gamma\gamma} \rangle +
        \langle\mathcal{M}^{(0)}_{q\bar{q}\gamma\gamma}
        |\mathcal{M}^{(2)}_{q\bar{q}\gamma\gamma} \rangle +
        \langle\mathcal{M}^{(2)}_{q\bar{q}\gamma\gamma}
        |\mathcal{M}^{(0)}_{q\bar{q}\gamma\gamma}
        \rangle \label{eq:sqamNNLO} \, ,
\end{align}
where we dropped the dependence on $\epsilon, s, x$ on the
r.h.s.\ terms. The squared amplitude coefficients are not yet averaged
over the initial spin polarisations and colours, which introduces a
factor $1/4 N^2_c$. The squared amplitudes should also be divided by
the number of identical particles in the final state, and we therefore
need to multiply by an additional factor $1/2$.  The explicit analytic
expressions in \eq{sqamNNLO}, in terms of logarithms  and classical
polylogarithms can be found in Ref.~\cite{Anastasiou:2002zn}. The IR
pole terms are expressed in terms of Catani's ${\boldsymbol
I}^{(1)}$ and ${\boldsymbol I}^{(2)}$ operators. The
translation into the $\overline{\text{MS}}$ ${\boldsymbol Z}$-factors can be
found in Ref.~\cite{Broggio:2014hoa} for a generic QCD process with
coloured particles in the final state.  For our convenience one of the
authors of Ref.~\cite{Anastasiou:2002zn} directly provided us with the explicit
expressions in \eq{sqamNNLO} in the \form format.

With this in mind we can now define the hard function for the diphoton
production process as
\begin{align}
	\label{eq:hardfunc}
	H(s,x,\mu) \aligneq \frac{16 \pi^2 \alpha}{N^2_c}\bigg[&H^{(0)}(s,x) + \bigg(\frac{\alpha_s}{4 \pi}\bigg) H^{(1)}(s,x,\mu)
          + \bigg(\frac{\alpha_s}{4 \pi}\bigg)^2 H^{(2)}(s,x,\mu) + \mathcal{O}(\alpha^3_s)\bigg]\, ,
\end{align}
where the prefactor $1/N^2_c$ (with $N_c=3$) accounts for the colour
average of the two initial-state quarks.  The hard function coefficients on
the r.h.s.\ of \eq{hardfunc} can be expressed by interfering the
IR-renormalised amplitudes in \eqs{ren1}{ren2}
with their complex conjugates. Unfortunately, the amplitudes are not
explicitly provided in Ref.~\cite{Anastasiou:2002zn}, but we can
express the results in terms of the squared amplitudes in \eq{sqamNNLO}.
The additional prefactor $1/8$ (average over initial spin
polarisations and symmetry factor for
identical final-state particles) is included in the perturbative
coefficients. We find
\begin{align}
	H^{(0)}(x) \aligneq&\; \frac{1}{8} \,
        \mathcal{A}^{\text{LO}}_{q\bar{q}\gamma\gamma}(0, s, x)\, ,
        \\ H^{(1)}(s,x,\mu) \aligneq&\; \frac{1}{8} \,
        \lim_{\epsilon\to 0} \big
            [\mathcal{A}^{\text{NLO}}_{q\bar{q}\gamma\gamma}(\epsilon,
              s, x) + 2
              \text{Re}\big[{\boldsymbol{Z}}^{(1)}(\epsilon,s,\mu)\big]\mathcal{A}^{\text{LO}}_{q\bar{q}\gamma\gamma}(\epsilon,
              s, x)\big] \, , \\ H^{(2)}(s,x,\mu) \aligneq&\;
            \frac{1}{8} \, \lim_{\epsilon\to 0}
            \big[\mathcal{A}^{\text{NNLO}}_{q\bar{q}\gamma\gamma}(\epsilon,
              s, x) +
              2\text{Re}\big[{\boldsymbol{Z}}^{(2)}(\epsilon,s,\mu)\big]\mathcal{A}^{\text{LO}}_{q\bar{q}\gamma\gamma}(\epsilon,
              s, x) \, \nonumber \\ &\; +
              |{\boldsymbol{Z}}^{(1)}(\epsilon,s,\mu)|^2\mathcal{A}^{\text{LO}}_{q\bar{q}\gamma\gamma}(\epsilon,
              s, x) + 2
              \text{Re}\big[{\boldsymbol{Z}}^{(1)}(\epsilon,s,\mu)\big]
              \mathcal{A}^{\text{NLO}}_{q\bar{q}\gamma\gamma}(\epsilon,
              s, x)\big]\, ,
\end{align}
where all of the hard function coefficients on the r.h.s.\ are real and
finite.  We report here the explicit expressions for the first two
perturbative coefficients of the hard function:
\begin{align}
	H^{(0)}(x) \aligneq&\;  N_c \, Q_q^4 \, \bigg( \frac{1}{x}  + \frac{1}{1-x} - 2 \bigg) \, , \\
	H^{(1)}(s,x,\mu) \aligneq&\; (N_c^2-1) \, Q_q^4\, \bigg[\bigg(\frac{1}{x} + \frac{1}{1-x} -2 \bigg) \bigg(- \ln^2\big(s/\mu^2\big) + 3\, \ln\big(s/\mu^2\big) - 7 + \frac{7\pi^2}{6}  \bigg)  \, \nonumber \\
	&\; + \bigg( \frac{2}{x} + \frac{1}{1-x} -1 \bigg)\ln^2(1-x) + \bigg(\frac{1}{x} + \frac{2}{1-x} -1 \bigg)\ln^2(x)\, \nonumber \\
	&\; + \bigg(\frac{3}{1-x} -1 \bigg)\ln(1-x)  + \bigg(\frac{3}{x} -1 \bigg)\ln(x) \bigg] \, ,
\end{align}
where $Q_q$ is the electric charge of the active quark.  Unfortunately,
the NNLO hard function is too lengthy to be included here. The result
is available upon request to the authors.

\newpage
\bibliographystyle{JHEP}
\bibliography{geneva}

\end{document}